\documentclass[traditabstract]{aa} 
\usepackage{graphicx,xcolor,ulem}
\usepackage{lscape}
\usepackage{units,natbib}
\usepackage{relsize}
\usepackage{amsmath}
\usepackage{txfonts}
\begin{document}

\newcommand{\com}[1]{\textbf{\color{blue}{#1}}}
\newcommand{\edit}[2]{\textbf{\color{red} {#1}}}
\newcommand{\so}[1]{\textbf{\color{red} \sout{#1}}}
\newcommand{\vs}{{\it vs\ }}
\newcommand{\kms}{km~s$^{-1}$}
\newcommand{\Msun}{M$_{\odot}$}
\newcommand{\msun}{M_{\odot}}
\newcommand{\myr}{M$_{\odot}$\,yr$^{-1}$}
\newcommand{\Teff}{$T_{\rm eff}$}
\newcommand{\FeH}{[Fe/H]}

\title {Li-rich K giants, dust excess, and binarity\thanks{Based on observations made with the Mercator Telescope, operated on the island of La Palma by the Flemish Community, at the Spanish Observatorio del Roque de los Muchachos of the Instituto de Astrofísica de Canarias.}$^,$\thanks{Individual radial velocities are only available in electronic form
at the CDS via anonymous ftp to cdsarc.u-strasbg.fr (130.79.128.5)
or via http://cdsweb.u-strasbg.fr/cgi-bin/qcat?J/A+A/}}

   \author{A. Jorissen
          \inst{1}
          \and H. Van Winckel
          \inst{2}
          \and L. Siess
          \inst{1}
          \and A. Escorza
          \inst{1,2}
          \and D. Pourbaix
          \inst{1}
          \and S. Van Eck
          \inst{1}
  }

  \institute{Institut d'Astronomie et d'Astrophysique, Universit\'e Libre
    de Bruxelles, ULB, CP. 226, Boulevard du Triomphe, 1050 Brussels, Belgium \\
    \email{ajorisse@ulb.ac.be}
    \and
    Instituut voor Sterrenkunde, KU Leuven, Celestijnenlaan 200D bus 2401, B-3001 Leuven, Belgium}

\date{Received ...; accepted ...}

\abstract {
The origin of the Li-rich K giants is still highly debated. Here, we investigate the incidence of binarity among this family from a nine-year radial-velocity monitoring of 
a sample of 11 Li-rich K giants using the HERMES spectrograph attached to the 1.2m Mercator telescope.  A sample of 13 non-Li-rich giants (8 of them being surrounded by dust according to IRAS, WISE, and ISO data) was monitored alongside.

When compared to the binary frequency in a reference sample of 190 K giants  (containing 17.4\% of definite spectroscopic binaries -- SB --  and  6.3\% of  possible spectroscopic binaries -- SB?), the binary frequency appears normal among the Li-rich giants (2/11 definite binaries plus 2 possible binaries, or 18.2\% SB + 18.2\% SB?), after taking account of the small sample size through the hypergeometric probability distribution. Therefore, there appears to be no causal relationship between Li enrichment and binarity. 
Moreover, there is no correlation between Li enrichment and the presence of circumstellar dust,
and the only correlation that could be found between Li enrichment and rapid rotation is that the most Li-enriched K giants appear to be fast-rotating stars.
However, among the dusty K giants, the binary frequency is much higher (4/8 definite binaries plus 1 possible binary). The remaining 3 dusty K giants suffer from a radial-velocity jitter, as is expected for the most luminous K giants, which these are. 
 }


\maketitle

%

\section{Introduction}
\label{Sect:intro}

The first Li-rich K giant was discovered by \citet{Wallerstein1982}, rapidly followed by many others \citep[see][and references therein]{BharatKumar2018}. Recently, large surveys like Gaia-ESO, LAMOST, and GALAH further increased the number of known Li-rich K giants \citep{Casey2016,Casey2019,Smiljanic2018,Singh2019,Deepak2019}. 
As an illustration, the number and frequency of Li-rich stars (i.e. with $\log \epsilon({\rm Li}) \ga 1.5$ in a scale where $\log \epsilon({\rm H}) = 12$)
among G and K giants evolved from 10/644 \citep[1.5\%;][]{Brown1989}, 3/400 \citep[0.75\%;][]{Lebzelter2012}, 23/8535 \citep[0.27\%;][]{Martell2013}, 15/2000 \citep[0.75\%;][]{Kumar2011,Kumar2015}, 9/1175 \citep[0.76\%;][]{Casey2016}, 335/51982 \citep[0.64\%;][]{Deepak2019} up to a record high of 2330/305793 \citep[0.76\%;][]{Casey2019}. It therefore seems that the frequency of Li-rich stars among G-K giants is about 0.7\%. Li-rich giants represent a puzzle in the framework of stellar evolution since Li is predicted to disappear when the star ascends the red giant branch (RGB). This decrease in Li abundance on the RGB results from Li depletion by nuclear burning during the pre-main sequence and main sequence phases, and its subsequent dilution by the first dredge-up during the ascent of the RGB. Moreover, thermohaline mixing causes additional Li destruction at the RGB bump \citep{Charbonnel2010,Angelou2015,Lattanzio2015,
Charbonnel2019}. An alternative model for Li enrichment at the RGB bump was proposed by \citet{Denissenkov2004} and suggests that rapid rotation (due to the presence of a stellar companion and the resulting tidal locking) could trigger extra-mixing leading to Li production. \citet{Casey2019} put forward the same scenario without restricting its operation solely to the RGB bump.

As a result, starting from the present interstellar-medium abundance of $\log \epsilon({\rm Li}) = 3.3$, the Li abundance 
after the first dredge-up is expected to be lower than about 1.5 to 1.8 in Population~I K giants. However, some K giants do not conform to these predictions, and some of these Li-rich K giants have Li abundances  that are even larger than the present interstellar-medium value \citep[e.g.][]{daSilva1995,deLaReza1995,deLaReza1996,Balachandran2000,
Ruchti2011,Casey2016,Takeda2017,BharatKumar2018}.

A number of 
K giants also exhibit  rapid
rotation rates that some authors \citep[e.g.][]{deMedeiros1996,deMedeiros2000,Drake2002,Carlberg2012,Smiljanic2018,Charbonnel2019} claimed to be correlated with high lithium
abundances. This is especially well illustrated by Fig.~3 of \citet{deMedeiros2000} which reveals that among 20 Li-rich stars ($\log \epsilon({\rm Li}) \ge 2.0$), 15 are rapid rotators ($V_{\rm rot} \sin i \ge 5$~\kms). The situation is even more extreme for the 14 stars with $\log \epsilon({\rm Li}) \ge 2.5$, among which 12 are rapid rotators.

At about the same time, K giants with IR excesses were reported \citep{Zuckerman1995,Plets1997}, based on data from the {\it Infrared Astronomy Satellite} \citep[IRAS;][]
{Neugebauer1984}, which surveyed the sky at 12, 25, 60, and
100~$\mu$m. 
Infrared excess is not expected in
giant stars prior to the late asymptotic giant branch (AGB) phase. 
\citet{Gregorio-Hetem1993}, \citet{deLaReza1997}, \citet{Castilho1998}, \citet{Fekel1998},  \citet{Jasniewicz1999}, \citet{Reddy2002}, and \citet{Reddy2005} suggested that 
several of the Li-rich K giants are surrounded by dust shells as they exhibit IR excesses, sometimes starting at 12~$\mu$m and sometimes at much longer wavelengths (60~$\mu$m). These early studies were based on IRAS data.

Various hypotheses have been proposed to explain the
combination of high Li abundances, rapid rotation rates, and IR excesses,
including the accretion of giant planets  \citep[e.g.][]{Siess1999,Casey2016}, or a sudden transport of matter that could eject a dusty shell \citep{delaReza2015}.
In both cases, based on
dust shell evolutionary models, the IR excess and a large amount of Li in RGB stars
should be transient phenomena that would last for a few $10^4$  years. Observations suggest that, if dust
shell production is a common by-product of Li enrichment mechanisms, the IR excess
stage should be very short-lived \citep{Rebull2015,delaReza2015}.
Indeed, even the early studies investigating the possible correlation between large Li abundances and dust excesses did not deliver strong evidence on a purely statistical basis. For instance,
\citet{Fekel1998} found 6
giants with greater-than-typical lithium abundances out of 39
giants with IR excess, which they
point out is similar to the fraction of stars with enhanced Li found in normal field giants. \citet{Jasniewicz1999} identified 8 Li-rich
stars out of 29 stars with IR excesses, but no correlation between Li abundance and IR
excess. 
\citet{Lebzelter2012}  report on 3 Li-rich giants (out of
more than 400 studied), none of which have IR excesses
suggestive of mass loss.  

The spatial resolution of IRAS was relatively low (a few arcminutes), making the identification of the optical counterpart difficult, especially in regions with
high source density or high background due to the  so-called IR cirruses \citep{Jura1999a}. Background galaxies with a large amount of interstellar material \citep[like hyperluminous IR galaxies;][]{Rowan-Robinson2000} could also be responsible for an apparent IR excess if they happen to lie along the same line of sight as the target star \citep[see for instance the case of HD~24124 described by][]{Kim2001}.

To clear these ambiguities, the reality of the IR excesses around (Li-rich) K giants was  later re-examined \citep[][]{Kim2001,Kumar2015,Rebull2015} with much better quality data from either the {\it Infrared Space Observatory} \citep[ISO;][]{Kessler1996} or the {\it Wide-field
Infrared Survey Explorer} (WISE) spacecraft, delivering four IR magnitudes  \citep[at 3.35, 4.6, 11.6, and 21.1~$\mu$m;][]{Wright2010,Cutri2013} at a higher spatial resolution and with a better sensitivity than IRAS. The  correlation between Li  richness and dust excess remained weak, at least for these comparatively short IR bands.   
\citet{Kumar2015} for instance report on a
search for IR excesses (combining WISE and IRAS data) in 2000 K-type giants. None of the far-IR
excess sources studied by these latter authors are lithium-rich, and of the 40 Li-rich sources, only seven show IR excess.
\citet{Rebull2015} performed a similar study and conclude that, intriguingly,  the largest IR excesses
all appear in Li-rich K giants, though very few Li-rich K giants have IR excesses (large or small). According to \citet{Rebull2015}, these largest IR excesses
also tend to be found in the fastest rotators, and there is no correlation of IR excess with the carbon isotopic ratio $^{12}$C/$^{13}$C.

Since fast rotation in Li-rich giants could be caused by either planet ingestion or synchronisation within a binary system, a search for either stellar or planetary companions around  these giants would be worthwhile. 
Similarly, a link between dust excess and binarity has been established very convincingly for post-AGB   
 \citep{VanWinckel2003} and post-RGB \citep{Kamath2016} stars.
It would therefore be of interest to check whether this correlation also holds for K giants with a dust excess.
Strangely enough, there has been no convincing estimate so far of either the binary frequency among Li-rich K giants or among dusty K giants, since the studies of  \citet{Ruchti2011} and \citet{Fekel1998}, respectively, relied on very scarce data. \citet{deMedeiros1996} claimed that their CORAVEL monitoring of Li-rich giants did not reveal any abnormal frequency of binaries, but they did not substantiate their claim by publishing the individual radial velocities (RVs).

In the present study, we therefore aim to check whether or not dust-rich K giants and Li-rich K giants have a  higher-than-normal binary frequency.
The paper is organised as follows. The studied samples are described in Sect.~\ref{Sect:sample}, along with the main properties of their stellar members (spectral energy distribution (SED) and location in the Hertzsprung--Russell (HR) diagram). The RV monitoring is presented in Sect.~\ref{Sect:RV}. Orbital elements and the binary frequency are presented in Sects.~\ref{Sect:orbits} and \ref{Sect:binary}, respectively, and conclusions are drawn in Sect.~\ref{Sect:conclusion}.

\setlength{\tabcolsep}{5pt}
\begin{table}
\caption[]{\label{Tab:sample}
The stellar sample S1. Stars with a confirmed IR excess appear in bold face. The column labelled $T_{\rm dust}$ lists the dust temperature derived by \citet{Kim2001} from their ISO data and based on a detached-shell model, except for HDE~233517 for which this information comes from the analysis by \citet{Jura2003} under the assumption of a flared disk. The column labelled `NLTE' indicates whether NLTE corrections have been applied to the Li abundance, whereas column $V_{\rm rot} \sin i$ lists a proxy for the rotational velocity (see text).
}
\begin{tabular}{lllllllllllllll}
\hline\\
HD & $\log \epsilon({\rm Li})$ & NLTE   & $V_{\rm rot} \sin i$ & \multicolumn{2}{c}{IR excess} & \multicolumn{1}{c}{$T_{\rm dust}$}\\
\cline{5-6} && &(km/s) & short & far & (K)\\
   \hline\\
   \noalign{\hfill Li-rich K giants\hfill \phantom{0}}\\
   \hline\\
 787    & 1.99 &  y & 1.5  &   n & n \\
 6665   & 2.93 & y & 2.3 &  n & -\\ 
 9746   & 3.44 & y & 5.5 &    n & -\\ 
 {\bf 30834} & 1.98 & y &  0. &   n &y$^f$\\ 
39853 & 2.75 & y  & 1.2 &   n & n \\
 40827 & 2.05 & y & 0. &          n & -\\ 
 63798 & 2.00 & y & 1.2 &  n & - \\ 
 90633  & 2.18 & y & 0. &   n &-\\ 
 112127 & 2.95 & y &0. &    n & - \\ 
 116292 & 1.65 & y & 0. &    n & - \\
 {\bf 233517}   & 3.95 & y & 19.7 & y & y & 70$^g$\\ 
   \hline\\
   \noalign{\hfill non-Li-rich K giants\hfill \phantom{0}}\\
   \hline\\
6       & 0.37$^c$  & n         & 0. &  n & n \\
 {\bf 3627}  & 0.20$^c$ & n & 0.  &      n & y$^f$ \\
 21078  & 1.30 & n &0. &  n & -\\  
 27497 & $<0.40$ & n & 0. &   n & -\\ 
 31553 & 0.5$^b$ & n & 0.       &   n & -  \\
 34043 &$ <-0.43$ & n & 0. &   n & -  \\
 43827 & $<0.00$ & n & 0. &   n & -\\
 108471 & - & - & 0. &   n & - \\ 
 {\bf 119853} & $<0.4^b$ & n & 0. &   n & y$^d$ & 68\\ 
{\bf 153687} & 0.10 & n & 1.2 &   n & y$^{d,f}$ & 74\\ 
{\bf 156115} & $<-1.0^b$ & n &  1.2 &  n & y$^d$ & 38\\
{\bf 212320}  & 0.87$^a$ & y & 0.8 &    n & y$^d$ & 46\\
{\bf 221776}  & $<-1^b$ & n & 1.0 &    y & y$^d$ & 59
\medskip\\
\hline\\
\noalign{$^a$ \citet{Liu2014}; $^b$ \citet{Fekel1998}; $^c$ \citet{Luck2015};  
$^d$ excess confirmed by ISO data  \citep{Kim2001};
$^f$ excess confirmed by AKARI data\citep{Murakami2007}; $^g$ Temperature at the outer  edge of a flared  disk \citep{Jura2003}.
}

\end{tabular}
\end{table}

\begin{table*}
\caption{\label{Tab:parallax}
Fundamental parameters of the S1 stars. The column labelled  $T_{\rm colour} (V-K)_0$ has been used to draw the HR diagram of Fig.~\protect\ref{Fig:HRD}, $E_{B-V}$ is the reddening obtained from  the SED fits presented in Figs.~\ref{Fig:SED3627} -- \ref{Fig:SED233517}.}
\begin{tabular}{rrlllcllrlccrllllll}
\hline\\
HD      &       \multicolumn{1}{c}{$\varpi$}  &$V$ & $K$ & $E_{B-V}$ & $(V-K)_0$ &       $BC_K$ &        \multicolumn{1}{c}{$M_{K_0}$} & $M_{\rm bol}$ & \multicolumn{1}{c}{$\log(L/L_{\odot}$) }& $    T_{\rm colour}$ & $T_{\rm eff}$         & [Fe/H] & Ref. & $\log g$\\
       &        (Gaia DR2)                                              & & & & & & & & &$(V-K)_0$ &(lit.) &\multicolumn{1}{c}{(lit.)}  & & (HRD)\\
\hline\\ 
6 &  $6.85\pm0.10$ &    6.31    &3.91   &0.02 & 2.34    &2.0 &  -1.92 &  0.08    & 1.86  & 4751& 4690&-0.03&     Ta13            & 2.52  \\      
787     & $5.08\pm0.22$ & 5.29 &1.86 && 3.43    & 2.5&-4.61     &       -2.11   &2.74 &       4021    & 4181& &Bl98 \\
3627 &  $30.86\pm0.05$ &  3.28  & 0.47  &0.06 & 2.66    &2.4&    -2.10  &          0.30   &        1.78 &         4488 &  4360&0.04&      McW90 &  2.20 \\                 
6665    &       $2.85\pm0.06$ & 8.44    &5.54   &&2.90& 2.3     & -2.18 &         0.17    &       1.83    & 4318& 4700&&  Ku11 \\
9746  & $6.35\pm0.09$&  6.22    &2.91   &&3.31  & 2.4   & -3.08&        -0.63   &       2.15    &4080   &4425   &&Br89\\ 
21078  &        $12.50\pm0.11$ &        7.98    &5.83   &&2.15& 1.9     &\phantom{-}1.31        & 3.21    & 0.61  &4936   &5068   &&Fe98 \\
27497  & $8.54\pm0.09$& 5.77    &3.81   &&1.96& 1.9&    -1.53&   0.37   & 1.75    &5141   &5180& +0.14&   He07\\ 
30834  & $6.33\pm0.27$& 4.79    &1.39   &0.11 & 3.10&   2.5&-4.64       & -2.14&          2.75&   4194    &4130   &-0.37&McW90            &1.38   \\      
31553  & $6.18\pm0.31$ &        5.82    &3.02   &&2.80  &2.3    &-3.02  & -0.72&   2.19&  4386    &4731   &&Fe98\\ 
34043 &         $5.49\pm0.25$   & 5.50  & 2.54&&        2.96    &2.4    &       -3.76   & -1.36   & 2.44  &4279   &4300   &&Fe98\\ 
39853 & $4.82\pm0.19$ & 5.64    &1.98   &&3.66  &2.6    &-4.61  & -2.01 & 2.70    &3919   &3900   &&Gr89\\ 
40827 & $7.29\pm0.04$&  6.32    &4.15   &&2.17  &1.9&   -1.54   & 0.36  & 1.75    &4916   &4575   &&Br89\\ 
43827&  $5.70\pm0.13$&  5.16    &2.26   &&2.90  &2.3    & -3.96 & -1.61 & 2.54    &4318   &4415   &&Fe98\\ 
63798  &$5.26\pm0.04$&  6.50&   4.49&&  2.01    &1.9    & -1.90 & 0.00  & 1.90    & 5085  &5000   &&Mi06\\ 
90633&$8.82\pm0.02$&    6.32    &4.01   &&2.31& 2.0     & -1.26 & 0.74  & 1.60    &4781   &4600   &&Mi06\\ 
108741& $2.01\pm0.05$&  9.49    &6.93   &&2.56  &2.1    &       -1.56   &       0.54    &       1.68    &4562\\         
112127  &$7.79\pm0.04$ &        6.88&   4.12    &&2.76& 2.5$^a$ & -1.42  &       1.08    &       1.46    &4414   &4340   &&Br89\\ 
116292 &        $11.10\pm0.11$ &        5.37    &3.12   &&2.25  &2.0&    -1.65   &       0.35     &      1.76    &4837&  4940&-0.07&     He07\\ 
119853& $9.38\pm0.23$ & 5.50 & 3.38&0.00 &2.12  &1.9&    -1.76 &                  -0.14 &                 1.84 &                 4967&   5136    &-&Fe98         & 2.68  \\              
153687 &        $9.31\pm0.17$ &4.83&    1.27    &0.10&3.29      &2.6    & -3.92&   -1.32  &       2.42&   4088    &3980   &-0.12&McW90    &1.37   \\      
156115 &        $3.87\pm0.11$&  6.58    &1.69   &0.42&3.73      &2.8    & -5.52   & -2.72 &       2.98    &3890&  3547    &-&Fe98 &0.67 \\        
212320 &        $7.00\pm0.10$   &5.92   &3.89&0.17      &1.55   &1.9    & -1.95   & -0.05 & 1.92  &5080$^b$&      5030    &-0.27&He07             &2.80         \\      
221776 & $4.03\pm0.08$  &6.20   &2.07&0.10      &3.87&  2.7     & -4.94 & -2.24&   2.79   &3838&  3964    & - &Fe98       &1.09   \\      
233517& $1.14\pm0.06$&  9.71    &6.64&0.03      &3.00   &2.4    &       -3.09   &       -0.70   &       2.18    &4251&  4475&-0.37&     Ba00                 &1.72   \\              
\hline
\end{tabular}

Notes:  a: HD 112127 is a carbon star; hence $BC_K$  from $J-K$ and Eq.~1 of \citet{Kerschbaum2010}.\\
b: This value of the colour temperature  is obtained from the non-dereddened $V-K$ index, as the dereddened index would yield instead an unrealistically high temperature of 5660~K. 

References: Ba00: \citet{Balachandran2000}, Bl98: \citet{Blackwell1998}, Br89: \citet{Brown1989}, Fe98: \citet{Fekel1998},  He07: \citet{Hekker2007}, Ku11: \citet{Kumar2011},  McW90: \citet{McWilliam1990}, Mi06: \citet{Mishenina2006}, Ta13:  \citet{Tautvaisiene2013}

\end{table*}

\section{The stellar samples and their properties}
\label{Sect:sample}

\subsection{Sample synopsis}
\label{Sect:synopsis}

Three samples are considered in this paper. A small sample of 11 Li-rich and 13 non-Li-rich K giants (hereafter sample S1)  extracted from \citet{Charbonnel2000} and \citet{Kumar2011} is described in Sect.~\ref{Sect:sample1} (Tables~\ref{Tab:sample} and \ref{Tab:parallax}). The RVs for these stars were extensively  monitored with the HERMES spectrograph \citep{Raskin2011} for 10 years, starting in April 2009 (Sect.~\ref{Sect:RV}). Sample S1, which can also be divided into 8 dusty and 16 non-dusty K giants, served to test the possible correlation between binarity and IR excess due to dust.

A second, more extended sample of 56 Li-rich giants (hereafter sample S2) was collected from the work of 
\citet{BharatKumar2015}
and \citet{Charbonnel2019} to infer the binary frequency from the Gaia DR2 RV standard deviations (Sect.~\ref{Sect:methodology}). Seven stars have been excluded from the samples of these latter authors because they have no RV available in Gaia DR2 (namely HD~35410, HD~71129, HD~86634,  HD~138289, HD~138905, HD~183912, and HD~188114). There are 8 stars in common between samples S1 and S2. The properties of this sample are presented in Sect.~\ref{Sect:sampleS2} and Table~\ref{Tab:sample2}.

Finally, sample R contains 190 K giants selected from the Kepler data set (Sect.~\ref{Sect:reference}) and monitored with the HERMES spectrograph since April 2016 (Table~\ref{Tab:binary_R}). Here, it serves as a reference sample against which the binary frequency of sample S1 can be compared.

\subsection{HERMES sample S1}
\label{Sect:sample1}

\subsubsection{Selection and spectral energy distributions}
\label{Sect:SED}

The sample of 11 Li-rich giants was selected from the list of \citet{Charbonnel2000}, with two additions from \citet{Kumar2011}, HD~63798 and HD~90633. These Li-rich stars are listed in the first part of Table~\ref{Tab:sample}, and are a mixture of stars on the RGB (some close to the bump), in the red clump and along the early AGB (as shown in Sect.~\ref{Sect:HRD}). In this sample, HD~112127 has sometimes been classified as a carbon star of type  R   \citep{Barnbaum1996}.
Table~\ref{Tab:sample} also provides a proxy for the rotational velocity  $V_{\rm rot} \sin i$. It is equal to $(\sigma^2_{\rm CCF} - \sigma^2_0)^{1/2}$, where $\sigma_{\rm CCF}$ is  the width of a Gaussian fitted to the cross-correlation function used to derive the RV (Sect.~\ref{Sect:RV}), and $\sigma_0 = 3.5$~\kms\  is the instrumental width  corresponding to the resolution of 86\ts000 for the HERMES spectrograph \citep{Raskin2011}.  There is a clear tendency for the most Li-enriched K giants to be the fastest rotating stars, as already pointed out by \citet{Drake2002} and \citet{Smiljanic2018}. 

On top of the Li-rich K giants, we selected 13 stars from 
the list of K giants with suspected IR excesses (as initially inferred from the IRAS data) from  \citet{Fekel1998}, itself being  a subsample of \citet{Zuckerman1995}, who list the corresponding  IRAS fluxes (their Table~1). 
These K giants with a suspected IR excess are listed in the second part of Table~\ref{Tab:sample}. Because an IR excess has in the end not been confirmed for all of them,  this second part of the sample is denoted by `non-Li-rich stars', so as to avoid any ambiguity in the terminology.

Given the difficulties in establishing firm IR excesses using IRAS data alone (as explained in Sect.~\ref{Sect:intro}), we sought confirmation from other IR data, namely WISE, AKARI  \citep{Murakami2007} and ISO \citep{Kim2001}. 
To this end, we built the spectral energy distributions (SEDs) using the   tool described by \citet{Escorza2017}
to retrieve broadband photometry from Simbad, and find the extinction on the line of sight (see more on this below). The IRAS fluxes listed as upper limits were removed from the SED, but the reality of any dust excess in cases like HD~119853 (Fig.~\ref{Fig:SED119853}) or HD~156115 (Fig.~\ref{Fig:SED156115}) with just one (firm, i.e. non-upper limit) IRAS band (marginally) deviating from the MARCS model would still be questionable were it not confirmed by the ISO data of  \citet{Kim2001}.
For the sake of concision, Figs.~\ref{Fig:SED3627} -- \ref{Fig:SED233517} present only those SEDs exhibiting a confirmed IR excess, whereas Fig.~\ref{Fig:SED6} presents one SED without IR excess for comparison (HD~6). These plots show the best-fitting reddened MARCS model \citep[red curves;][]{Gustafsson2008} on top of the observed photometry, while the unreddened MARCS model is displayed in black. Since the fitting process is ill-behaved (because of a strong coupling between the effective temperature $T_{\rm eff}$ and the reddening $E_{B-V}$), it was necessary to obtain a first guess of $T_{\rm eff}$  from the literature and listed in Table~\ref{Tab:parallax} (a more detailed description of the contents of this Table is given in Sect.~\ref{Sect:HRD}). Moreover, fixing the gravity $\log g$ also helped to obtain reasonable fits. A first estimate of the gravity was obtained from the location of  the star in the HR diagram (see Sect.~\ref{Sect:HRD}), before applying the de-reddening correction.

This way, a satisfactory SED fit could be obtained, often with the minimum possible reddening (listed as well in Table~\ref{Tab:parallax}), and with a de-reddened $(V-K)_0$ colour temperature \citep[derived from the reddening ][]{Cardelli1989,Bessel1998} in good agreement with the literature $T_{\rm eff}$ value.
An exception is the  star HD~212320 (HIP~110532) whose de-reddened $(V-K)_0$ index of 1.55 yields an uncomfortably warm temperature of 5660~K, as compared to literature estimates of 5030~K \citep{Hekker2007} or 4825~K \citep{Stock2018}. Therefore, in the remainder of this paper, we adopt for HD~212320 the non-dereddened colour temperature of 5080~K, in better  agreement with the spectroscopic temperatures.

Those stars with a confirmed IR excess are listed in bold face in Table~\ref{Tab:sample}. We see that, among our sample of Li-rich giants, only two are simultaneously dust-rich (HD~30834 and HDE~233517). 

\subsubsection{Origin of the dust excess observed in sample S1}
\label{Sect:dust}

The IR excess observed in the dusty K giants is peculiar in that it involves very cool dust \citep[$T_{\rm dust} < 100$~K according to the model of][and \citealt{Kim2001}]{Jura1999a}; the excess is often restricted to wavelengths longwards of 60~$\mu$m. Except for HDE~233517, where the IR excess starts at 10~$\mu$m (Fig.~\ref{Fig:SED233517}), dusty K giants strongly differ from mass-losing AGB stars where dust features are seen from 10~$\mu$m onwards \citep{Waters1999}. When mass loss stops, their dust-emission peak progressively moves towards longer wavelengths as the dust shell becomes detached and cools down. In non-binary post-AGB stars with an expanding dust shell, IR excesses may start from close to 10~$\mu$m up to about 20~$\mu$m \citep{VanWinckel2003,Gezer2015}.  When the star becomes surrounded by a (proto-)planetary nebula (PPN or PN), dust emission peaks around 30~$\mu$m \citep[e.g. PPN SAO~34504 and PN NGC~7027, where dust emission peaks at 30 and 33~$\mu$m, respectively;][]{Waters1999}.  For dusty K giants, where the dust emission peaks longwards of 60~$\mu$m, the dust shell is detached, indicating that the mass-loss process is no longer active. The dust temperatures 
derived by \citet{Kim2001} from their ISO data and based on the hypothesis of a detached shell are listed in Table~\ref{Tab:sample}.

For HDE~233517, the large inferred dust mass has been suggested to be either the result of the disintegration of comets \citep{ Jura1999a,Fisher2003} or of the ingestion of a large planetary body or low-mass star  \citep{Jura2003}. The presence of warm dust and of a flared disk inferred by  \citet{Jura2003} is reminiscent of the situation prevailing among binary post-AGB stars \citep{Kluska2019}. Although  our HERMES data do not flag  HDE~233517 as a binary system, Gaia DR2 data suggest the opposite (see Sect.~\ref{Sect:binary}).

\subsubsection{Dusty and Li-rich K giants of sample S1 in the Hertzsprung-Russell diagram}
\label{Sect:HRD}

\begin{figure}
    \includegraphics[width=9.5cm]{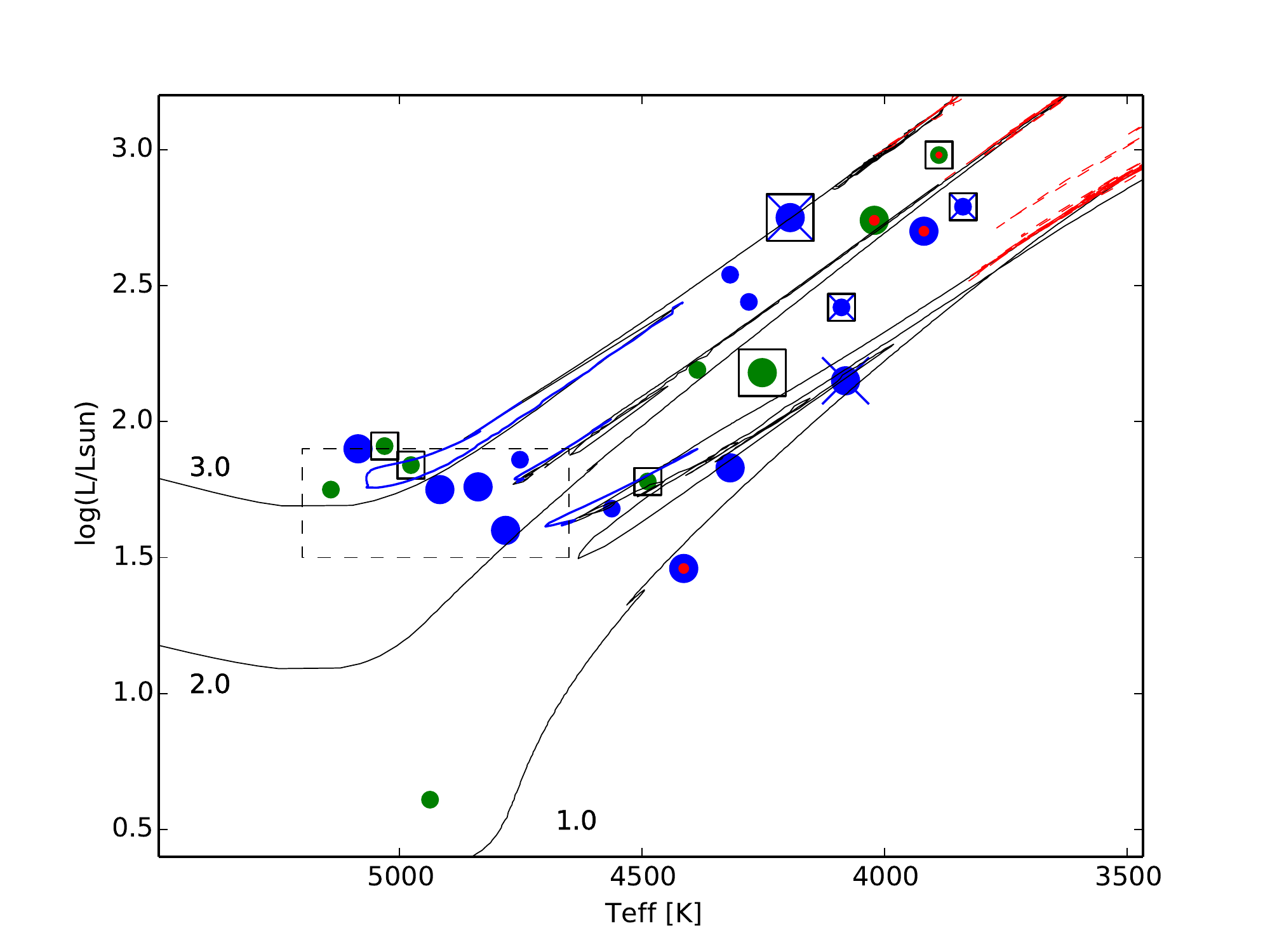}
    \caption{Li-rich (large filled dots) and dust-rich (open squares) K giants from sample S1 in the HR diagram. Small filled dots are non-Li-rich K giants. Green symbols denote SB, as discussed in Sect.~\ref{Sect:binary}. Small red dots indicate the presence of regular small-amplitude variations, whereas large crosses indicate irregular small-amplitude variations (RV jitter). STAREVOL tracks \protect\citep{Siess2000,Siess2008} are labelled according to their initial mass, and correspond to solar metallicity. Tracks evolving along the core He-burning phase  are depicted in blue, whereas tracks along the AGB are displayed in red. The red clump as defined by \protect\nocite{Deepak2019} Deepak \& Reddy (2019; their Fig.~1) is represented by the dashed rectangle. The bump is well visible on the 1~M$_\odot$ track as the hook feature located to the lower right of the red clump region.}
    \label{Fig:HRD}
\end{figure}

The HR diagram of the S1 stars was constructed in the following way. First, the colour temperature was computed from the de-reddenned $V-K$ index (with $V$ and $K$ taken from the Simbad database) using the \citet{Bessel1998} calibration (relation `abcd' in their Table~7). The colour excess $E_{\rm B-V}$ used to de-redden the  $V-K$ index is the one obtained from the SED fit (Table~\ref{Tab:parallax}), and the $E_{\rm V-K}/E_{\rm B-V}$ ratio was taken from  \citet{Cardelli1989}. The de-reddening correction (including both circumstellar and interstellar contributions) was computed only for those stars showing the presence of circumstellar dust. For the other stars, no interstellar de-reddening correction was applied since the stars are relatively nearby (closer than 350~pc, except for HDE~233517, for which the colour excess has been derived), meaning that the interstellar  reddening is considered negligible.  The bolometric correction for the $K$ band has been taken from \citet{Bessel1998} for  oxygen-rich stars, and from \citet{Kerschbaum2010} for the carbon star HD~112127 (their Eq.~1). The luminosity is then derived from the bolometric magnitude combined with the Gaia DR2 parallax \citep{Gaia2018}. There is no need to use Bayesian estimates for the distance because all the targets stars are close enough for the ratio $\varpi/\sigma_\varpi$ to be sufficiently large ($\ge 19$) to avoid biasing the distance when inverting the parallax.

For the sake of homogeneity, the temperature used to construct the HR diagram is the colour temperature, except for HD~212320 (for the reason discussed in Sect.~\ref{Sect:SED}) where $T_{\rm eff}$ is taken instead from \citet{Hekker2007}.

Before discussing the HR diagram thus constructed (Fig.~\ref{Fig:HRD}), we first compare in Table~\ref{Tab:comparison} the stellar parameters listed in Table~\ref{Tab:parallax} with their Bayesian estimates provided by \citet{Stock2018} for the three stars in common between the two samples, namely HD~27497 (= HIP 20268), HD~116292 (= HIP~65301) and HD~ 212320 (= HIP~110532). The luminosities are in good agreement, and the temperatures as well (except for HD 212320).

Evolutionary tracks for stars of solar metallicity (given the small  metallicity range of our target stars as revealed by Table~\ref{Tab:parallax}) from the STAREVOL code \citep{Siess2000,Siess2008} are displayed as well in Fig.~\ref{Fig:HRD}.

Some Li-rich K giants are located in the red clump (dashed rectangle in Fig.~\ref{Fig:HRD}), but several are also spread along the giant branch (either RGB or E-AGB), namely HD~787, HD~9746, HD~30834, HD~39853, and HD~233517.
The HR diagram of Fig.~\ref{Fig:HRD} is similar to that of \citet{Charbonnel2000} with its extension along the giant branch, and both contrast with the well-documented claim (based on asteroseismologic data) by \citet{Singh2019} that {all} Li-rich stars are restricted to the red clump. 
Nevertheless, \citet{Casey2019} and \citet{Deepak2019} disagree with that claim, quoting a frequency of 75 - 85\%  Li-rich stars located in the red clump (not 100\%).
The reason for this difference between \citet{Singh2019} and all other studies most likely resides in the fact that \citet{Singh2019} restricted their study to {super}-Li-rich stars with $\log \epsilon({\rm Li}) \ge 3.0$. As confirmed by all the other studies quoted above \citep[see especially Fig.~3b of][]{Deepak2019}, super-Li-rich stars are largely confined to the red clump. This leaves room then for a limited fraction of Li-rich stars (of the order of 15 - 25\%) that may still be found along the RGB.

\subsection{Reference sample R}
\label{Sect:reference}

\begin{figure}
\includegraphics[width=9cm]{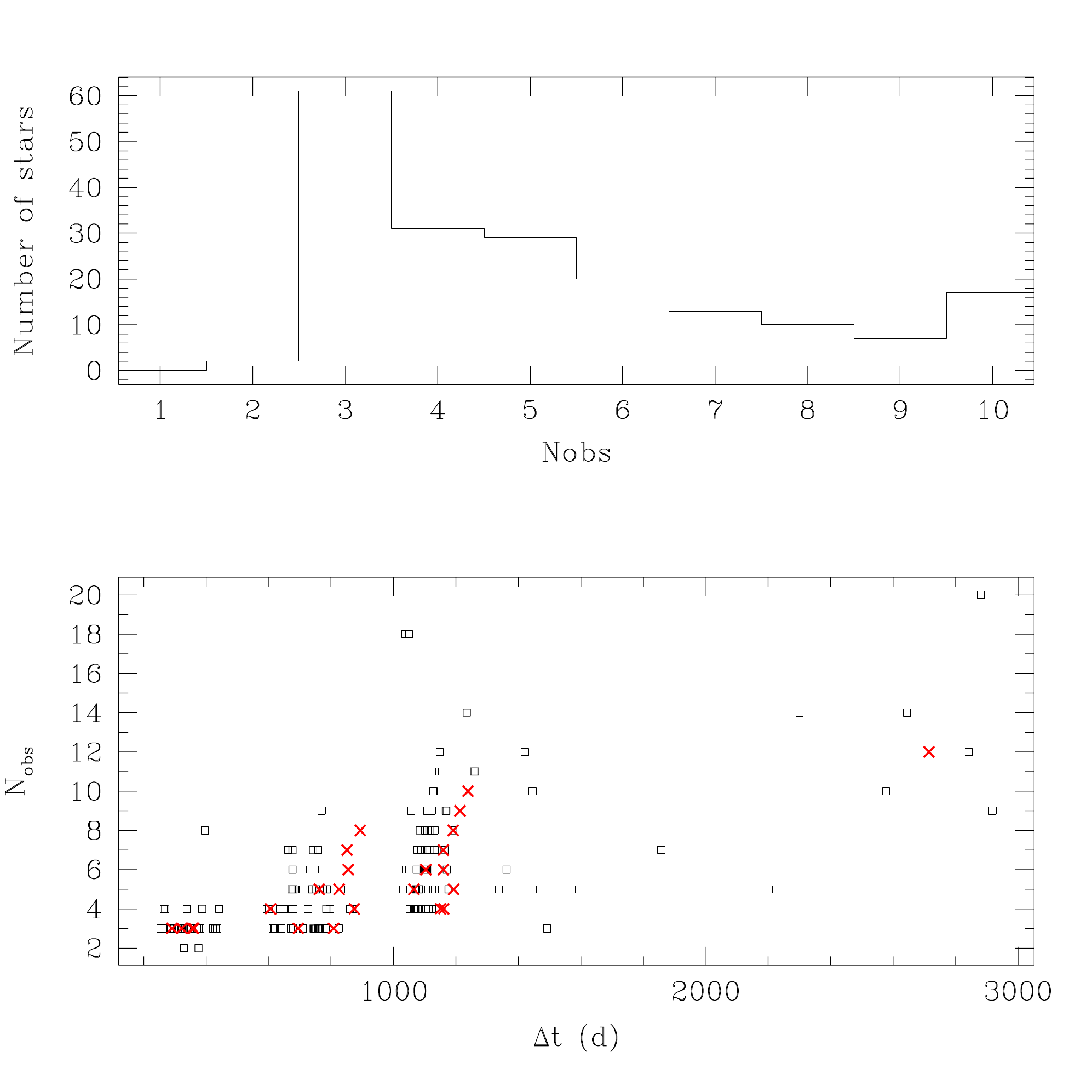}
\caption{\label{Fig:DeltatNobs}
Top panel: Distribution of the number of observations per star in sample R. Bottom panel: Number of HERMES observations vs. time span of the RV observations for the comparison sample of Kepler/CoRoT giants (sample R; open squares). Most of the stars have three observations spanning 300 to 900~d. Red crosses correspond to the re-sampled data of sample S1 (denoted `sample S1') with $N_{\rm obs}$ and $\Delta t$ modified to mimic the distribution of sample R (see Sect.~\ref{Sect:bin2}). 
}
\end{figure}

To evaluate whether or not the binary frequencies in our dusty and Li-rich samples of K giants are peculiar, it is necessary to first evaluate this frequency in a reference sample observed as well with the HERMES spectrograph. This reference sample (listed in Table~\ref{Tab:binary_R}; the full content of this table will be described in Sect.~\ref{Sect:methodology}) contains the 160 brightest K giants (i.e. with $V$ magnitudes in the range 8.6 to 11.4) observed with the Kepler satellite \citep{Koch2010}, and with known evolutionary status \citep[from][]{Mosser2014}, along with 30 K giants from the CoRoT\footnote{CoRoT (Convection, Rotation and planetary Transits) is a mini satellite developed by the French Space agency CNES in collaboration with the Science Programmes of ESA, Austria, Belgium, Brazil, Germany and Spain.}  
 \citep{Baglin2006} sample.  Figure~\ref{Fig:DeltatNobs} presents the number of HERMES observations versus time-span for this reference sample of Kepler/CoRoT giants (referred to as `sample R' below). The RV observations cover the period from April 2016 to August 2019. 
The original purpose of this HERMES observing program is to compare the evolutionary properties of K giants with their orbital properties. The sample will be fully described in a forthcoming paper especially devoted to that issue (Jorissen et al., in prep.). Here, we simply perform a first evaluation of the binary frequency (Sect.~\ref{Sect:bin2}) in sample R along the guidelines described in Sect.~\ref{Sect:methodology}.

\begin{table*}
\caption{\label{Tab:comparison}
Comparison between the stellar parameters derived in this study (Table~\ref{Tab:parallax}) with their Bayesian estimates from \citet{Stock2018}.
} 
\begin{tabular}{rrccccccccccc}
\hline
HD & HIP & &$T_{\rm colour}$ & $T_{\rm eff}$ & \multicolumn{2}{c}{$\log (L/L_\odot)$} & \multicolumn{2}{c}{Mass} & \multicolumn{2}{c}{$\log g$} \\
      &          &    & (K)  & (K) &      &     & (M$_\odot$) & ($M_\odot$)\\
      &          &    & (1)  & (2) & (1) & (2) & (1) & (2) & (1) & (2) \\
      \hline\\
27497 & 20268 & RGB & 5141 & 5134 & 1.75 & 1.78 & 3.0 & 2.64 & 3.0 & 2.88\\
           &            & HB   &  5141 & 5074 & 1.75 & 1.79& 3.0 & 2.24 &  3.0 & 2.77
\medskip\\
116292 & 65301 & RGB & 4837 & 4896 & 1.76 & 1.84 & 2.5 & 2.23 &2.77 &2.66\\
           &            & HB     &  4837 & 4908 & 1.76  & 1.84 & 2.0& 1.93 & 2.67& 2.61
\medskip\\  
212320 & 110532 & RGB &5660 & 4814 & 1.92 & 1.98 &- & 1.89&-&2.43\\
             &              & HB   &5660  & 4825 &1.92 &  1.96 &- & 1.64 &-&2.37
\medskip\\     
\hline  
\end{tabular}

Notes: (1): This work; (2): \citet{Stock2018}. There are two different estimates, one corresponding to the location on the RGB, and the other to the location on the Horizontal Branch (HB, or `red clump').
\end{table*}

\section{Radial-velocity observations}
\label{Sect:RV}

The RV monitoring of samples S1 and R  was performed with
the HERMES spectrograph attached to the 1.2 m Mercator
telescope from the KU Leuven installed
at the Roque de los Muchachos Observatory (La
Palma, Spain). The spectrograph is fully described in \citet{Raskin2011}. 
The fibre-fed HERMES spectrograph is designed
to be optimised both in stability as well as in efficiency, and
samples the whole optical range from 380 to 900~nm in one
shot, with a spectral resolution of about 86\ts000 for the high-resolution
science fibre. This fibre has a 2.5 arcsec aperture
on the sky and the high resolution is reached by mimicking
a narrow slit using a two-sliced image slicer.
The RV is measured by  cross-correlating the observed spectrum with a spectral mask constructed on an Arcturus spectrum \citep[e.g.][]{Jorissen2016}.
The HERMES/Mercator combination is precious because
it guarantees regular telescope time. This is needed
for our monitoring programme and the operational agreement
reached by all consortium partners (KU Leuven, Universit\'e
libre de Bruxelles, Royal Observatory of Belgium,
Landessternwarnte Tautenburg) is optimised to allow efficient
long-term monitoring, which is mandatory for this
programme. On average, 250 nights per year are available
for the monitoring (spread equally on HERMES-consortium
time, and on KU Leuven time), and the observation sampling
is adapted to the known variation timescale. During
these nights, about 300 target stars are monitored, addressing
several science cases \citep{Gorlova2013} and delivering significant results concerning different families of long-period binaries such as Ba and CH stars \citep[e.g.][]{Jorissen2016,Jorissen2019, Escorza2019}, post-AGB binaries \citep[e.g.][]{Manick2017,Oomen2018}, or sub-dwarf B stars \citep[e.g.][]{Vos2015}.
Most importantly, the long-term stability of the RVs has been assessed from the monitoring of RV standard stars from the list of Udry et al. (1999).
From this sample of RV standard stars, the long-term velocity stability is estimated as 55~m/s \citep[see][for details; also Sect.~\ref{Sect:methodology}]{Jorissen2016}. 

The RV monitoring of sample S1 began in April 2009, with the regular science operations
of HERMES. The individual RVs are only available online at the CDS, Strasbourg. 
 The RV curves for all target stars are presented in Figs.~\ref{Fig:6} -- \ref{Fig:233517}. Some clearly allow an orbit to be computed, as done in Sect.~\ref{Sect:orbits}.

\section{Orbital elements}
\label{Sect:orbits}

The available orbital solutions in sample~S1 are displayed in Figs.~\ref{Fig:Orb_787} -- \ref{Fig:Orb_212320}, and the orbital elements are listed in Table~\ref{Tab:orbits}. Notes about individual stars are listed in Appendix~\ref{Appendix:orbits}.

The dispersions of the $O-C$ residuals are listed in Table~\ref{Tab:binary}, and they are all compatible with the HERMES accuracy, except for HD~156115, for which they amount to 0.38~\kms. A look at Fig.~\ref{Fig:Orb_156115} reveals that the large residuals are caused by an oscillation superimposed on the long-term orbit. A similar behaviour is observed for HD~787 as well (Fig.~\ref{Fig:Orb_787}), even though it is not apparent in the value of the $O-C$ dispersion (Table~\ref{Tab:binary}). 
The resulting orbital solutions for the HD~787 and HD~156115 residuals are listed in Table~\ref{Tab:orbits}b under the heading Aa+Ab to distinguish them from the main AB orbit. 
The ratio of the outer to inner  periods is 8.3 and 7.6 for HD~787 and HD~156115, respectively. These ratios are compatible in principle with the usual requirement for orbital stability in a hierarchical triple system \citep[e.g.][]{Tokovinin2014}. The inner  periods are of the order of 1.5~years. For these small-amplitude variations to be due to orbital motion would require a (brown dwarf) companion with  a mass larger than 0.015~\Msun\  in the case of HD~156115 (adopting a mass of 1~\Msun\ for component A according to its location in the HR diagram; see Fig.~\ref{Fig:HRD}). The companion could be even less massive in the case of HD~787 ($\ge 0.005$~\Msun\ adopting a mass of 2.5~\Msun\ for component A), but the mass function is not well constrained (Table~\ref{Tab:orbits}b). The dispersion of the $O-C$ residuals for the Aa+Ab orbit of HD~156115 amounts to 0.26~\kms, indicating that the Keplerian solution is not of excellent quality, since that dispersion could not be lowered to values typical of the HERMES accuracy. For HD~787, the $O-C$ dispersion is 64~m~s$^{-1}$.  

Similarly small-amplitude variations are also found in HD~39854 (Table~\ref{Tab:orbits}b), and a circular Keplerian orbit with a semi-amplitude of 0.2~\kms\ could match them. As listed in Table~\ref{Tab:orbits}, the current data do not allow us to rule out one of two possible periods  (106 or 282~d).

We stress that, although not found for all long-period binary giants (see for example HD~3627, Fig.~\ref{Fig:Orb_3627}), the short-term oscillations reported above are found in many of them. They were reported as well  in the K giants HE~0017+0055 \citep{Jorissen2016b}, HE~1120-2122 and 
HD~76396 \citep{Jorissen2016}, and HD~175370 \citep{Hrudkova2017}.
Except for the latter star, there was no compelling evidence to reject the possibility that the small-amplitude, approximately one-year RV oscillations could be caused by stellar pulsation\footnote{However, the Kepler data for HD~175370 do not support the pulsation hypothesis, and for this reason the presence of a giant-planet companion around that star was privileged by \citet{Hrudkova2017}. See also \citet{Hatzes2018} for a recent discussion about the difficulty of distinguishing Jupiter-like planets from intrinsic pulsations in the RV variations of K giants.}. This hypothesis gains further support from the fact that all three stars in our sample which show such small-amplitude variations (namely HD~787, HD~39854, and HD~156115) are among the most luminous stars in our sample, with $\log L/L_{\odot} > 2.5$ (small red dots in Fig.~ \ref{Fig:HRD}).  Incidentally, the other stars at similarly large luminosities but not showing regular, short-amplitude variations exhibit instead irregular RV jitter, another signature of envelope instability (blue crosses in Fig.~ \ref{Fig:HRD}).

\begin{table*}
\renewcommand{\tabcolsep}{3pt}
    \caption{Orbital elements. Among the astrometric orbital elements, $a_0$ is the semi-major axis of the photocentric orbit.}
    \label{Tab:orbits}
       a. Spectroscopic elements\medskip\\
    \begin{tabular}{rrrrrrrrrrrrrrr}
    \hline\\
    HD & \multicolumn{1}{c}{$P$} & \multicolumn{1}{c}{$e$} & \multicolumn{1}{c}{$\omega$} & \multicolumn{1}{c}{$V_0$} & \multicolumn{1}{c}{$K$} & \multicolumn{1}{c}{$T$} & \multicolumn{1}{c}{$a_1 \sin i$} & \multicolumn{1}{c}{$f(M)$}\\
       & \multicolumn{1}{c}{(d)} & & \multicolumn{1}{c}{($^\circ$)} & \multicolumn{1}{c}{(\kms)} & \multicolumn{1}{c}{(\kms)} & (JD-2\ts400\ts000) & \multicolumn{1}{c}{(Gm)} & \multicolumn{1}{c}{(M$_\odot$)} \\
       \hline
787AB & $>4196$ & 0.57: &   &   &   &   &   &   & \\
3627 & 27807: & 0.5: &   &   &   &   &   &   \\
$^c$  & 20158 & 0.34 & 356 & -8.5 & 4.0 & 15568 & 1042.7 &  0.111\\
21078 & $263.697\pm0.007$ & $0.6379\pm0.0004$ & $214.61\pm0.06$ & $49.83\pm0.01$ & $19.94\pm0.02$ & $56216.99\pm0.02$ & $55.68\pm0.07$ & $0.0989\pm0.0003$ \\
27497 & $976.4\pm0.5$ & $0.292\pm0.003$ & $89.5\pm0.5$ & $4.07\pm0.01$ & $4.82\pm0.01$ & $58244\pm1$ & $62.0\pm0.3$ & $0.0099\pm0.0001$ \\
 $^d$ & $976.3\pm0.4$ & $0.297\pm0.007$ & $88.1\pm1.7$ & $4.63\pm0.03$ & $4.77\pm0.04$ & $53357.4\pm3.8$ & $61.1\pm0.5$ & $0.0096\pm0.00025$ \\
31553 & $3279\pm52$ & $0.56\pm0.01$ & $78.8\pm0.5$ & $-3.26\pm0.03$ &  $3.57\pm0.06$ & $55517\pm30$ & $133.6\pm5.6$ & $0.0088\pm0.0005$ \\
 156115AB & $4169\pm11$ & $0.469\pm0.002$ & $217.6\pm0.2$ & $-12.66\pm0.01$ & $7.17\pm0.02$ & $60524\pm11$ & $363\pm2$ & $0.110\pm0.001$ &        \\
212320 & $1691.9\pm1.7$ & $0.226\pm0.002$ & $57.7\pm0.6$ & $-3.49\pm0.01$ & $4.87\pm0.01$ & $58224\pm3$ & $110.4\pm0.3$ & $0.0187\pm0.0001$ \\
        \hline\\
    \end{tabular}
    
    b. Small-amplitude variations
      \medskip\\
        \begin{tabular}{rllrrrrrrrrrrrr}
    \hline\\
    HD & \multicolumn{1}{c}{$P$} & \multicolumn{1}{c}{$e$} & \multicolumn{1}{c}{$\omega$} & \multicolumn{1}{c}{$V_0$} & \multicolumn{1}{c}{$K$} & \multicolumn{1}{c}{$T$} & \multicolumn{1}{c}{$a_1 \sin i$} & \multicolumn{1}{c}{$f(M)$}\\
       & \multicolumn{1}{c}{(d)} & & \multicolumn{1}{c}{($^\circ$)} & \multicolumn{1}{c}{(\kms)} & \multicolumn{1}{c}{(\kms)} & (JD-2\ts400\ts000) & \multicolumn{1}{c}{(Gm)} & \multicolumn{1}{c}{(M$_\odot$)} \\
       \hline    
    787Aa+Ab & $508.4\pm4.8$ & $0.49\pm0.11$ & $100\pm18$ & $-0.009\pm0.008$ & $0.09\pm0.02$ & $57005\pm16$ & $0.53\pm0.16$ & $(2.3\pm1.6)\times10^{-8}$\\
    39853$^a$ & 106 & 0.0 & -- & 81.6 & 0.20 & &  0.29 & $0.9\times10^{-8}$ &\\
 & 282 & 0.0 & -- & 81.6 & 0.18 & &  0.71 & $1.7\times10^{-7}$ &\\
    156115Aa+Ab & $548.6\pm0.9$ & $<0.02$ & -- & $0.008\pm0.006$ & $0.40\pm0.01$ & $57082.8\pm8^b$ & $3.04\pm0.07$ & $(3.7\pm0.3)\times10^{-6}$ \\
 \hline\\
\end{tabular}

       c. Astrometric elements
       \medskip\\
\begin{tabular}{rrrrrrrrrrl}
\hline\\
HD & \multicolumn{1}{c}{$\Omega$} & \multicolumn{1}{c}{$i$} & \multicolumn{2}{c}{$a_0$} & \multicolumn{1}{c}{$\varpi$} & \multicolumn{1}{c}{$\mu_{\alpha} \cos \delta$} & \multicolumn{1}{c}{$\mu_{\delta}$} & $M_1$ & $M_2$ \\
\cline{4-5}
& \multicolumn{1}{c}{($^\circ$)} & 
\multicolumn{1}{c}{($^\circ$)} & \multicolumn{1}{c}{(mas)}  & \multicolumn{1}{c}{(Gm)} & \multicolumn{1}{c}{(mas)} & (mas yr$^{-1}$)  & (mas yr$^{-1}$) & (M$_{\odot}$)  & (M$_{\odot}$) \\
\hline\\
 21078 & $110\pm18$ & $91\pm13$ & $5.1\pm0.5$ & $57\pm11$ & $13.3\pm1.1$ & $216.0\pm0.8$ & $114.4\pm1.0$ & 1.2 & 0.71 & solution coherent with location in HRD\\
       &       &  &  & & & & & 1.5 & 0.81\\
       &       &  &  & & & & & 2.0 & 0.96\\
\hline
\end{tabular}

a) HD 39853: Keplerian orbits uncertain (pulsations instead?)\\
b) HD 156115Aa: epoch of maximum velocity\\
c) Orbit from SB9 catalogueue \citep{Pourbaix2004} and \citet{Bakos1976}\\
d) From \citet{Griffin2013}

\end{table*}

\begin{table*}
\renewcommand{\tabcolsep}{2pt}
\caption[]{Binary properties of sample S1 ordered by  increasing RV  standard deviation $\sigma_j(RV)$. Columns are as in Table~\ref{Tab:sample2}, and include the properties of the \citet{Famaey2005} data listed in columns labelled $RV_F$ (average RV) and $N$obsF (number of measurements). The error associated with $RV_F$ is the statistical error on the mean (i.e. $\sigma_F(RV)/\sqrt{N{\rm obsF}}$, where $\sigma_F(RV)$ is the standard deviation of the RV values). The column labelled `SB' marks the final decision about binarity from HERMES data, but also from \citet{Famaey2005}, Pulkovo \citep{Gontcharov2006}, \citet{Fekel1998}, and Gaia DR2 \citep{Gaia2018} data (see text). Values in bold face identify fulfilled binary criteria.
\label{Tab:binary}
}
{\fontsize{8}{10}\selectfont
\begin{tabular}{rrcrclllrrcrclllll}
\hline\\
HD &  $\sigma_j(RV)$ & $\sigma(O-C)$ & \multicolumn{1}{c}{$\overline{RV}_H$} & $N$obs$_H$ & $\Delta t$  & $\chi_j^2/\nu_j$ & Prob. & $F2_j$ & \multicolumn{1}{c}{$\overline{RV}_F$} &  $N$obs$_F$ & \multicolumn{1}{c}{$\overline{RV}_G$} & $|\overline{RV}_G - \overline{RV}_H|$ & $\epsilon(RV_G)$ & SB & Type \\ 
    & (\kms) & (\kms) & (\kms) & & (d) & &RV var & & \multicolumn{1}{c}{(\kms)} & &\multicolumn{1}{c}{(\kms)}&\multicolumn{1}{c}{(\kms)} &\multicolumn{1}{c}{(\kms)}\\
   \hline\\
   \noalign{\hfill Li-rich\hfill \phantom{0}}\\
   \hline\\
40827  & 0.05 & - &  31.55&  23 & 855 & 0.56 & 0.05 & -1.67 & $31.58\pm0.20$ & 2 & $31.87\pm0.18$  & 0.32 & 0.26 & N & Li\\
116292 & 0.06 & - & -25.97&  45 & 1193 & 0.81 & 0.19& -0.88 & - & - &  $-25.85\pm0.11$ & $0.12$ & $0.29$ & N & Li\\
6665   & 0.06 & - & -21.72&  29 & 1213 & 0.84& 0.30 & -0.53 & - & - &$-21.55\pm0.19$ & $0.17$&$0.27$ & N & Li \\
63798  & 0.07 & - &   8.28&  27 & 875 & 0.99 & 0.52 & 0.04 & - &  - & $8.38\pm0.15$ & $0.10$&$0.25$ & N & Li\\
90633  & 0.07 & - & -25.81&  89 &2716 & 1.00 & 0.52 & 0.05 &  $-26.08\pm0.20$ & 2 & $-25.78\pm0.13$ & $0.03$&$0.26$ & N & Li \\
112127 & 0.09 & - &   6.17&  89 &2967 & 1.80 & \bf {1.0} & 4.38 & $5.83\pm0.28$  & 3 & $5.76\pm0.14$ & $0.41$&$0.25$ & Y? & Li, long-term \\
          & &         &         &          &       &         &           &        &   &   & &    & & &velocity drift?\\
233517 & 0.10 & - &  46.71&  31 & 851 & 1.73 & 0.99 & 2.43 & - & - & $\bf {45.92\pm1.53}$  &  \bf {0.79}&0.31 &  Y & Li, dust\\
30834  & 0.12 & - & -17.09&  79 &9758 & 1.98 & \bf {1.0} & 4.84 &  $-17.24\pm0.08$ & 11 & $-17.27 \pm0.33$ & $0.15$&$0.37$ & N & Li, dust, RV jitter\\
9746   & 0.15 & - & -42.44&  63 &11159 & 2.09 & \bf {1.0} & 4.71 & $-42.64\pm0.34$ & 6 &  $-42.12\pm0.14$ & $0.32$&$0.28$ & N & Li, RV jitter\\
39853  & 0.30 & - &  81.62&  53 &2714& 18.9 & \bf{1.0} &25.5 &  - & - &  $81.50\pm0.19$ &  $0.12$&$0.32$ & Y? & Li, small-amplitude\\
          & &         &         &          &       &         &           &        &   &   & &    & & &variations\\
787    & 0.96 & 0.08 &  -6.04&  44 &2847 & 186.8 & \bf {1.0} & 65.7 & - & - & $-6.29\pm0.14$ &$0.25$&$0.35$ & Y & Li, long-term SB  \\
          & &         &         &          &       &         &           &        &   &   & &    & & & + small-amplitude\\
            &        &         &         &          &       &         &           &        &  &  &   &    & & & variations\\
\hline\\
\noalign{\hfill not Li-rich\hfill \phantom{0}}\\
\hline\\
108741 & 0.05 & - & -1.12 &  26 & 762 & 0.54 & 0.03 & -1.88 & - & - & $-0.96\pm0.21$&  $0.16$&$0.31$ & N & - \\
6      & 0.07 & - &  14.97&  23 &826  & 0.88 & 0.37 & -0.33 & - & - & $14.98\pm0.14$ & $0.01$&$0.25$ & N & -\\
119853 & 0.07 & - &  -8.67&  97 &3643 & 1.11 & 0.78 & 0.77 & - & - & $-8.38\pm0.13$ & $0.29$&$0.28$ & Y? & dust, long-term velocity\\ 
              &      &         &         &          &       &         &           &        & &   &   &    & & &drift \\
43827  & 0.08 & - &  -8.00&  41 &2714 & 1.47& 0.97 &1.91 &  - & - & $-7.93\pm0.21$& $0.07$&$0.32$ & N & -\\
34043  & 0.09 & - &  -3.01&  55 &2714 & 1.75 & 0.99 & 3.27 & $ -3.38\pm0.22$  & 10 & $-3.07 \pm0.14$ & $0.06$&$0.31$ & N & -\\
153687 & 0.12 & - &  -7.63&  29 &1791 & 2.82 & \bf {1.0} & 4.74 & - & - & $-8.26\pm0.30$ & 0.63&$0.37$ & N & dust, RV jitter\\
221776 & 0.15 & - & -13.69&  79 &9755 & 4.39 & \bf {1.0} & 12.0 & $-13.74\pm0.11$ & 6 & $-13.46\pm0.17$ & $0.23$&$0.29$ & N & dust, RV jitter\\
31553  & 1.85 & 0.06 &  -4.02&  45 &8163 & 699 & \bf {1.0} & 111 & - & - &  $-1.11\pm0.36$ & $\bf {0.78}$&$0.27$ & Y & -\\
27497 & 3.06 & 0.06 &   4.98&  55 &2865 & 1908& \bf {1.0}& 178 & - & - &  $\bf {6.80\pm1.41}$ & $\bf {3.75}$&$0.27$ & Y & - \\
3627   & 3.12 & 0.11$^a$ &  -6.85&  65 &12015 & 853& \bf {1.0}& 144 &  $\bf {-11.23\pm0.17}$ & 19 & - & - & - & Y & dust \\
212320 & 3.50 & 0.09 &  -4.16&  57 &6832 &2420 & \bf {1.0} & 197 & - & - & $\bf {-6.81\pm0.54}$ &$\bf {2.65}$&$0.26$ & Y & dust\\ 
156115 & 5.09 & 0.38 & -13.74&  85 &3732 & 5289 & \bf {1.0} & 319 & - & - &  $-8.16\pm0.34$ &$\bf {5.58}$&$0.29$ &  Y & dust, also\\ 
            &        &         &         &          &       &         &         &   &        &   &   &    & & &small-amplitude\\ 
              &                          &         &         &          &       &     &     &           &        &   &   &    & & &variations\\
21078  &13.22 & 0.21 &  46.88&  44 &1239 & 35662& \bf {1.0} & 444 & - & - &  $\bf {47.47\pm1.67}$ & $\bf {0.59}$&$0.26$ & Y & also an astrometric\\
               &     &         &         &          &       &         &           &     &    &   &   &    & & &   binary\\          
\hline
\end{tabular}
}

Remark: 
a: HD 3627: the $O-C$ residuals are for the orbit with 19 CORAVEL velocities and 38 HERMES velocities.
\end{table*}

\section{Binary frequency}
\label{Sect:binary}

\begin{figure}
\includegraphics[width=9cm]{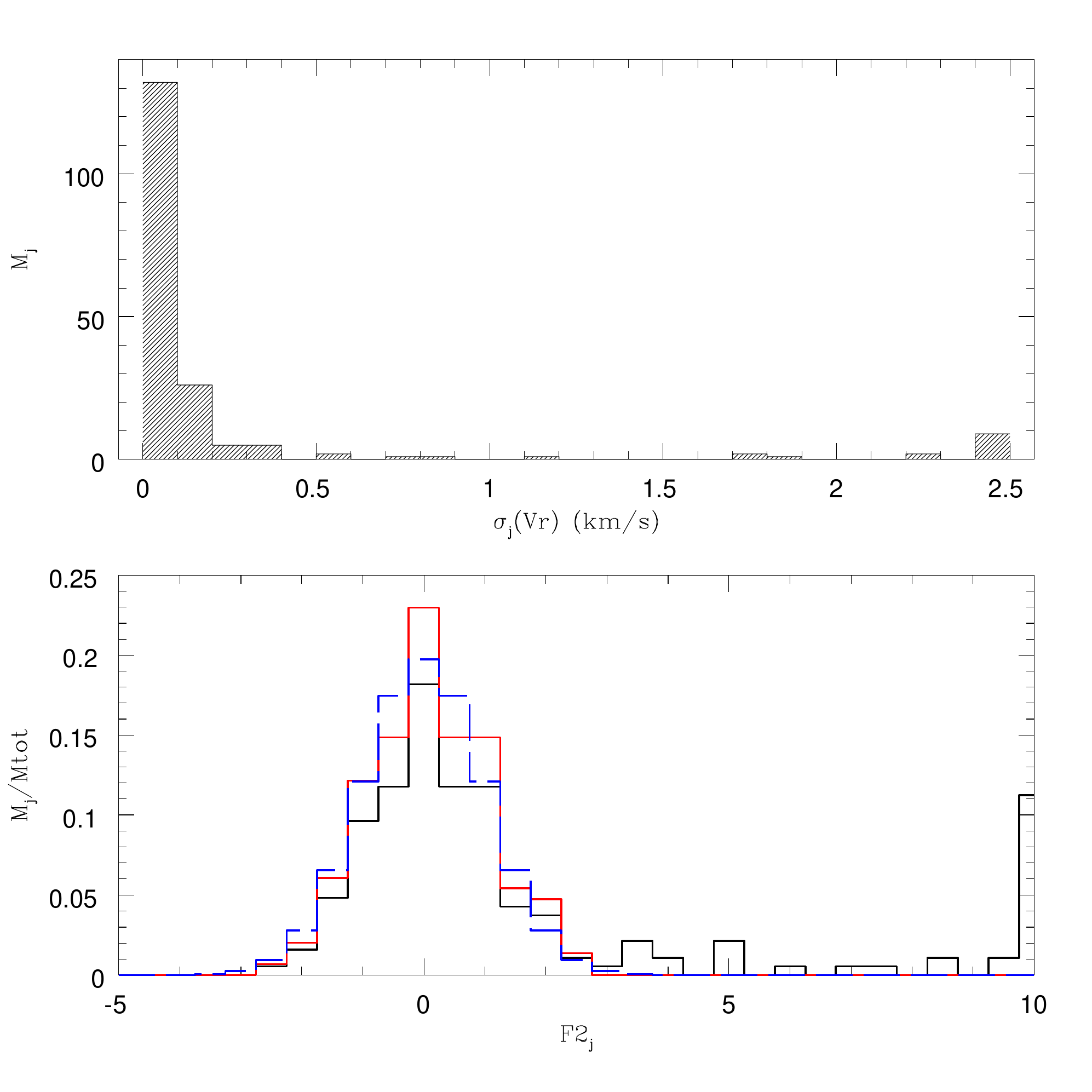}
\caption{\label{Fig:F2_Kepler}
Top panel: Distribution of  HERMES $\sigma_j(RV)$ per star $j$ of the reference sample (R) of CoRoT/Kepler K giants. Bottom:  $F2_j$ distribution for the same sample. In black is drawn the full sample, and in red the sample with binaries removed ($F2 \ge 3.0$), which leads to an almost perfect match between the binary-free observed $F2$ distribution (red curve) and the expected  ${\cal N}(0,1)$ normal-reduced distribution (blue curve).
}
\end{figure}

\begin{figure}
\includegraphics[width=9cm]{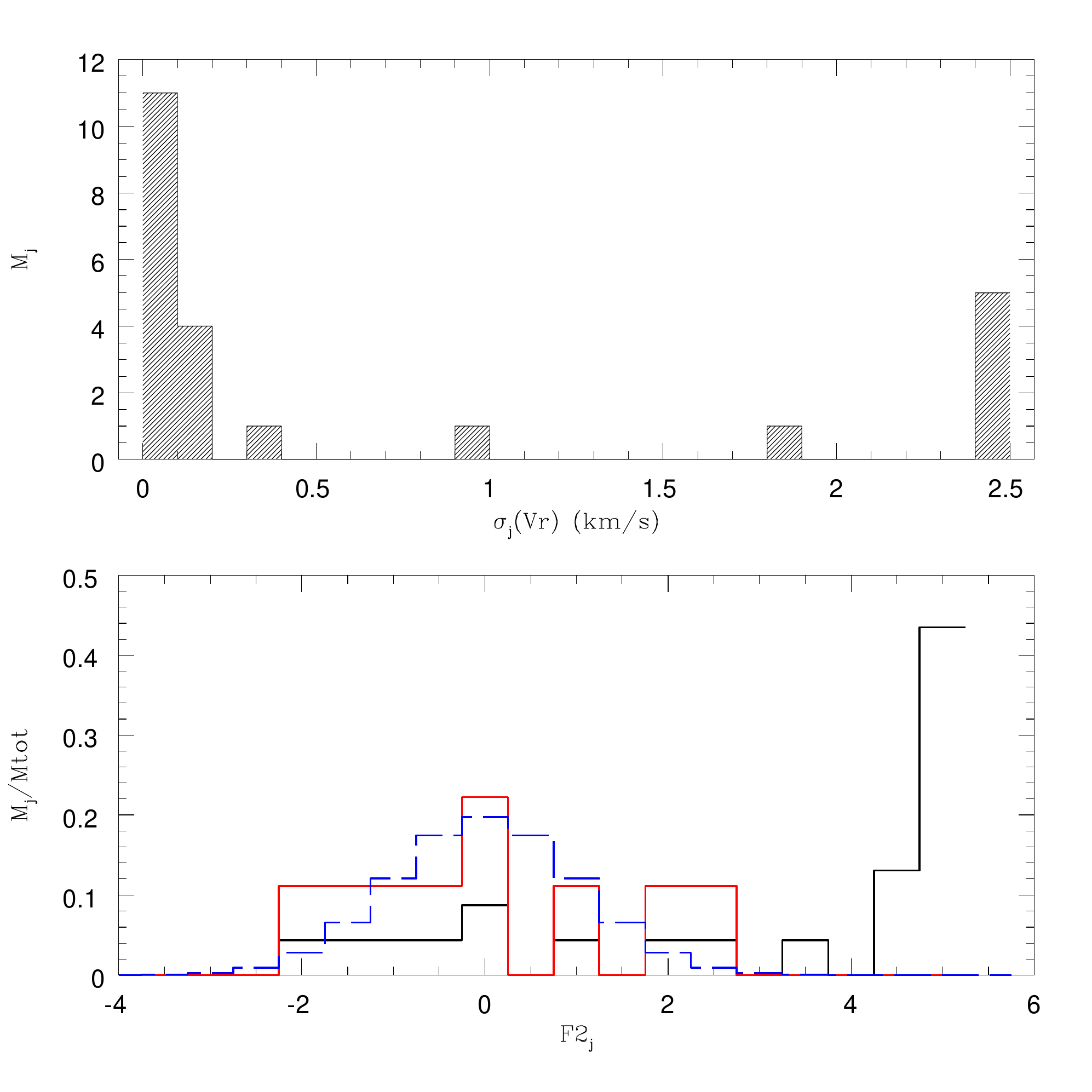}
\caption{\label{Fig:F2_K_Li}
Same as Fig.~\ref{Fig:F2_Kepler} but for sample S1.
}
\end{figure}

\begin{figure}
    \includegraphics[width=9.5cm]{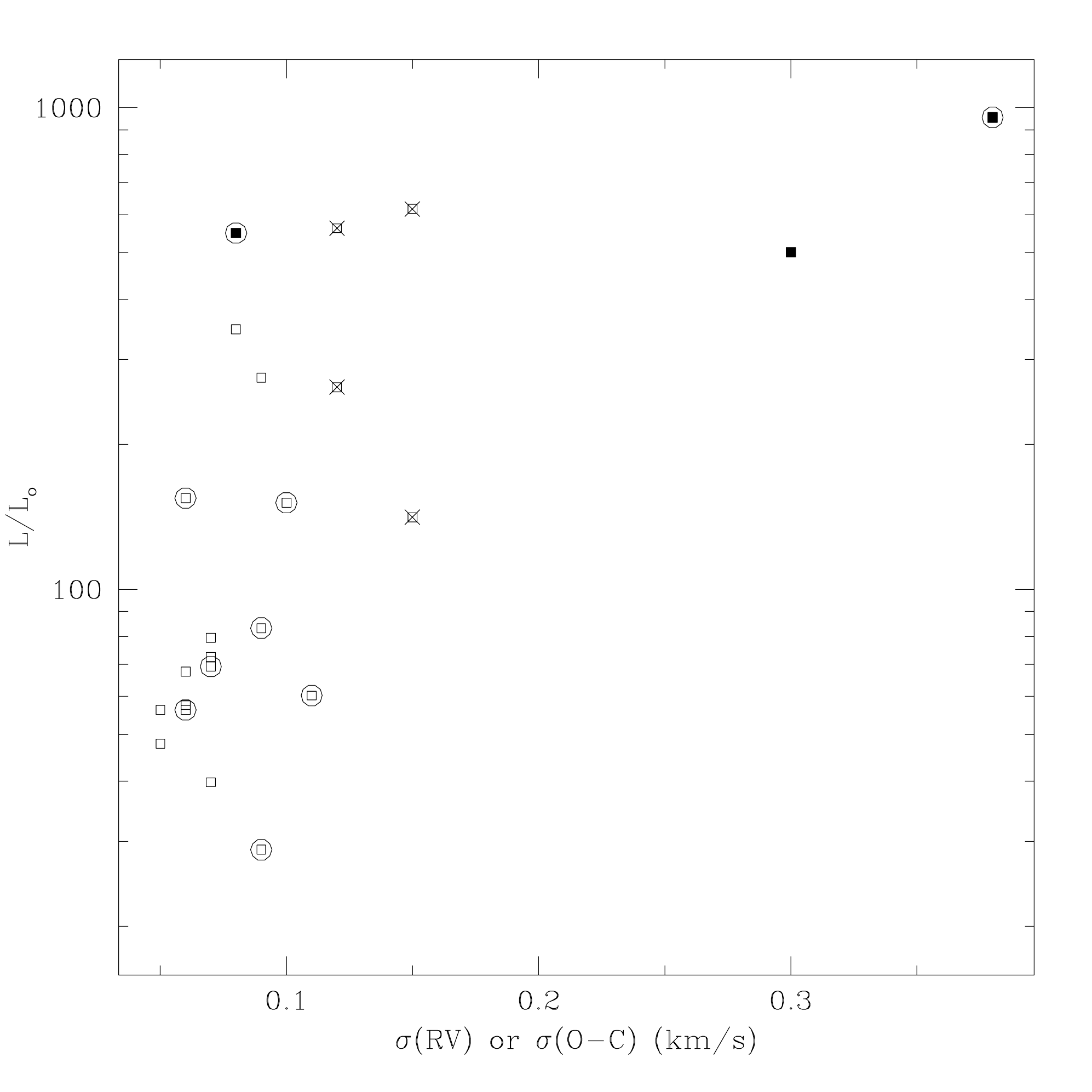}
    \caption{Radial-velocity standard deviation ($\sigma$(RV), or  $\sigma$(O-C) in case of orbital SBs) vs. luminosity for sample S1. Symbols are as follows: filled squares show stars with small-amplitude variations, large open circles signify SBs, crosses represent stars showing jitter, open squares show stars with no special features. We note that the jitter stars and those with small-amplitude RV variations are among the most luminous ones, as expected.}
    \label{Fig:jitter}
\end{figure}

\subsection{Methodology}
\label{Sect:methodology}

The binary frequency has been derived by combining different strategies depending on the considered sample: (i) from the HERMES data alone (samples S1 and R), (ii) by combining HERMES with Gaia DR2 data (samples S1 and R), (iii) by combining HERMES data with other data from the literature \citep[sample S1;][]{Fekel1998,Famaey2005,Gontcharov2006}, and finally (iv) from Gaia DR2 RV data (samples S1, S2, and R).

We explain each of these methods in turn in the remainder of this section. Method (i) is simply based on the evaluation of the $F2$ statistic. This statistic has been defined by \citet{Wilson-hilferty1931} as a way to approximate the reduced $\chi^2$ statistic by a reduced normal distribution ${\cal N}(0,1)$ (of mean zero and standard deviation unity), thus making it independent from the number of degrees of freedom $\nu_j$:
\begin{equation}
\label{Eq:F2}
F2_j = \sqrt{\frac{9\nu_j}{2}} \left(\left(\frac{\chi_j^2}{\nu_j}\right)^{1/3} + \frac{2}{9\nu_j} - 1\right),
\end{equation}
where index $j$ runs over the $M$ different stars ($1 \le j \le M$) and
\begin{equation}
\label{Eq:chi2}
\chi_j^2 = \mathlarger{\sum}_{i=1}^{N{\rm obs}_j} \left( \frac{(RV_{i,j} - \overline{RV}_j)}{\epsilon}\right)^2,
\end{equation}
where index $i$ runs over the $N$obs$_j$ RV observations $RV_{i,j}$ of star $j$. There is only one constraint (the average velocity $\overline{RV}_j$) among the $N$obs$_j$ variables $RV_{i,j}$, and therefore $\nu_j = N$obs$_j-1$.
In the above expressions, $\epsilon$ (the same for all HERMES observations) is the long-term stability of the spectrograph. As stated in Sect.~ \ref{Sect:RV}, the long-term RV standard deviation of a sample of RV standard stars amounts to 0.055~\kms\ \citep{Jorissen2016}, which would be the natural choice for $\epsilon$. However, adopting that value in  Eq.~\ref{Eq:chi2} would not make the $F2$ distribution  (derived from Eq.~\ref{Eq:F2}) of the reference sample R compatible with a reduced normal distribution  ${\cal N}(0,1)$ as it should after removing the binaries, defined as those stars with $F2 \ge 3.0$ (this threshold is similar to the usual `$3\sigma$' cut-off, since by construction $F2$ behaves similarly to a reduced normal distribution). We therefore proceeded by trial and error to find the proper $\epsilon$ value ensuring that the $F2$ distribution behaves as a reduced normal distribution  ${\cal N}(0,1)$. Figure~\ref{Fig:F2_Kepler} reveals that 0.070~\kms\ is required\footnote{instead of  0.055~\kms\ as stated in Sect.~ \ref{Sect:RV}} to ensure the best match between a ${\cal N}(0,1)$ distribution (blue dashed line) and the binary-free observed distribution (red curve). The same $\epsilon$ value was adopted for sample S1 (Fig.~\ref{Fig:F2_K_Li}). However, because of the smaller sample size, the resulting $F2$ distribution does not match the expected ${\cal N}(0,1)$ distribution as accurately as it does for the Kepler reference sample R of Fig.~\ref{Fig:F2_Kepler}.

The problem with this method based on the $F2$ distribution is that it does not distinguish RV variations associated with orbital motion from variations associated with RV jitter in K giants. Thus, stars flagged as having a probability of unity of being RV variables are not necessarily spectroscopic binaries. Therefore, a visual inspection of the RV data is necessary to identify those RV variables where no obvious evidence of orbital motion is present. Those are flagged as `jitter' in Table~\ref{Tab:binary} for sample S1. This RV jitter has been reported by for example  \citet{Mayor1984}, \citet{Carney2003}, and \citet{Hekker2008}  to increase with luminosity. This is also the case in sample S1, as revealed by Fig.~\ref{Fig:HRD} where stars with RV jitter are identified with crosses. These stars are indeed restricted to the highest luminosities in the HR diagram, as confirmed from Fig.~\ref{Fig:jitter}.

Visual inspection of the RV curve also revealed that a drift seems to be present in the RV data of some of the S1 stars (like HD~119853; Fig.~\ref{Fig:119853}). To set this visual feeling on objective grounds, methods (ii) and (iii) were used, that is, comparing the HERMES average velocity for a given target with other data sets, most often the \citet{Famaey2005} or the Gaia DR2 RV sets. Since the former covers epochs not overlapping with the time-span of the HERMES observations, any long-term drift will be easily detected as an offset between the two data sets.  
However, merging different RV sets requires careful {\it a priori} evaluation of  possible zero-point offsets between them. 
The HERMES data set has been tied to the IAU system thanks to the use of  \citet{Udry1999} RV standard stars. The \citet{Famaey2005} data set, which uses CORAVEL velocities \citep{Baranne1979}, is also tied to the IAU system, as checked by \nocite{Gontcharov2006} Gontcharov (2006;  see his Tables~2 and 3). Therefore, there should be no offset between the HERMES and 
\citet{Famaey2005} RV sets, meaning that the latter may be used to detect long-term trends.  Table~\ref{Tab:binary} (for sample S1) lists the average velocity and its associated error $\sigma = \sigma_F(RV)/\sqrt{N_F}$ for the \citet{Famaey2005} RVs whenever available.

 \citet{deMedeiros1999} also studied several stars that are in common with our sample. Nevertheless, we have not used any of the results found by these latter authors in our study, because (i) their study relies on older CORAVEL measurements that were not yet tied to the IAU system (and indeed there appears to be an offset between the HERMES system and this old CORAVEL system, which we did not try to correct at the required accuracy level; see also Sect.~2.1 and Fig.~1 of \citealt{Escorza2019}), and (ii) they do not add  any new information.

Similarly, the offset between the average HERMES and Gaia DR2 velocities has been computed and is listed in Table~\ref{Tab:binary} as $|\overline{RV}_G - \overline{RV}_H|$. However, this offset should  first be compared to the uncertainty on $\overline{RV}_G$ before being used to identify a velocity drift. This uncertainty $\epsilon(RV_G)$ was computed in the following manner.  Figure~3 of \citet{Katz2017} presents the expected accuracy of the average Gaia RV as a function of the number of transits used to compute it, and the stellar magnitude $G_{\rm RVS}$ in the {\it Radial Velocity Spectrometer} pass band. For sample S1, an average of  eight RV transits were used to derive $RV_G$. Therefore, the corresponding curve has been approximated by  a Lagrange polynomial of degree 5 in $G_{\rm RVS}$, as follows:
\begin{eqnarray}
\label{Eq:eps_RV}
\epsilon(RV_G) &= & - 0.429 + 1.019\; G_{\rm RVS} \nonumber\\
&&  -4.456\;  10^{-1}\;G_{\rm RVS}^2\; + \; 8.542\;  10^{-2}\;  G_{\rm RVS}^3 \nonumber\\
& & -\; 7.629\;  10^{-3}\;  G_{\rm RVS}^4\; 
+\;2.626\;  10^{-4} \; G_{\rm RVS}^5. 
\end{eqnarray} 
For sample R, the $G_{\rm RVS}$ magnitude was computed from the Gaia $G$ magnitude using the $G-G_{\rm RVS}$ calibration from Table~2 of \citet{Jordi2018}, namely
\begin{equation}
G-G_{\rm RVS} = 0.4433 + 1.9100\; (r-i) - 0.6984\; (r-i)^2 + 0.0787\; (r-i)^3,
\end{equation}
with the $r-i$ index derived from the Sloan Digital Sky Survey Data Release 12 \citep{Alam2015}.  
For sample S1, the $G_{\rm RVS}$ magnitude was computed from the Gaia $G$ magnitude using a slightly different $G-G_{\rm RVS}$ calibration from Table~2 of \citet{Jordi2018}, namely
\begin{equation}
G-G_{\rm RVS} = 0.017 + 1.0810\; (V-I) - 0.1694\; (V-I)^2 + 0.0075\; (V-I)^3,
\end{equation}
with the Johnson - Cousins $V-I$ index derived from the Hipparcos catalogue \citep{ESA1997}. 
If neither $V-I$ nor $r-i$ were available, the typical value $G-G_{\rm RVS} = 1 $ was used instead. A drift was considered as very likely when the $| \overline{RV}_G -  \overline{RV}_H|$ offset was found to be at least twice larger than the uncertainty  $\epsilon(RV_G)$. These situations are outlined in bold face in Tables~\ref{Tab:binary_R} and \ref{Tab:binary} corresponding to samples R and S1, respectively. All the other targets  have $| \overline{RV}_G -  \overline{RV}_H| \le 0.3$~\kms\  \citep[already discussed by][]{Jorissen2019}, with the exception of HD~112127 where it amounts to 0.41~\kms. Table~\ref{Tab:binary} provides as well the average velocity $ \overline{RV}_F$ of the \citet{Famaey2005} data set, and again any offset $| \overline{RV}_F -  \overline{RV}_H| \ge 2 \; \sigma_F/\sqrt{N_F}$ is outlined in boldface.

The above estimate of $\epsilon(RV_G)$ is also used in method (iii) to flag binarity when the Gaia DR2 RV standard deviation (in column $ \overline{RV}_G$) is at least twice larger than  $\epsilon(RV_G)$. As above for method (ii), these values are outlined in bold face in Tables~\ref{Tab:binary_R} and \ref{Tab:binary}.

Coming back to method (ii), that is, detecting binarity from the offset between HERMES $ \overline{RV}_H$ and Gaia DR2 $ \overline{RV}_G$, we stress that the efficiency of this method is largely dependent upon the period and amplitude of the binary system and upon the HERMES and Gaia DR2 respective time spans. Gaia DR2 measurements span from 25 July 2014 to 23 May 2016 ($56\ts864 \le (JD - 2\ts400\ts000) \le 57\ts532$, corresponding to 662~d). For sample R, Gaia DR2 and HERMES time ranges are disjointed as the HERMES observations started in May 2016 (JD~2\ts457\ts509) and the current data set extends up to August 2019  (JD~2\ts458\ts727). This situation is the most favourable to detect a RV drift between the two data sets which come in succession. A long-period binary will be detected if it causes an offset between the average Gaia DR2 and HERMES velocities that exceeds $2\times0.3$~\kms.

For sample S1, the situation is not as clear cut because the HERMES S1 time-span (April 2009 -- August 2019) encompasses the Gaia DR2 epochs. Hence, the possible offset between their respective average RVs will strongly depend on their  time sampling and their distribution over the orbital cycle. It is therefore impossible to make  definite predictions in this case.

In summary, to flag a star as binary, either a satisfactory orbit is computable, or at least two of the following criteria must be satisfied:
\begin{itemize}
\item[(i)] The probability that the star has a variable (HERMES) RV is strictly larger than 0.99;
\item[(ii)]a drift is seen in the HERMES data, and is confirmed by a significant offset  $|\overline{RV}_F - \overline{RV}_H| \ge 2 \; \sigma_F/\sqrt{N_F}$, or  $|\overline{RV}_G - \overline{RV}_H| \ge 2 \; \epsilon(RV_G)$;
\item[(iii)] the standard deviation of the Gaia DR2 RV is larger than $2 \; \epsilon(RV_G)$.
\end{itemize}

Regarding criterion (iii), which was the only one available to detect binaries in sample~S2, its efficiency has been evaluated by \citet{Jorissen2020}. Although that diagnostic offers a good detection efficiency for systems with orbital periods up to 1000~d \citep[Fig. 6 of][]{Jorissen2020}, its efficiency decreases for longer-period binaries.

Based on the extensive data available for stars of sample S1, we present in Table~\ref{Tab:binary} (in column labelled SB) our final diagnostic regarding the single or binary nature of the S1 targets, and in Table~\ref{Tab:binary_R} we do the same for the targets of sample R. Section~\ref{Sect:individualS1}  presents additional comments on some S1 targets.

We defer the evaluation of the binary frequencies in these two samples to Sect.~\ref{Sect:binfreqS1_R}, 
and the discussion as to whether or not the binary frequencies among Li-rich and dusty K giants are compatible with that in the reference sample R to Sect.~\ref{Sect:bin2}. 

\subsection{Comments on individual stars in sample S1}
\label{Sect:individualS1}

Not mentioned in Table~\ref{Tab:binary} are the  RV data
from \citet{Fekel1998} which confirm the absence of any long-term drift  for
HD~6, HD~153687,  and HD~221776 -- and thus their non-binary nature. HD~153687 and HD~221776 are nevertheless flagged as RV variables by the $F2$ criterion, but a look at their RV curves (Figs.~\ref{Fig:153687} and \ref{Fig:221776}) reveals no orbital signature. These stars are therefore  flagged as `RV jitter'. Similar RV jitter is observed for HD~9746 and HD~30834 (Figs.~\ref{Fig:9746} and \ref{Fig:30834}), and it is noteworthy that all four jitter stars have their $F2$ values just above the threshold value flagging them as RV variables ($F2$ around 4 for HD~9746, HD~30834 and HD~153687,  and 12 for HD~221776). These $F2$ values nevertheless remain much smaller than those of the genuine SB ($F2$ in excess of 25).  
\citet{Fekel1998} also  report scattered velocities for HD~30834  ($-16.7, -17.1, -17.2$, and $-17.6$~\kms, as compared to $-17.1\pm0.1$~\kms\  from HERMES). 

HD~34043 has a $F2$ of 3.27 and a probability of 0.99 of having a variable RV. Although not formally flagged as a RV variable, HD~34043 is very close to the adopted threshold, and may be considered as a further case of RV jitter (Fig.~\ref{Fig:34043}).

With $\sigma(RV) = 0.30$~\kms\ and $F2 = 25.5$, HD~39853 is formally flagged as RV variable from the HERMES data; however, its RV curve does not reveal any clear long-term drift (Top panel of Fig.~\ref{Fig:39853a}). The Pulkovo catalogue provides a velocity of $81.3\pm0.4$~\kms\ as compared to $81.60\pm0.3$~\kms\  for HERMES, confirming the absence of large-amplitude variations.
However, a periodogram analysis based on the phase-dispersion-minimisation method of \citet{Stellingwerf1978}
(Fig.~\ref{Fig:39853b}) reveals the existence of a periodic signal of small amplitude ($K \sim 0.2$~\kms), with a period that could either be 106 or  282~d,  and with zero eccentricity (the 282~d signal is represented  on the upper panel of Fig.~\ref{Fig:39853b}). This would imply very small mass functions of either $8.8\times10^{-8}$~M$_\odot$ (for the 106~d period) or $1.7\times10^{-7}$~M$_\odot$ (for the 282~d period). If these variations were indeed associated to a Keplerian motion (rather than to envelope pulsations), they would correspond to a companion of 6 to 7 Jupiter masses (assuming a stellar mass of 1.5~M$_\odot$ and an orbit seen edge-on). 

The HERMES data reveal a long-term drift for HD~119853 (Fig.~\ref{Fig:119853}) and therefore this star should be included in the category of possible binaries (labelled `SB?'). This conclusion is supported by the velocity listed in the Pulkovo catalogue \citep [$-9.7\pm0.8$~\kms;][]{Gontcharov2006}, 1$\sigma$ away from the HERMES average velocity (-8.67~\kms).  The velocities ($-10.2$ and $-8.7$~\kms) measured by \citet{Fekel1998} also support the hypothesis that HD~119853 is a binary system. 

Hints that HD~43827 (Fig.~\ref{Fig:43827}) could have a variable velocity are provided by the Pulkovo catalogue ($-6.5\pm0.8$~\kms\  as compared to $-8.0$~\kms\  from HERMES and to $-8.2$ and $-8.5$~\kms\ from \citealt{Fekel1998}). However, since these variations could also be caused by an intrinsic RV jitter, as for the other stars discussed above, HD~ 43827 is not included in our list of binaries, especially since the HERMES velocities alone are characterised by a small standard deviation of 0.08~\kms.

HD 233517 is special in sample S1 since it is the only star for which the two binary criteria involving Gaia DR2 data are positive (and may thus be flagged as a binary according to the criteria described in Sect.~\ref{Sect:methodology}), whereas the HERMES $F2$ criterion is not met, although it is not missed by a significant margin, as $F2 = 0.99$. Given the 
very specific SED of that star (Fig.~\ref{Fig:SED233517}), reminiscent of dusty post-AGB systems surrounded by a circumbinary disc (Sect.~\ref{Sect:sample1}), this binary classification is not surprising.

\subsection{Binary frequency in samples R and S1} 
\label{Sect:binfreqS1_R}

In samples R and S1, there are 33/190 (17.4\%) and 8/24 (33.3\%) stars, respectively,  satisfying the conditions of Sect.~ \ref{Sect:methodology} for definite SBs, as summarised on line `Total (S1)' in Table~\ref{Tab:correlation}.
We slightly relaxed the above criteria to define possible SBs when just one of the above conditions (i) -- (iii) is satisfied. For sample R, this yields 12/190 (or 6.3\%) supplementary `SB?'. For sample S1, this yields 8/24 stars, from which we nevertheless exclude  4 stars which we flag as `jitter', a category made possible by the large number of data points that do not reveal any orbital trend, despite a probability of unity for the RV to be variable (see e.g. Figs.~\ref{Fig:30834} and \ref{Fig:221776} for HD~30834 and HD~221776, respectively). In sample R, this category cannot be defined because the number of data points is not large enough. In conclusion, the reference sample R contains 17.4\% SB and 6.3\% SB? candidates, whereas sample S1 contains in total 33.3\% SB and another 33.3\% SB?, from which 16.7\% must nevertheless be subtracted for they seem to suffer from RV jitter.

It is interesting to note that, among the 45 SB+SB? candidates in sample R, 37 were seen by HERMES and 8 (not detected by HERMES) were added by Gaia DR2. Method (ii) based on the HERMES--Gaia DR2 offset detected 26 binaries (or 58\% of the total number of binaries), whereas method (iii) based on Gaia DR2 $\sigma(RV_G)$ alone detected 14 RV variables (31\% of the total number of binaries). The latter frequency corresponds to the detection efficiency of method (iii); it will be very important to correct in this way the binary frequency detected by method (iii) on sample S2, for which no HERMES data are available yet (Sect.~\ref{Sect:sampleS2}).

\subsection{Binary frequency in sample S2}
\label{Sect:sampleS2}

In order to expand the sample of Li-rich stars on which the binary frequency is being tested, we applied criterion (iii) of Sect.~\ref{Sect:methodology}, that is, using only the standard deviation of the Gaia DR2 RVs, to a more extended sample of 56 Li-rich stars, as described in Sect.~\ref{Sect:synopsis}.
This extended sample of Li-rich stars is listed in Table~\ref{Tab:sample2},
 which provides the Gaia DR2 RV along with its standard deviation $\sigma(RV)$ and the expected uncertainties $\epsilon(RV)$ when $N_{\rm RV} = 8$ or 40 RV measurements are available, computed along the guidelines explained in Sect.~\ref{Sect:methodology}. As before, a star is flagged as a binary when its $\sigma(RV)$ is larger than $2\;\epsilon(RV)$ with $\epsilon(RV)$ selected according to the number $N_{\rm RV}$ of Gaia DR2 RV observations available.

There are 6 Gaia DR2 RV binaries out of 56 targets. Since the Gaia DR2 method detects only binaries with short periods \citep[as shown by Fig.~ 6 of][]{Jorissen2020}, a correction factor (1/0.31, as derived at the end of Sect.~\ref{Sect:binfreqS1_R}) has to be applied to get the global binary frequency.  
After this correction, a 34.6\% frequency of binaries is obtained among the Li-rich sample S2. Although this method is the least efficient used so far (for the reason indicated above), it is noteworthy that it provides a binary frequency similar to that found for Li-rich K giants in sample S1 
(36.4\%  for SB+SB? as listed on line `Li-rich' for sample S1 in Table~\ref{Tab:correlation}). Therefore, sample S2 does not alter the conclusion about the normality of the binary frequency among Li-rich K giants that is  presented in Sect.~\ref{Sect:bin2}.

\begin{table}
\caption[]{\label{Tab:S1'}
Classification of stars from Table~\ref{Tab:binary} (listed in the same order) in samples S1 and S1'.
}
\begin{tabular}{rllcrl}
\hline\\
HD     & S1 & \multicolumn{3}{c}{S1'} \\
\cline{3-5}
           &  type &  type &  \multicolumn{1}{c}{Prob} & F2 \\
            &      &             &   \multicolumn{1}{c}{RV Var} \\
\hline\\
\noalign{\hfill Li-rich\hfill \phantom{0}}\\
\hline\\
40827  & cst & cst       &0.00 & -3.0 \\
116292 & cst & cst      & 0.00 & -2.6 \\
6665    & cst & cst       & 0.31 & -0.5\\
63798  & cst & cst       & 0.03 & -1.9\\
90633  & cst & cst      & 0.94 & 1.6\\
112127 & SB? & cst   & 0.58 & 0.2\\
233517 & SB  & SB   & (0.92) & (1.4) & Gaia DR2 SB\\
30834 & jitter & cst & 0.04 & -1.8\\ 
9746   & jitter & cst & 0.51 & 0.02\\
39853  & SB? & SB? & 1.00 & 4.9\\
787    & SB & SB & 1.00 & 11.3\\
\hline\\
\noalign{\hfill not Li-rich\hfill \phantom{0}}\\
\hline\\
108741 & cst & cst & 0.28 & -0.6\\
6       & cst & cst & 0.42 & -0.2 \\
119853 & SB? & cst & 0.54 & 0.1\\ 
43827 & cst & cst & 0.18 & -0.9 \\  
34043  & cst & cst & 0.97  & 1.9 \\ 
153687 & jitter & SB? & 1.00 & 2.7 \\
221776 & jitter & cst &  0.62   & 0.3 \\
31553  & SB & SB & 1.00 & 23.8\\
27497 & SB & SB & 1.00 & 28.4\\
3627   & SB & SB & 1.00 & 8.0\\
212320 & SB & SB & 1.00 & 14.0\\ 
156115 & SB & SB & 1.00 & 34.0\\
21078  & SB & SB & 1.00 & 169.8\\
\hline
\end{tabular}
\end{table}

\begin{figure}
\includegraphics[width=9cm]{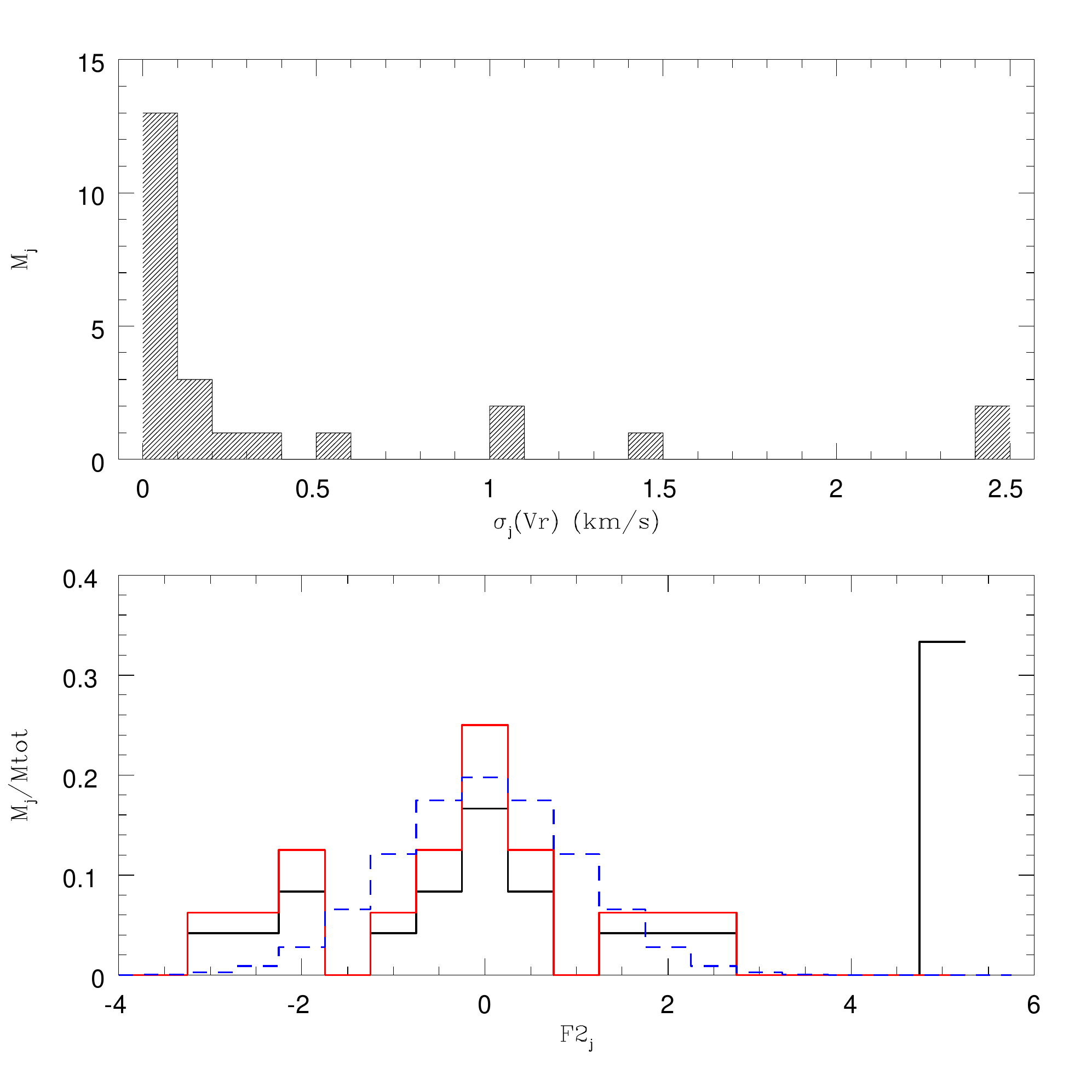}
\caption{\label{Fig:F2_K_fake}
Same as Fig.~\ref{Fig:F2_K_Li} but for S1 resampled 
 (S1') so as to mimic the R-sample in terms of $\Delta t$ and $N$obs.
}
\end{figure}

\subsection{Comparison of the binary frequencies (Li versus non-Li K giants, and dusty versus non-dusty K giants)}
\label{Sect:bin2}

In this section, we evaluate whether or not the binary frequencies detected among the families of Li-rich K giants and dusty K giants are typical among K giants. Two methods are used for that purpose. Method (a) is solely based on the data collected for sample S1 (hence it is referred to as the `internal method'), estimating the binary frequency directly from the whole sample, irrespective of the dusty or Li subtypes, whereas method (b) uses as reference the binary frequency obtained in the reference sample R (hence its name `external method'). Each method has its specific advantages and disadvantages. For that reason, both methods are used in the following. In method (a), we compare the binary frequency among  Li-rich and non-Li stars, and among dusty and non-dusty stars.  
In addition to the small sample size, the other potential problem with method (a) is the fact that the binary frequency in the comparison sample of non-Li-rich stars might not be normal. This could arise if the dusty K giants are not equally balanced between Li-rich and non-Li-rich stars, since dusty K giants might include a large proportion of binary stars. 

Method (b) does not suffer from these flaws since it is based on the much larger and unbiased reference sample R containing 190 K giant stars; however, it relies on far fewer RV measurements (just a few; see Fig.~\ref{Fig:DeltatNobs}) and therefore its efficiency in detecting binary systems could possibly be much lower than method (a). A way to partially circumvent this weakness of method (b) is to compare the binary frequencies in samples with the same measurement-sampling properties in terms of number of measurements $N$obs and of their time span $\Delta t$. Therefore, we built a new sample S1' from sample S1, ensuring that the sampling properties of S1' are very similar to those of R. We first select pairs ($N$obs,  $\Delta t$), represented by 
red crosses in Fig.~\ref{Fig:DeltatNobs}, distributed similarly to those of sample R (open squares  in Fig.~\ref{Fig:DeltatNobs}), and assign them to S1 targets, in such a way that the actual time span of the S1 measurements of the considered star is larger than or equal to the assigned $\Delta t$. Subsequently, along the time span $\Delta t$, we randomly select $N$obs$-2$ measurements among the initial 
ones from S1 (fixing the extreme measurements to preserve the $\Delta t$ assignment). These measurements form a subset of the initial S1 ones, and define sample S1'. 
The stars considered as binaries in sample S1' according to the rules defined in Sect.~\ref{Sect:methodology} are shown in Table~\ref{Tab:S1'}. They are not exactly the same as in sample S1.  As in  S1, HD~233517 is considered a SB  according to the Gaia DR2 $\sigma(RV)$ criterion.  
The $F2$ distribution for sample S1' is illustrated in Fig.~\ref{Fig:F2_K_fake}. 
As for sample S1, a binary star requires the RV-variability probability to be $>0.99$ (with $F2$ in the range 2.7 -- 5 for SB? and $F2 > 5$ for SB). 

The small number of observations per star available in sample R forbids us from distinguishing between `jitter' stars and `SB?' stars. Therefore, category `jitter' is no longer considered in method (b).

\begin{table}
\caption[]{\label{Tab:contingency} Illustration of a $2\times2$ contingency table, where $a$, $b$, $c$, $d$ are integers denoting the number of individuals in the respective cells.
}
\begin{tabular}{l|cc|l}
 &  SB & non-SB & Total \\
 \hline\\
Li-rich& $a$& $c$& $N_{x1} = a + c$\\
no-Li & $b$& $d$ & $N_{x2} = b + d$\\
\hline\\
Total & $N_{y1} = a + b$  & $N_{y2} = c + d$ & $N = a+b+c+d$\\
\end{tabular}
\end{table}

\begin{table*}
\caption{Contingency tables testing the correlation between binarity frequency and dusty or Li-rich nature. See text (Sect.~\ref{Sect:bin2}) for the meaning of the symbols. Methods (a) and (b) differ by  the way the binary frequencies are compared: method (a) makes an internal comparison (i.e. dust vs. no-dust, Li-rich vs. no-Li, and therefore the significance given in the last column refers to such an internal comparison), whereas method (b) compares them with that in the reference sample R. Therefore, for method (b),  the significance in the last column refers to the contingency tables constructed from the considered row and the last row referring to sample R. Method (a) applies to sample S1 whereas method (b) applies to sample S1'.  }
\label{Tab:correlation}
\begin{tabular}{r|rrrrr|r|rrrrr|c|r}
\noalign{\hspace*{\fill} \bf{Method (a) on S1}\hspace*{\fill}}\\
             && const. & jitter & SB?  & SB &  \multicolumn{1}{c}{Total} && const. & jitter & SB?  & SB &  \multicolumn{1}{c}{Total} & Significance of $H_0$  \\
             \cline{3-6}\cline{9-12}
       &&     \multicolumn{4}{c}{counts $n$} &&&   \multicolumn{4}{c}{Observed $p$ (\%)} & &(internal)\\
  \hline\\
  dust          && 0 &   3 & 1 &  4 
  & 8  &&    0.0 & 37.5 & 12.5 & 50.0  & 100\% &\\
  no-dust            && 9 & 1 & 2 & 4 & 16 && 56.3 & 6.2 & 12.5 & 25.0 & 100\% & \raisebox{1.5ex}[0pt]{1.6\%}\\
  \hline\\
    Li-rich      && 5 & 2 & 2 & 2 & 11  && 45.4 & 18.2 & 18.2 & 18.2 & 100\%\\
   no-Li           &&  4 & 2 & 1 & 6 & 13  && 30.8 & 15.4 & 7.7 & 46.2  & 100\% & \raisebox{1.5ex}[0pt]{54.3\%}\\
   \hline\\
{\bf Total  $N_y$ (S1)}     & &  9 & 4 & 3 & 8  & 24 &&   37.5 & 16.7 & 12.5 & 33.3 & Expected $p$ (\%)  \\
   \hline\\
   \noalign{\hspace*{\fill} \bf{Method (b) on S1'}\hspace*{\fill}}\\
             && const. & & SB?  & SB &  \multicolumn{1}{c}{Total} && const. & & SB?  & SB &  \multicolumn{1}{c}{Total} & Significance of $H_0$  \\
             \cline{3-6}\cline{9-12}
       &&     \multicolumn{4}{c}{counts $n$} &&&   \multicolumn{4}{c}{Observed $p$ (\%)} & &(external w.r.t. R)\\
  \hline\\  
     {\bf Total  $N_y$ (S1')}     & &  14 &  & 2 & 8  & 24 &&   58.3 &  & 8.3 & 33.4 &  100\%&  13.0\% \\
\hline
   \hline\\ 
  dust           && 3 &&  1 & 4  & 8 &&    37.5 && 12.5 & 50.0  & 100\% &  3.3\%\\
      \hline\\
    no-dust          && 11 & &1 & 4  & 16 &&68.75 && 6.25 & 25.0 & 100\% & 63.6\%\\
     \hline
     \hline\\
    Li-rich      && 8 &  & 1 & 2  & 11 &&  72.7 && 9.1 & 18.2 & 100\% & 73.3\%\\
        \hline\\
   no-Li           &&  6 &  & 1 & 6 & 13  && 46.2 & & 7.6 & 46.2  & 100\% & 2.4\%\\
\hline   \hline\\
 {\bf Total $N_y$ (R)}     && 145 & & 12 & 33 & 190    && 76.3 && 6.3 & 17.4 & Expected $p$ (\%)\\
   \hline\\
\end{tabular}
\end{table*}

Both methods (a) and (b) rely on the same premise, namely the `test of equality of two proportions' based on a `contingency table'. A basic description of the principles of this method may be found in for example section 14.4 of \textit{Numerical Recipes} \citep{Press2007} or in \citet{Agresti2012}. The situation is as follows.   Each  target star  has  two  different properties associated with it (in our case Li-rich or not-Li-rich and  binary or non-binary in the simplest situation), and we want to know whether knowledge of one property gives us any demonstrable advantage in predicting the value of the other property. In other words, we want to know whether or not these quantities are `associated' or `correlated'. For such a pair of variables, the data can be displayed as a `contingency table', that is, a table whose rows are labelled by the value of one property, and the columns by the value of the other property, and whose entries are non-negative integers giving the number of observed targets for each combination of row and column. We first present the method in the simplified situation expressed above where one property is  Li-rich (or not) and the other  is SB (or not). This classification scheme defines a $2\times2$ contingency table, as shown in Table~\ref{Tab:contingency}, where the Li property corresponds to the rows of the contingency table (identified by the subscript $x$ in what follows; each row containing $N_x$ stars), whereas the SB property  corresponds to columns (denoted by subscript $y$ in what follows).

The null hypothesis,
\begin{equation}
H_0 :  p_1 = p_2
,\end{equation}
tests  the equality of the two proportions $p_1 =  a/N_{x1}  = p_2 = b/N_{x2}$, with $a$ and $b$ defined in Table~\ref{Tab:contingency}. In our specific situation, this means testing whether the frequency of SB may be considered the same in the Li-rich and non-Li categories. The hypergeometric distribution law then allows one to evaluate the probability of encountering, under the hypothesis that $H_0$ holds true, a distribution as deviant as the one actually observed. The probability of obtaining the value $a$ under $H_0$ is then expressed by
\begin{equation}
\label{Eq:P(a)}
P(a) = \frac{C_{a+c}^a\;C_{b+d}^b}{C_{a+b+c+d}^{a+b}} = \frac{(a+c)!\;(b+d)!\;(a+b)!\;(c+d)!}{a!\;b!\;c!\;d!\;(a+b+c+d)!} ,
\end{equation}
where $C_n^m = \frac{n!}{(n-m)!\;m!}$ is the number of combinations of $m$ objects among $n$ ($> m$). If the probability of obtaining the value  $a$ or a value more deviant than $a$ is low, this means that the observed counts are not
consistent with the null hypothesis of equal proportions, and the null hypothesis must be rejected (at a significance level that we specify below). If we assume that the table is ordered in such a way that the first sample is the least numerous (i.e. $N_{x1} \le N_{x2}$, or $a+c \le b+d$) and the first category is the least frequent (i.e. $N_{y1} \le N_{y2}$, or $a+b \le c+d$), then the number $a$ counting the number of individuals in the first sample and in the first category is necessarily somewhere between zero and the minimum of ($N_{x1}, N_{y1}$), or ($a+c, a+b$). If we further assume that $a < \mu_a$, where $\mu_a$ is the expected value for $a$, namely  $\mu_a = p_{y1} \times N_{x1} = (a+b) \; (a+c) / (a+b+c+d)$, then $p_1 < p_2$, and the significance level with which the null hypothesis may be rejected is obtained by summing the hypergeometric probabilities $P(n)$ of  Eq.~\ref{Eq:P(a)} for $n$ going from zero to $a$. More precisely, for a bilateral test, the null hypothesis may be rejected at a significance level (of approximately\footnote{Because of the discrete character of the hypergeometric distribution, this is only approximate, since the anomalous cases at each side of the distribution do not yield exactly the same tail probabilities.}) $\alpha$ if that probability sums up to $\alpha/2$. 
This is known as the Fisher exact test for a $2\times2$ contingency table \citep[e.g.][]{Agresti2012}.  In cases where the two samples are categorised on more than two parameters, as we do in Table~\ref{Tab:correlation} with the four categories `RV constant', `RV jitter', `SB?', and `SB' (method a; for method b, the category `RV jitter' does not exist as the number of data points is too small to distinguish between `RV jitter' and `SB?'), the Fisher test has to be generalised along the same principles as above, but this time using  the generalised hypergeometric distribution. However, the number of computations becomes  rapidly gigantic; fortunately, there are several  calculators available online for computing the significance level when rejecting  the null hypothesis for higher-order contingency tables. We used the one available at http://vassarstats.net/fisher2x4.html, based on the extension of the Fisher test by \citet{Freeman1951}.
We discuss the results of the tests in the following sections.

\subsubsection{Internal method (a)}  
\label{Sect:methoda}

The conclusions resulting from the statistical analysis with method (a) are twofold (they are presented in the upper part of Table~\ref{Tab:correlation}). First, regarding the Li-rich K giants, the comparison of Li-rich and non-Li K giants reveals no significant difference in the binary frequencies between these two groups (since the first-kind risk of erroneously rejecting the $H_0$ hypothesis while it is true is as large as 54.3\%). 
If anything, 
there even seems to be a deficit of SB  among Li-rich K giants, since there are 18.2\% large-amplitude binaries among those (2/11), as compared to 46.2\% (6/13) in the comparison sample of non Li-rich K giants. 
However, rather than resulting from an actual deficit of binaries among Li-rich stars (not confirmed by method b; see Sect.~\ref{Sect:methodb}), the above imbalance seems to be caused by the fact that the comparison sample of non-Li-rich stars contains several dusty K giants, and this class seems to host an anomalously large number of  binaries.
 Indeed, the analysis of the dusty versus non-dusty contingency table reveals a significant difference in their binary frequencies, with a first-kind risk when rejecting the $H_0$ hypothesis of only 1.6\%. Method (b) indeed reveals that this imbalance is caused by the presence of an unusually high frequency of binaries among dusty K giants as compared to the reference sample R (first-kind risk of 3.3\% only; Sect.~\ref{Sect:methodb}).

The difference between dusty and non-dusty K giants further extends to the jitter category, which is in excess among dusty K giants  (37.5\%, or 3 stars among 8), as compared to 6.25\% or 1/16 among the non-dusty K giants.  A likely explanation is that the dusty K giants are on average more luminous than the non-dusty ones and hence more prone to RV jitter, as jitter is known to increase with luminosity \citep[e.g.][]{Mayor1984,Carney2003,Hekker2008}. This hypothesis is indeed confirmed by the location of dusty and RV-jitter stars in the HR diagram, as discussed in Sect.~\ref{Sect:HRD} in relation with Fig.~\ref{Fig:HRD}.
Overall, the dusty K giants do not include a single case of constant RV (0/8), and this result is highly significant because 56\% (9/16) of the stars in the non-dusty sample are constant-RV stars.

\subsubsection{External method (b)}  
\label{Sect:methodb}

The results from the statistical analysis using method (b) are presented in the lower part of Table~\ref{Tab:correlation}. It should be read in the following way: $3\times2$ contingency tables were built by combining any of the first lines with the last one corresponding to sample R, and the resulting first-kind risk of rejecting the null hypothesis $H_0$ (that the binary frequency is the same in the two samples being compared) while it is true is listed in the last column.

As already hinted at in Sect.~\ref{Sect:methoda}, the comparison with the reference sample R now makes it quite clear that the frequency of binaries in the sample of dusty K giants is significantly higher than in the reference sample of K giants (with a first-kind risk when rejecting the $H_0$ hypothesis of 3.3\%), whereas the binary frequency among Li-rich giants is totally normal (with a first-kind risk as high as 73.3\%). The high frequency of binaries among non-Li-rich stars simply reflects the larger number of dusty stars among this subsample than among Li-rich giants.

We also note that the binary frequency (SB+SB?) in sample S1' is almost identical to that in sample S1, namely 10 and 11 stars, respectively, as seen by comparing the lines labelled `Total (S1)' and `Total (S1')' in Table~\ref{Tab:correlation}. 

\subsubsection{Comparison with literature}  

To conclude this section, it is worth mentioning that the overall binary frequency in  sample S1 of 24 K giants, namely 10/24 or 11/24 or 41.7 -- 45.8\% (Rows $N_{y1}$ of Table~\ref{Tab:correlation}) is higher than  that obtained by studying more extended samples, but this may simply reflect the relatively large fraction of dusty K giants in sample S1, and their associated binary nature.

In contrast, sample R has a binary frequency (17.4 -- 23.7\%) well in line with that derived by previous studies focusing on K giants. For instance, among 5643 field K giants, \citet{Famaey2005} found 14.5\% spectroscopic binaries (with a number of RV points per target of  only 2 or 3, comparable to that of sample R), a result in agreement with those of \citet{Harris1983} (15 to 20\% binaries) and of \citet{Mermilliod2007} ($16\pm2$\% of binaries after 3 RV measurements). 
The more recent analysis of GES data \citep{Merle2020}, with a SB1 fraction of $21\pm3$\% among K giants, confirms the results of all previous analyses, including our present result for sample R binary frequency.

\section{Conclusions}
\label{Sect:conclusion}

The main conclusion of our analysis is that the binary frequency among Li-rich K giants is normal when compared to that found under similar observing conditions in a reference sample of 190 K giants.
Therefore, the claim by \citet{Casey2019} that there is a link between binarity and Li excess through tidal effects is not supported by our study, since this would require a 100\% frequency of binaries among Li-rich stars, which is clearly not observed. There is also no correlation between Li-richness and the presence of circumstellar dust. The only correlation that could be found between Li enrichment and rapid rotation is that the most Li-enriched K giants appear to be fast-rotating stars.
The availability of Gaia parallaxes allowed us to locate the S1 targets in the HR diagram, from which it appears that:
\begin{itemize}
\item The only two long-period binary systems among Li-rich K giants (HD 787 and HD 233517) are not in the red clump but are at a much higher luminosity (either on the RGB or early-AGB). 
\item Despite the fact that our S1 sample contains several binary systems in the red clump, none  of those among the Li-rich K giants in the red clump are binaries.
\item Three among the four binary (non-Li-rich) stars in the red clump (HD~3627, HD~119853 and HD~212320) are surrounded by cool dust. The five other dusty stars are located along the giant branches (either RGB or early-AGB).  
\item The stars with RV jitter (crosses in Figs.~\ref{Fig:HRD} and \ref{Fig:jitter}) are among the most luminous in the sample, hinting at some intrinsic origin (perhaps non-radial pulsations) for this jitter \citep[see][for a recent discussion]{Hatzes2018}. Furthermore, those among the most luminous stars which do not exhibit RV jitter show small-amplitude regular RV variations (small red dots in Fig.~\ref{Fig:HRD}). HD~112127 is a special case since its RV variations are difficult to characterise, simultaneously exhibiting short-period, small-amplitude variations and a long-term velocity drift, sometimes interrupted by some sort of RV jitter (Fig.~\ref{Fig:112127}). 
\end{itemize}

We also find that  dusty K giants are either the most luminous among K giants along the RGB or early-AGB, or are binaries located in the red clump.

\begin{acknowledgement}
  Based on observations obtained with the HERMES spectrograph, which is supported by the Research Foundation - Flanders (FWO), Belgium, the Research Council of KU Leuven, Belgium, the Fonds National de la Recherche Scientifique (F.R.S.-FNRS), Belgium, the Royal Observatory of Belgium, the Observatoire de Gen\`eve, Switzerland and the Th\"uringer Landessternwarte Tautenburg, Germany.   
This work has made use of data from the European Space Agency (ESA) mission Gaia (https://www.cosmos.esa.int/gaia), processed by the Gaia Data Processing and Analysis Consortium (DPAC, https://www.cosmos. esa.int/web/gaia/dpac/consortium). Funding for the DPAC has been provided by national institutions, in particular the institutions participating in the Gaia Multilateral Agreement. This research has made use of the SIMBAD database, operated at CDS, Strasbourg, France. This research has been funded by the Belgian Science Policy Office under contract BR/143/A2/STARLAB, and by the F.W.O. under contract ZKD1501-00-W01. LS and DP are senior research associates at the F.R.S.-FNRS. 
\end{acknowledgement}

\bibliographystyle{aa}
\bibliography{biblio}

\begin{thebibliography}{117}
\expandafter\ifx\csname natexlab\endcsname\relax\def\natexlab#1{#1}\fi

\bibitem[{{Agresti}(2012)}]{Agresti2012}
{Agresti}, A. 2012, Categorical Data Analysis (3rd edition) (Wiley)

\bibitem[{{Alam} {et~al.}(2015){Alam}, {Albareti}, {Allende Prieto}, {Anders},
  {Anderson}, {Anderton}, {Andrews}, {Armengaud}, {Aubourg}, {Bailey}, {Basu},
  {Bautista}, {Beaton}, {Beers}, {Bender}, {Berlind}, {Beutler}, {Bhardwaj},
  {Bird}, {Bizyaev}, {Blake}, {Blanton}, {Blomqvist}, {Bochanski}, {Bolton},
  {Bovy}, {Shelden Bradley}, {Brandt}, {Brauer}, {Brinkmann}, {Brown},
  {Brownstein}, {Burden}, {Burtin}, {Busca}, {Cai}, {Capozzi}, {Carnero
  Rosell}, {Carr}, {Carrera}, {Chambers}, {Chaplin}, {Chen}, {Chiappini},
  {Chojnowski}, {Chuang}, {Clerc}, {Comparat}, {Covey}, {Croft}, {Cuesta},
  {Cunha}, {da Costa}, {Da Rio}, {Davenport}, {Dawson}, {De Lee}, {Delubac},
  {Deshpande}, {Dhital}, {Dutra-Ferreira}, {Dwelly}, {Ealet}, {Ebelke},
  {Edmondson}, {Eisenstein}, {Ellsworth}, {Elsworth}, {Epstein}, {Eracleous},
  {Escoffier}, {Esposito}, {Evans}, {Fan}, {Fern{\'a}ndez-Alvar}, {Feuillet},
  {Filiz Ak}, {Finley}, {Finoguenov}, {Flaherty}, {Fleming}, {Font-Ribera},
  {Foster}, {Frinchaboy}, {Galbraith-Frew}, {Garc{\'\i}a},
  {Garc{\'\i}a-Hern{\'a}ndez}, {Garc{\'\i}a P{\'e}rez}, {Gaulme}, {Ge},
  {G{\'e}nova-Santos}, {Georgakakis}, {Ghezzi}, {Gillespie}, {Girardi},
  {Goddard}, {Gontcho}, {Gonz{\'a}lez Hern{\'a}ndez}, {Grebel}, {Green},
  {Grieb}, {Grieves}, {Gunn}, {Guo}, {Harding}, {Hasselquist}, {Hawley},
  {Hayden}, {Hearty}, {Hekker}, {Ho}, {Hogg}, {Holley-Bockelmann}, {Holtzman},
  {Honscheid}, {Huber}, {Huehnerhoff}, {Ivans}, {Jiang}, {Johnson},
  {Kinemuchi}, {Kirkby}, {Kitaura}, {Klaene}, {Knapp}, {Kneib}, {Koenig},
  {Lam}, {Lan}, {Lang}, {Laurent}, {Le Goff}, {Leauthaud}, {Lee}, {Lee},
  {Licquia}, {Liu}, {Long}, {L{\'o}pez-Corredoira}, {Lorenzo-Oliveira},
  {Lucatello}, {Lundgren}, {Lupton}, {Mack}, {Mahadevan}, {Maia}, {Majewski},
  {Malanushenko}, {Malanushenko}, {Manchado}, {Manera}, {Mao}, {Maraston},
  {Marchwinski}, {Margala}, {Martell}, {Martig}, {Masters}, {Mathur},
  {McBride}, {McGehee}, {McGreer}, {McMahon}, {M{\'e}nard}, {Menzel},
  {Merloni}, {M{\'e}sz{\'a}ros}, {Miller}, {Miralda-Escud{\'e}}, {Miyatake},
  {Montero-Dorta}, {More}, {Morganson}, {Morice-Atkinson}, {Morrison},
  {Mosser}, {Muna}, {Myers}, {Nand ra}, {Newman}, {Neyrinck}, {Nguyen},
  {Nichol}, {Nidever}, {Noterdaeme}, {Nuza}, {O'Connell}, {O'Connell},
  {O'Connell}, {Ogando}, {Olmstead}, {Oravetz}, {Oravetz}, {Osumi}, {Owen},
  {Padgett}, {Padmanabhan}, {Paegert}, {Palanque-Delabrouille}, {Pan},
  {Parejko}, {P{\^a}ris}, {Park}, {Pattarakijwanich}, {Pellejero-Ibanez},
  {Pepper}, {Percival}, {P{\'e}rez-Fournon}, {Ṕrez-Ra`fols}, {Petitjean},
  {Pieri}, {Pinsonneault}, {Porto de Mello}, {Prada}, {Prakash},
  {Price-Whelan}, {Protopapas}, {Raddick}, {Rahman}, {Reid}, {Rich}, {Rix},
  {Robin}, {Rockosi}, {Rodrigues}, {Rodr{\'\i}guez-Torres}, {Roe}, {Ross},
  {Ross}, {Rossi}, {Ruan}, {Rubi{\~n}o-Mart{\'\i}n}, {Rykoff},
  {Salazar-Albornoz}, {Salvato}, {Samushia}, {S{\'a}nchez}, {Santiago},
  {Sayres}, {Schiavon}, {Schlegel}, {Schmidt}, {Schneider}, {Schultheis},
  {Schwope}, {Sc{\'o}ccola}, {Scott}, {Sellgren}, {Seo}, {Serenelli}, {Shane},
  {Shen}, {Shetrone}, {Shu}, {Silva Aguirre}, {Sivarani}, {Skrutskie},
  {Slosar}, {Smith}, {Sobreira}, {Souto}, {Stassun}, {Steinmetz}, {Stello},
  {Strauss}, {Streblyanska}, {Suzuki}, {Swanson}, {Tan}, {Tayar}, {Terrien},
  {Thakar}, {Thomas}, {Thomas}, {Thompson}, {Tinker}, {Tojeiro}, {Troup},
  {Vargas-Maga{\~n}a}, {Vazquez}, {Verde}, {Viel}, {Vogt}, {Wake}, {Wang},
  {Weaver}, {Weinberg}, {Weiner}, {White}, {Wilson}, {Wisniewski},
  {Wood-Vasey}, {Ye`che}, {York}, {Zakamska}, {Zamora}, {Zasowski}, {Zehavi},
  {Zhao}, {Zheng}, {Zhou}, {Zhou}, {Zou}, \& {Zhu}}]{Alam2015}
{Alam}, S., {Albareti}, F.~D., {Allende Prieto}, C., {et~al.} 2015, \apjs, 219,
  12

\bibitem[{{Angelou} {et~al.}(2015){Angelou}, {D'Orazi}, {Constantino},
  {Church}, {Stancliffe}, \& {Lattanzio}}]{Angelou2015}
{Angelou}, G.~C., {D'Orazi}, V., {Constantino}, T.~N., {et~al.} 2015, \mnras,
  450, 2423

\bibitem[{{Asplund} {et~al.}(2009){Asplund}, {Grevesse}, {Sauval}, \&
  {Scott}}]{Asplund2009}
{Asplund}, M., {Grevesse}, N., {Sauval}, A.~J., \& {Scott}, P. 2009, \araa, 47,
  481

\bibitem[{{Baglin} {et~al.}(2006){Baglin}, {Auvergne}, {Barge}, {Deleuil},
  {Catala}, {Michel}, {Weiss}, \& {COROT Team}}]{Baglin2006}
{Baglin}, A., {Auvergne}, M., {Barge}, P., {et~al.} 2006, in ESA Special
  Publication, Vol. 1306, The CoRoT Mission Pre-Launch Status - Stellar
  Seismology and Planet Finding, ed. M.~{Fridlund}, A.~{Baglin}, J.~{Lochard},
  \& L.~{Conroy}, 33

\bibitem[{{Bakos}(1976)}]{Bakos1976}
{Bakos}, G.~A. 1976, \jrasc, 70, 23

\bibitem[{{Balachandran} {et~al.}(2000){Balachandran}, {Fekel}, {Henry}, \&
  {Uitenbroek}}]{Balachandran2000}
{Balachandran}, S.~C., {Fekel}, F.~C., {Henry}, G.~W., \& {Uitenbroek}, H.
  2000, \apj, 542, 978

\bibitem[{{Baranne} {et~al.}(1979){Baranne}, {Mayor}, \&
  {Poncet}}]{Baranne1979}
{Baranne}, A., {Mayor}, M., \& {Poncet}, J.~L. 1979, Vistas in Astronomy, 23,
  279

\bibitem[{{Barnbaum} {et~al.}(1996){Barnbaum}, {Stone}, \&
  {Keenan}}]{Barnbaum1996}
{Barnbaum}, C., {Stone}, R. P.~S., \& {Keenan}, P.~C. 1996, \apjs, 105, 419

\bibitem[{{Bessel} {et~al.}(1998){Bessel}, {Castelli}, \& {Plez}}]{Bessel1998}
{Bessel}, M.~S., {Castelli}, F., \& {Plez}, B. 1998, \aap, 333, 231

\bibitem[{{Bharat Kumar} {et~al.}(2015){Bharat Kumar}, {Reddy},
  {Muthumariappan}, \& {Zhao}}]{BharatKumar2015}
{Bharat Kumar}, Y., {Reddy}, B.~E., {Muthumariappan}, C., \& {Zhao}, G. 2015,
  \aap, 577, A10

\bibitem[{{Bharat Kumar} {et~al.}(2018){Bharat Kumar}, {Singh}, {Eswar Reddy},
  \& {Zhao}}]{BharatKumar2018}
{Bharat Kumar}, Y., {Singh}, R., {Eswar Reddy}, B., \& {Zhao}, G. 2018, \apjl,
  858, L22

\bibitem[{{Blackwell} \& {Lynas-Gray}(1998)}]{Blackwell1998}
{Blackwell}, D.~E. \& {Lynas-Gray}, A.~E. 1998, \aaps, 129, 505

\bibitem[{{Brown} {et~al.}(1989){Brown}, {Sneden}, {Lambert}, \&
  {Dutchover}}]{Brown1989}
{Brown}, J.~A., {Sneden}, C., {Lambert}, D.~L., \& {Dutchover}, Edward, J.
  1989, \apjs, 71, 293

\bibitem[{{Cardelli} {et~al.}(1989){Cardelli}, {Clayton}, \&
  {Mathis}}]{Cardelli1989}
{Cardelli}, J.~A., {Clayton}, G.~C., \& {Mathis}, J.~S. 1989, \apj, 345, 245

\bibitem[{{Carlberg} {et~al.}(2012){Carlberg}, {Cunha}, {Smith}, \&
  {Majewski}}]{Carlberg2012}
{Carlberg}, J.~K., {Cunha}, K., {Smith}, V.~V., \& {Majewski}, S.~R. 2012,
  \apj, 757, 109

\bibitem[{{Carney} {et~al.}(2003){Carney}, {Latham}, {Stefanik}, {Laird}, \&
  {Morse}}]{Carney2003}
{Carney}, B.~W., {Latham}, D.~W., {Stefanik}, R.~P., {Laird}, J.~B., \&
  {Morse}, J.~A. 2003, \aj, 125, 293

\bibitem[{{Casey} {et~al.}(2019){Casey}, {Ho}, {Ness}, {Hogg}, {Rix},
  {Angelou}, {Hekker}, {Tout}, {Lattanzio}, {Karakas}, {Woods}, {Price-Whelan},
  \& {Schlaufman}}]{Casey2019}
{Casey}, A.~R., {Ho}, A. Y.~Q., {Ness}, M., {et~al.} 2019, \apj, 880, 125

\bibitem[{{Casey} {et~al.}(2016){Casey}, {Ruchti}, {Masseron}, {Randich},
  {Gilmore}, {Lind}, {Kennedy}, {Koposov}, {Hourihane}, {Franciosini}, {Lewis},
  {Magrini}, {Morbidelli}, {Sacco}, {Worley}, {Feltzing}, {Jeffries},
  {Vallenari}, {Bensby}, {Bragaglia}, {Flaccomio}, {Francois}, {Korn},
  {Lanzafame}, {Pancino}, {Recio-Blanco}, {Smiljanic}, {Carraro}, {Costado},
  {Damiani}, {Donati}, {Frasca}, {Jofr{\'e}}, {Lardo}, {de Laverny}, {Monaco},
  {Prisinzano}, {Sbordone}, {Sousa}, {Tautvai{\v s}ien{\.e}}, {Zaggia},
  {Zwitter}, {Delgado Mena}, {Chorniy}, {Martell}, {Silva Aguirre}, {Miglio},
  {Chiappini}, {Montalban}, {Morel}, \& {Valentini}}]{Casey2016}
{Casey}, A.~R., {Ruchti}, G., {Masseron}, T., {et~al.} 2016, \mnras, 461, 3336

\bibitem[{{Castilho} {et~al.}(1998){Castilho}, {Gregorio-Hetem}, {Spite},
  {Spite}, \& {Barbuy}}]{Castilho1998}
{Castilho}, B.~V., {Gregorio-Hetem}, J., {Spite}, F., {Spite}, M., \& {Barbuy},
  B. 1998, \aaps, 127, 139

\bibitem[{{Charbonnel} \& {Balachandran}(2000)}]{Charbonnel2000}
{Charbonnel}, C. \& {Balachandran}, S.~C. 2000, \aap, 359, 563

\bibitem[{{Charbonnel} \& {Lagarde}(2010)}]{Charbonnel2010}
{Charbonnel}, C. \& {Lagarde}, N. 2010, \aap, 522, A10

\bibitem[{{Charbonnel} {et~al.}(2019){Charbonnel}, {Lagarde}, {Jasniewicz},
  {North}, {Shetrone}, {Krugler Hollek}, {Smith}, {Smiljanic}, {Palacios}, \&
  {Ottoni}}]{Charbonnel2019}
{Charbonnel}, C., {Lagarde}, N., {Jasniewicz}, G., {et~al.} 2019, \aap, in
  press (arXiv:1910.12732)

\bibitem[{{Cutri}(2013)}]{Cutri2013}
{Cutri}, R.~M.~e. 2013, VizieR Online Data Catalog II/328, II/328

\bibitem[{{da Silva} {et~al.}(1995){da Silva}, {de La Reza}, \&
  {Barbuy}}]{daSilva1995}
{da Silva}, L., {de La Reza}, R., \& {Barbuy}, B. 1995, \apjl, 448, L41

\bibitem[{{de La Reza} \& {da Silva}(1995)}]{deLaReza1995}
{de La Reza}, R. \& {da Silva}, L. 1995, \apj, 439, 917

\bibitem[{{de La Reza} {et~al.}(1996){de La Reza}, {Drake}, \& {da
  Silva}}]{deLaReza1996}
{de La Reza}, R., {Drake}, N.~A., \& {da Silva}, L. 1996, \apjl, 456, L115

\bibitem[{{de la Reza} {et~al.}(1997){de la Reza}, {Drake}, {da Silva},
  {Torres}, \& {Martin}}]{deLaReza1997}
{de la Reza}, R., {Drake}, N.~A., {da Silva}, L., {Torres}, C.~A.~O., \&
  {Martin}, E.~L. 1997, \apjl, 482, L77

\bibitem[{{de la Reza} {et~al.}(2015){de la Reza}, {Drake}, {Oliveira}, \&
  {Rengaswamy}}]{delaReza2015}
{de la Reza}, R., {Drake}, N.~A., {Oliveira}, I., \& {Rengaswamy}, S. 2015,
  \apj, 806, 86

\bibitem[{{De Medeiros} {et~al.}(2000){De Medeiros}, {do Nascimento},
  {Sankarankutty}, {Costa}, \& {Maia}}]{deMedeiros2000}
{De Medeiros}, J.~R., {do Nascimento}, J.~D., J., {Sankarankutty}, S., {Costa},
  J.~M., \& {Maia}, M.~R.~G. 2000, \aap, 363, 239

\bibitem[{{De Medeiros} \& {Mayor}(1999)}]{deMedeiros1999}
{De Medeiros}, J.~R. \& {Mayor}, M. 1999, \aaps, 139, 433

\bibitem[{{De Medeiros} {et~al.}(1996){De Medeiros}, {Melo}, \&
  {Mayor}}]{deMedeiros1996}
{De Medeiros}, J.~R., {Melo}, C.~H.~F., \& {Mayor}, M. 1996, \aap, 309, 465

\bibitem[{{Deepak} \& {Reddy}(2019)}]{Deepak2019}
{Deepak} \& {Reddy}, B.~E. 2019, \mnras, 484, 2000

\bibitem[{{Denissenkov} \& {Herwig}(2004)}]{Denissenkov2004}
{Denissenkov}, P.~A. \& {Herwig}, F. 2004, \apj, 612, 1081

\bibitem[{{Drake} {et~al.}(2002){Drake}, {de la Reza}, {da Silva}, \&
  {Lambert}}]{Drake2002}
{Drake}, N.~A., {de la Reza}, R., {da Silva}, L., \& {Lambert}, D.~L. 2002,
  \aj, 123, 2703

\bibitem[{{ESA}(1997)}]{ESA1997}
{ESA}. 1997, {The HIPPARCOS and TYCHO catalogues}, Vol. 1200 (ESA Special
  Publication)

\bibitem[{{Escorza} {et~al.}(2017){Escorza}, {Boffin}, {Jorissen}, {Van Eck},
  {Siess}, {Van Winckel}, {Karinkuzhi}, {Shetye}, \& {Pourbaix}}]{Escorza2017}
{Escorza}, A., {Boffin}, H.~M.~J., {Jorissen}, A., {et~al.} 2017, \aap, 608,
  A100

\bibitem[{{Escorza} {et~al.}(2019){Escorza}, {Karinkuzhi}, {Jorissen}, {Siess},
  {Van Winckel}, {Pourbaix}, {Johnston}, {Miszalski}, {Oomen}, {Abdul-Masih},
  {Boffin}, {North}, {Manick}, {Shetye}, \& {Miko{\l}ajewska}}]{Escorza2019}
{Escorza}, A., {Karinkuzhi}, D., {Jorissen}, A., {et~al.} 2019, \aap, 626, A128

\bibitem[{{Famaey} {et~al.}(2005){Famaey}, {Jorissen}, {Luri}, {Mayor}, {Udry},
  {Dejonghe}, \& {Turon}}]{Famaey2005}
{Famaey}, B., {Jorissen}, A., {Luri}, X., {et~al.} 2005, \aap, 430, 165

\bibitem[{{Fekel} \& {Watson}(1998)}]{Fekel1998}
{Fekel}, F.~C. \& {Watson}, L.~C. 1998, \aj, 116, 2466

\bibitem[{{Fisher} {et~al.}(2003){Fisher}, {Telesco}, {Pi{\~n}a}, \&
  {Knacke}}]{Fisher2003}
{Fisher}, R.~S., {Telesco}, C.~M., {Pi{\~n}a}, R.~K., \& {Knacke}, R.~F. 2003,
  \apjl, 586, L91

\bibitem[{{Frankowski} {et~al.}(2007){Frankowski}, {Jancart}, \&
  {Jorissen}}]{Frankowski2007}
{Frankowski}, A., {Jancart}, S., \& {Jorissen}, A. 2007, \aap, 464, 377

\bibitem[{{Freeman} \& {Halton}(1951)}]{Freeman1951}
{Freeman}, G.~H. \& {Halton}, J. 1951, Biometrika, 38, 141

\bibitem[{{Gaia Collaboration} {et~al.}(2018){Gaia Collaboration}, {Brown},
  {Vallenari}, {Prusti}, {de Bruijne}, {Babusiaux}, {Bailer-Jones}, {Biermann},
  {Evans}, {Eyer}, {Jansen}, {Jordi}, {Klioner}, {Lammers}, {Lindegren},
  {Luri}, {Mignard}, {Panem}, {Pourbaix}, {Randich}, {Sartoretti}, {Siddiqui},
  {Soubiran}, {van Leeuwen}, {Walton}, {Arenou}, {Bastian}, {Cropper},
  {Drimmel}, {Katz}, {Lattanzi}, {Bakker}, {Cacciari}, {Casta{\~n}eda},
  {Chaoul}, {Cheek}, {De Angeli}, {Fabricius}, {Guerra}, {Holl}, {Masana},
  {Messineo}, {Mowlavi}, {Nienartowicz}, {Panuzzo}, {Portell}, {Riello},
  {Seabroke}, {Tanga}, {Th{\'e}venin}, {Gracia-Abril}, {Comoretto},
  {Garcia-Reinaldos}, {Teyssier}, {Altmann}, {Andrae}, {Audard},
  {Bellas-Velidis}, {Benson}, {Berthier}, {Blomme}, {Burgess}, {Busso},
  {Carry}, {Cellino}, {Clementini}, {Clotet}, {Creevey}, {Davidson}, {De
  Ridder}, {Delchambre}, {Dell'Oro}, {Ducourant},
  {Fern{\'a}ndez-Hern{\'a}ndez}, {Fouesneau}, {Fr{\'e}mat}, {Galluccio},
  {Garc{\'\i}a-Torres}, {Gonz{\'a}lez-N{\'u}{\~n}ez}, {Gonz{\'a}lez-Vidal},
  {Gosset}, {Guy}, {Halbwachs}, {Hambly}, {Harrison}, {Hern{\'a}ndez},
  {Hestroffer}, {Hodgkin}, {Hutton}, {Jasniewicz}, {Jean-Antoine-Piccolo},
  {Jordan}, {Korn}, {Krone-Martins}, {Lanzafame}, {Lebzelter}, {L{\"o}ffler},
  {Manteiga}, {Marrese}, {Mart{\'\i}n-Fleitas}, {Moitinho}, {Mora}, {Muinonen},
  {Osinde}, {Pancino}, {Pauwels}, {Petit}, {Recio-Blanco}, {Richards},
  {Rimoldini}, {Robin}, {Sarro}, {Siopis}, {Smith}, {Sozzetti}, {S{\"u}veges},
  {Torra}, {van Reeven}, {Abbas}, {Abreu Aramburu}, {Accart}, {Aerts},
  {Altavilla}, {{\'A}lvarez}, {Alvarez}, {Alves}, {Anderson}, {Andrei},
  {Anglada Varela}, {Antiche}, {Antoja}, {Arcay}, {Astraatmadja}, {Bach},
  {Baker}, {Balaguer-N{\'u}{\~n}ez}, {Balm}, {Barache}, {Barata}, {Barbato},
  {Barblan}, {Barklem}, {Barrado}, {Barros}, {Barstow}, {Bartholom{\'e}
  Mu{\~n}oz}, {Bassilana}, {Becciani}, {Bellazzini}, {Berihuete}, {Bertone},
  {Bianchi}, {Bienaym{\'e}}, {Blanco-Cuaresma}, {Boch}, {Boeche}, {Bombrun},
  {Borrachero}, {Bossini}, {Bouquillon}, {Bourda}, {Bragaglia}, {Bramante},
  {Breddels}, {Bressan}, {Brouillet}, {Br{\"u}semeister}, {Brugaletta},
  {Bucciarelli}, {Burlacu}, {Busonero}, {Butkevich}, {Buzzi}, {Caffau},
  {Cancelliere}, {Cannizzaro}, {Cantat-Gaudin}, {Carballo}, {Carlucci},
  {Carrasco}, {Casamiquela}, {Castellani}, {Castro-Ginard}, {Charlot},
  {Chemin}, {Chiavassa}, {Cocozza}, {Costigan}, {Cowell}, {Crifo}, {Crosta},
  {Crowley}, {Cuypers}, {Dafonte}, {Damerdji}, {Dapergolas}, {David}, {David},
  {de Laverny}, {De Luise}, {De March}, {de Martino}, {de Souza}, {de Torres},
  {Debosscher}, {del Pozo}, {Delbo}, {Delgado}, {Delgado}, {Di Matteo},
  {Diakite}, {Diener}, {Distefano}, {Dolding}, {Drazinos}, {Dur{\'a}n},
  {Edvardsson}, {Enke}, {Eriksson}, {Esquej}, {Eynard Bontemps}, {Fabre},
  {Fabrizio}, {Faigler}, {Falc{\~a}o}, {Farr{\`a}s Casas}, {Federici},
  {Fedorets}, {Fernique}, {Figueras}, {Filippi}, {Findeisen}, {Fonti},
  {Fraile}, {Fraser}, {Fr{\'e}zouls}, {Gai}, {Galleti}, {Garabato},
  {Garc{\'\i}a-Sedano}, {Garofalo}, {Garralda}, {Gavel}, {Gavras}, {Gerssen},
  {Geyer}, {Giacobbe}, {Gilmore}, {Girona}, {Giuffrida}, {Glass}, {Gomes},
  {Granvik}, {Gueguen}, {Guerrier}, {Guiraud}, {Guti{\'e}rrez-S{\'a}nchez},
  {Haigron}, {Hatzidimitriou}, {Hauser}, {Haywood}, {Heiter}, {Helmi}, {Heu},
  {Hilger}, {Hobbs}, {Hofmann}, {Holland}, {Huckle}, {Hypki}, {Icardi},
  {Jan{\ss}en}, {Jevardat de Fombelle}, {Jonker}, {Juh{\'a}sz}, {Julbe},
  {Karampelas}, {Kewley}, {Klar}, {Kochoska}, {Kohley}, {Kolenberg},
  {Kontizas}, {Kontizas}, {Koposov}, {Kordopatis}, {Kostrzewa-Rutkowska},
  {Koubsky}, {Lambert}, {Lanza}, {Lasne}, {Lavigne}, {Le Fustec}, {Le
  Poncin-Lafitte}, {Lebreton}, {Leccia}, {Leclerc}, {Lecoeur-Taibi},
  {Lenhardt}, {Leroux}, {Liao}, {Licata}, {Lindstr{\o}m}, {Lister}, {Livanou},
  {Lobel}, {L{\'o}pez}, {Managau}, {Mann}, {Mantelet}, {Marchal}, {Marchant},
  {Marconi}, {Marinoni}, {Marschalk{\'o}}, {Marshall}, {Martino}, {Marton},
  {Mary}, {Massari}, {Matijevi{\v{c}}}, {Mazeh}, {McMillan}, {Messina},
  {Michalik}, {Millar}, {Molina}, {Molinaro}, {Moln{\'a}r}, {Montegriffo},
  {Mor}, {Morbidelli}, {Morel}, {Morris}, {Mulone}, {Muraveva}, {Musella},
  {Nelemans}, {Nicastro}, {Noval}, {O'Mullane}, {Ord{\'e}novic},
  {Ord{\'o}{\~n}ez-Blanco}, {Osborne}, {Pagani}, {Pagano}, {Pailler},
  {Palacin}, {Palaversa}, {Panahi}, {Pawlak}, {Piersimoni}, {Pineau}, {Plachy},
  {Plum}, {Poggio}, {Poujoulet}, {Pr{\v{s}}a}, {Pulone}, {Racero}, {Ragaini},
  {Rambaux}, {Ramos-Lerate}, {Regibo}, {Reyl{\'e}}, {Riclet}, {Ripepi}, {Riva},
  {Rivard}, {Rixon}, {Roegiers}, {Roelens}, {Romero-G{\'o}mez}, {Rowell},
  {Royer}, {Ruiz-Dern}, {Sadowski}, {Sagrist{\`a} Sell{\'e}s}, {Sahlmann},
  {Salgado}, {Salguero}, {Sanna}, {Santana-Ros}, {Sarasso}, {Savietto},
  {Schultheis}, {Sciacca}, {Segol}, {Segovia}, {S{\'e}gransan}, {Shih},
  {Siltala}, {Silva}, {Smart}, {Smith}, {Solano}, {Solitro}, {Sordo}, {Soria
  Nieto}, {Souchay}, {Spagna}, {Spoto}, {Stampa}, {Steele},
  {Steidelm{\"u}ller}, {Stephenson}, {Stoev}, {Suess}, {Surdej}, {Szabados},
  {Szegedi-Elek}, {Tapiador}, {Taris}, {Tauran}, {Taylor}, {Teixeira},
  {Terrett}, {Teyssand ier}, {Thuillot}, {Titarenko}, {Torra Clotet}, {Turon},
  {Ulla}, {Utrilla}, {Uzzi}, {Vaillant}, {Valentini}, {Valette}, {van Elteren},
  {Van Hemelryck}, {van Leeuwen}, {Vaschetto}, {Vecchiato}, {Veljanoski},
  {Viala}, {Vicente}, {Vogt}, {von Essen}, {Voss}, {Votruba}, {Voutsinas},
  {Walmsley}, {Weiler}, {Wertz}, {Wevers}, {Wyrzykowski}, {Yoldas},
  {{\v{Z}}erjal}, {Ziaeepour}, {Zorec}, {Zschocke}, {Zucker}, {Zurbach}, \&
  {Zwitter}}]{Gaia2018}
{Gaia Collaboration}, {Brown}, A.~G.~A., {Vallenari}, A., {et~al.} 2018, \aap,
  616, A1

\bibitem[{{Gezer} {et~al.}(2015){Gezer}, {Van Winckel}, {Bozkurt}, {De Smedt},
  {Kamath}, {Hillen}, \& {Manick}}]{Gezer2015}
{Gezer}, I., {Van Winckel}, H., {Bozkurt}, Z., {et~al.} 2015, \mnras, 453, 133

\bibitem[{{Gontcharov}(2006)}]{Gontcharov2006}
{Gontcharov}, G.~A. 2006, Astronomy Letters, 32, 759

\bibitem[{{Gorlova} {et~al.}(2013){Gorlova}, {Van Winckel}, {Vos},
  {{\O}stensen}, {Jorissen}, {Van Eck}, \& {Ikonnikova}}]{Gorlova2013}
{Gorlova}, N., {Van Winckel}, H., {Vos}, J., {et~al.} 2013, in EAS Publications
  Series, Vol.~64, EAS Publications Series, ed. K.~{Pavlovski}, A.~{Tkachenko},
  \& G.~{Torres}, 163--170

\bibitem[{{Gregorio-Hetem} {et~al.}(1993){Gregorio-Hetem}, {Castilho}, \&
  {Barbuy}}]{Gregorio-Hetem1993}
{Gregorio-Hetem}, J., {Castilho}, B.~V., \& {Barbuy}, B. 1993, \aap, 268, L25

\bibitem[{{Griffin}(2013)}]{Griffin2013}
{Griffin}, R.~F. 2013, The Observatory, 133, 212

\bibitem[{{Gustafsson} {et~al.}(2008){Gustafsson}, {Edvardsson}, {Eriksson},
  {J{\o}rgensen}, {Nordlund}, \& {Plez}}]{Gustafsson2008}
{Gustafsson}, B., {Edvardsson}, B., {Eriksson}, K., {et~al.} 2008, \aap, 486,
  951

\bibitem[{{Harris} \& {McClure}(1983)}]{Harris1983}
{Harris}, H.~C. \& {McClure}, R.~D. 1983, \apjl, 265, L77

\bibitem[{{Hatzes} {et~al.}(2018){Hatzes}, {Endl}, {Cochran}, {MacQueen},
  {Han}, {Lee}, {Kim}, {Mkrtichian}, {D{\"o}llinger}, {Hartmann},
  {Karjalainen}, \& {Dreizler}}]{Hatzes2018}
{Hatzes}, A.~P., {Endl}, M., {Cochran}, W.~D., {et~al.} 2018, \aj, 155, 120

\bibitem[{{Hekker} \& {Mel{\'e}ndez}(2007)}]{Hekker2007}
{Hekker}, S. \& {Mel{\'e}ndez}, J. 2007, \aap, 475, 1003

\bibitem[{{Hekker} {et~al.}(2008){Hekker}, {Snellen}, {Aerts}, {Quirrenbach},
  {Reffert}, \& {Mitchell}}]{Hekker2008}
{Hekker}, S., {Snellen}, I.~A.~G., {Aerts}, C., {et~al.} 2008, \aap, 480, 215

\bibitem[{{Hrudkov{\'a}} {et~al.}(2017){Hrudkov{\'a}}, {Hatzes}, {Karjalainen},
  {Lehmann}, {Hekker}, {Hartmann}, {Tkachenko}, {Prins}, {Van Winckel}, {De
  Nutte}, {Dumortier}, {Fr{\'e}mat}, {Hensberge}, {Jorissen}, {Lampens},
  {Laverick}, {Lombaert}, {P{\'a}pics}, {Raskin}, {S{\'o}dor}, {Thoul}, {Van
  Eck}, \& {Waelkens}}]{Hrudkova2017}
{Hrudkov{\'a}}, M., {Hatzes}, A., {Karjalainen}, R., {et~al.} 2017, \mnras,
  464, 1018

\bibitem[{{Jancart} {et~al.}(2005){Jancart}, {Jorissen}, {Babusiaux}, \&
  {Pourbaix}}]{Jancart2005}
{Jancart}, S., {Jorissen}, A., {Babusiaux}, C., \& {Pourbaix}, D. 2005, \aap,
  442, 365

\bibitem[{{Jasniewicz} {et~al.}(1999){Jasniewicz}, {Parthasarathy}, {de
  Laverny}, \& {Th{\'e}venin}}]{Jasniewicz1999}
{Jasniewicz}, G., {Parthasarathy}, M., {de Laverny}, P., \& {Th{\'e}venin}, F.
  1999, \aap, 342, 831

\bibitem[{{Jordi}(2018)}]{Jordi2018}
{Jordi}, C. 2018, {Gaia DPAC report GAIA-C5-TN-UB-CJ-041}

\bibitem[{{Jorissen}(2020)}]{Jorissen2020}
{Jorissen}, A. 2020, Mem. Soc. Ast. Ital., in press

\bibitem[{{Jorissen} {et~al.}(2019){Jorissen}, {Boffin}, {Karinkuzhi}, {Van
  Eck}, {Escorza}, {Shetye}, \& {Van Winckel}}]{Jorissen2019}
{Jorissen}, A., {Boffin}, H.~M.~J., {Karinkuzhi}, D., {et~al.} 2019, \aap, 626,
  A127

\bibitem[{{Jorissen} {et~al.}(2016{\natexlab{a}}){Jorissen}, {Hansen}, {Van
  Eck}, {Andersen}, {Nordstr{\"o}m}, {Siess}, {Torres}, {Masseron}, \& {Van
  Winckel}}]{Jorissen2016b}
{Jorissen}, A., {Hansen}, T., {Van Eck}, S., {et~al.} 2016{\natexlab{a}}, \aap,
  586, A159

\bibitem[{{Jorissen} {et~al.}(2016{\natexlab{b}}){Jorissen}, {Van Eck}, {Van
  Winckel}, {Merle}, {Boffin}, {Andersen}, {Nordstr{\"o}m}, {Udry}, {Masseron},
  {Lenaerts}, \& {Waelkens}}]{Jorissen2016}
{Jorissen}, A., {Van Eck}, S., {Van Winckel}, H., {et~al.} 2016{\natexlab{b}},
  \aap, 586, A158

\bibitem[{{Jura}(1999)}]{Jura1999a}
{Jura}, M. 1999, \apj, 515, 706

\bibitem[{{Jura}(2003)}]{Jura2003}
{Jura}, M. 2003, \apj, 582, 1032

\bibitem[{{Kamath} {et~al.}(2016){Kamath}, {Wood}, {Van Winckel}, \&
  {Nie}}]{Kamath2016}
{Kamath}, D., {Wood}, P.~R., {Van Winckel}, H., \& {Nie}, J.~D. 2016, \aap,
  586, L5

\bibitem[{{Katz} \& {Brown}(2017)}]{Katz2017}
{Katz}, D. \& {Brown}, A.~G.~A. 2017, in SF2A-2017: Proceedings of the Annual
  meeting of the French Society of Astronomy and Astrophysics, Di

\bibitem[{{Kerschbaum} {et~al.}(2010){Kerschbaum}, {Lebzelter}, \&
  {Mekul}}]{Kerschbaum2010}
{Kerschbaum}, F., {Lebzelter}, T., \& {Mekul}, L. 2010, \aap, 524, A87

\bibitem[{{Kessler} {et~al.}(1996){Kessler}, {Steinz}, {Anderegg}, {Clavel},
  {Drechsel}, {Estaria}, {Faelker}, {Riedinger}, {Robson}, {Taylor}, \&
  {Xim{\'e}nez de Ferr{\'a}n}}]{Kessler1996}
{Kessler}, M.~F., {Steinz}, J.~A., {Anderegg}, M.~E., {et~al.} 1996, \aap, 315,
  L27

\bibitem[{{Kim} {et~al.}(2001){Kim}, {Zuckerman}, \& {Silverstone}}]{Kim2001}
{Kim}, S.~S., {Zuckerman}, B., \& {Silverstone}, M. 2001, \apj, 550, 1000

\bibitem[{{Kluska} {et~al.}(2019){Kluska}, {Van Winckel}, {Hillen}, {Berger},
  {Kamath}, {Le Bouquin}, \& {Min}}]{Kluska2019}
{Kluska}, J., {Van Winckel}, H., {Hillen}, M., {et~al.} 2019, \aap, in press

\bibitem[{{Koch} {et~al.}(2010){Koch}, {Borucki}, {Basri}, {Batalha}, {Brown},
  {Caldwell}, {Christensen-Dalsgaard}, {Cochran}, {DeVore}, {Dunham},
  {Gautier}, {Geary}, {Gilliland}, {Gould}, {Jenkins}, {Kondo}, {Latham},
  {Lissauer}, {Marcy}, {Monet}, {Sasselov}, {Boss}, {Brownlee}, {Caldwell},
  {Dupree}, {Howell}, {Kjeldsen}, {Meibom}, {Morrison}, {Owen}, {Reitsema},
  {Tarter}, {Bryson}, {Dotson}, {Gazis}, {Haas}, {Kolodziejczak}, {Rowe}, {Van
  Cleve}, {Allen}, {Chand rasekaran}, {Clarke}, {Li}, {Quintana}, {Tenenbaum},
  {Twicken}, \& {Wu}}]{Koch2010}
{Koch}, D.~G., {Borucki}, W.~J., {Basri}, G., {et~al.} 2010, \apjl, 713, L79

\bibitem[{{Kumar} {et~al.}(2015){Kumar}, {Reddy}, {Muthumariappan}, \&
  {Zhao}}]{Kumar2015}
{Kumar}, B.~Y., {Reddy}, B.~E., {Muthumariappan}, C., \& {Zhao}, G. 2015, \aap,
  577, A10

\bibitem[{{Kumar} {et~al.}(2011){Kumar}, {Reddy}, \& {Lambert}}]{Kumar2011}
{Kumar}, Y.~B., {Reddy}, B.~E., \& {Lambert}, D.~L. 2011, \apjl, 730, L12

\bibitem[{{Lattanzio} {et~al.}(2015){Lattanzio}, {Siess}, {Church}, {Angelou},
  {Stancliffe}, {Doherty}, {Stephen}, \& {Campbell}}]{Lattanzio2015}
{Lattanzio}, J.~C., {Siess}, L., {Church}, R.~P., {et~al.} 2015, \mnras, 446,
  2673

\bibitem[{{Lebzelter} {et~al.}(2012){Lebzelter}, {Uttenthaler}, {Busso},
  {Schultheis}, \& {Aringer}}]{Lebzelter2012}
{Lebzelter}, T., {Uttenthaler}, S., {Busso}, M., {Schultheis}, M., \&
  {Aringer}, B. 2012, \aap, 538, A36

\bibitem[{{Liu} {et~al.}(2014){Liu}, {Tan}, {Wang}, {Zhao}, {Sato}, {Takeda},
  \& {Li}}]{Liu2014}
{Liu}, Y.~J., {Tan}, K.~F., {Wang}, L., {et~al.} 2014, \apj, 785, 94

\bibitem[{{Luck}(2015)}]{Luck2015}
{Luck}, R.~E. 2015, \aj, 150, 88

\bibitem[{{Makarov} \& {Kaplan}(2005)}]{Makarov2005}
{Makarov}, V.~V. \& {Kaplan}, G.~H. 2005, \aj, 129, 2420

\bibitem[{{Manick} {et~al.}(2017){Manick}, {Van Winckel}, {Kamath}, {Hillen},
  \& {Escorza}}]{Manick2017}
{Manick}, R., {Van Winckel}, H., {Kamath}, D., {Hillen}, M., \& {Escorza}, A.
  2017, \aap, 597, A129

\bibitem[{{Martell} \& {Shetrone}(2013)}]{Martell2013}
{Martell}, S.~L. \& {Shetrone}, M.~D. 2013, \mnras, 430, 611

\bibitem[{{Mayor} {et~al.}(1984){Mayor}, {Imbert}, {Andersen}, {Ardeberg},
  {Benz}, {Lindgren}, {Martin}, {Maurice}, {Nordstrom}, \&
  {Prevot}}]{Mayor1984}
{Mayor}, M., {Imbert}, M., {Andersen}, J., {et~al.} 1984, \aap, 134, 118

\bibitem[{{McWilliam}(1990)}]{McWilliam1990}
{McWilliam}, A. 1990, \apjs, 74, 1075

\bibitem[{{Merle} {et~al.}(2016){Merle}, {Jorissen}, {Van Eck}, {Masseron}, \&
  {Van Winckel}}]{Merle2016}
{Merle}, T., {Jorissen}, A., {Van Eck}, S., {Masseron}, T., \& {Van Winckel},
  H. 2016, \aap, 586, A151

\bibitem[{{Merle} {et~al.}(2020){Merle}, {Van der Swaelmen}, {Van Eck},
  {Jorissen}, {Jackson}, {Traven}, {Zwitter}, {Pourbaix}, {Klutsch}, {Sacco},
  {Blomme}, {Masseron}, {Gilmore}, {Randich}, {Badenes}, {Bayo}, {Bensby},
  {Bergemann}, {Biazzo}, {Damiani}, {Feuillet}, {Frasca}, {Gonneau},
  {Jeffries}, {Jofr{\'e}}, {Morbidelli}, {Mowlavi}, {Pancino}, \&
  {Prisinzano}}]{Merle2020}
{Merle}, T., {Van der Swaelmen}, M., {Van Eck}, S., {et~al.} 2020, \aap, 635,
  A155

\bibitem[{{Mermilliod} {et~al.}(2007){Mermilliod}, {Andersen}, {Latham}, \&
  {Mayor}}]{Mermilliod2007}
{Mermilliod}, J.-C., {Andersen}, J., {Latham}, D.~W., \& {Mayor}, M. 2007,
  \aap, 473, 829

\bibitem[{{Mishenina} {et~al.}(2006){Mishenina}, {Bienaym{\'e}}, {Gorbaneva},
  {Charbonnel}, {Soubiran}, {Korotin}, \& {Kovtyukh}}]{Mishenina2006}
{Mishenina}, T.~V., {Bienaym{\'e}}, O., {Gorbaneva}, T.~I., {et~al.} 2006,
  \aap, 456, 1109

\bibitem[{{Mosser} {et~al.}(2014){Mosser}, {Benomar}, {Belkacem}, {Goupil},
  {Lagarde}, {Michel}, {Lebreton}, {Stello}, {Vrard}, {Barban}, {Bedding},
  {Deheuvels}, {Chaplin}, {De Ridder}, {Elsworth}, {Montalban}, {Noels},
  {Ouazzani}, {Samadi}, {White}, \& {Kjeldsen}}]{Mosser2014}
{Mosser}, B., {Benomar}, O., {Belkacem}, K., {et~al.} 2014, \aap, 572, L5

\bibitem[{{Murakami} {et~al.}(2007){Murakami}, {Baba}, {Barthel}, {Clements},
  {Cohen}, {Doi}, {Enya}, {Figueredo}, {Fujishiro}, {Fujiwara}, {Fujiwara},
  {Garcia-Lario}, {Goto}, {Hasegawa}, {Hibi}, {Hirao}, {Hiromoto}, {Hong},
  {Imai}, {Ishigaki}, {Ishiguro}, {Ishihara}, {Ita}, {Jeong}, {Jeong},
  {Kaneda}, {Kataza}, {Kawada}, {Kawai}, {Kawamura}, {Kessler}, {Kester},
  {Kii}, {Kim}, {Kim}, {Kobayashi}, {Koo}, {Kwon}, {Lee}, {Lorente}, {Makiuti},
  {Matsuhara}, {Matsumoto}, {Matsuo}, {Matsuura}, {M{\"U}ller}, {Murakami},
  {Nagata}, {Nakagawa}, {Naoi}, {Narita}, {Noda}, {Oh}, {Ohnishi}, {Ohyama},
  {Okada}, {Okuda}, {Oliver}, {Onaka}, {Ootsubo}, {Oyabu}, {Pak}, {Park},
  {Pearson}, {Rowan-Robinson}, {Saito}, {Sakon}, {Salama}, {Sato}, {Savage},
  {Serjeant}, {Shibai}, {Shirahata}, {Sohn}, {Suzuki}, {Takagi}, {Takahashi},
  {Tanab{\'E}}, {Takeuchi}, {Takita}, {Thomson}, {Uemizu}, {Ueno}, {Usui},
  {Verdugo}, {Wada}, {Wang}, {Watabe}, {Watarai}, {White}, {Yamamura},
  {Yamauchi}, \& {Yasuda}}]{Murakami2007}
{Murakami}, H., {Baba}, H., {Barthel}, P., {et~al.} 2007, \pasj, 59, S369

\bibitem[{{Neugebauer} {et~al.}(1984){Neugebauer}, {Habing}, {van Duinen},
  {Aumann}, {Baud}, {Beichman}, {Beintema}, {Boggess}, {Clegg}, {de Jong},
  {Emerson}, {Gautier}, {Gillett}, {Harris}, {Hauser}, {Houck}, {Jennings},
  {Low}, {Marsden}, {Miley}, {Olnon}, {Pottasch}, {Raimond}, {Rowan-Robinson},
  {Soifer}, {Walker}, {Wesselius}, \& {Young}}]{Neugebauer1984}
{Neugebauer}, G., {Habing}, H.~J., {van Duinen}, R., {et~al.} 1984, \apjl, 278,
  L1

\bibitem[{{Oomen} {et~al.}(2018){Oomen}, {Van Winckel}, {Pols}, {Nelemans},
  {Escorza}, {Manick}, {Kamath}, \& {Waelkens}}]{Oomen2018}
{Oomen}, G.-M., {Van Winckel}, H., {Pols}, O., {et~al.} 2018, \aap, 620, A85

\bibitem[{{Plets} {et~al.}(1997){Plets}, {Waelkens}, {Oudmaijer}, \&
  {Waters}}]{Plets1997}
{Plets}, H., {Waelkens}, C., {Oudmaijer}, R.~D., \& {Waters}, L.~B.~F.~M. 1997,
  \aap, 323, 513

\bibitem[{{Pourbaix} {et~al.}(2004){Pourbaix}, {Tokovinin}, {Batten}, {Fekel},
  {Hartkopf}, {Levato}, {Morrell}, {Torres}, \& {Udry}}]{Pourbaix2004}
{Pourbaix}, D., {Tokovinin}, A.~A., {Batten}, A.~H., {et~al.} 2004, \aap, 424,
  727

\bibitem[{{Press} {et~al.}(2007){Press}, {Teukolsky}, {Vetterling}, \&
  {Flannery}}]{Press2007}
{Press}, W.~H., {Teukolsky}, S.~A., {Vetterling}, W.~T., \& {Flannery}, B.
  2007, {Numerical Recipes: The Art of Scientific Computing} ({Cambridge
  University Press, 3rd edition})

\bibitem[{{Raskin} {et~al.}(2011){Raskin}, {van Winckel}, {Hensberge},
  {Jorissen}, {Lehmann}, {Waelkens}, {Avila}, {de Cuyper}, {Degroote},
  {Dubosson}, {Dumortier}, {Fr{\'e}mat}, {Laux}, {Michaud}, {Morren}, {Perez
  Padilla}, {Pessemier}, {Prins}, {Smolders}, {van Eck}, \&
  {Winkler}}]{Raskin2011}
{Raskin}, G., {van Winckel}, H., {Hensberge}, H., {et~al.} 2011, \aap, 526, A69

\bibitem[{{Rebull} {et~al.}(2015){Rebull}, {Carlberg}, {Gibbs}, {Deeb},
  {Larsen}, {Black}, {Altepeter}, {Bucksbee}, {Cashen}, {Clarke}, {Datta},
  {Hodgson}, \& {Lince}}]{Rebull2015}
{Rebull}, L.~M., {Carlberg}, J.~K., {Gibbs}, J.~C., {et~al.} 2015, \aj, 150,
  123

\bibitem[{{Reddy} \& {Lambert}(2005)}]{Reddy2005}
{Reddy}, B.~E. \& {Lambert}, D.~L. 2005, \aj, 129, 2831

\bibitem[{{Reddy} {et~al.}(2002){Reddy}, {Lambert}, {Hrivnak}, \&
  {Bakker}}]{Reddy2002}
{Reddy}, B.~E., {Lambert}, D.~L., {Hrivnak}, B.~J., \& {Bakker}, E.~J. 2002,
  \aj, 123, 1993

\bibitem[{{Rowan-Robinson}(2000)}]{Rowan-Robinson2000}
{Rowan-Robinson}, M. 2000, \mnras, 316, 885

\bibitem[{{Ruchti} {et~al.}(2011){Ruchti}, {Fulbright}, {Wyse}, {Gilmore},
  {Grebel}, {Bienaym{\'e}}, {Bland-Hawthorn}, {Freeman}, {Gibson}, {Munari},
  {Navarro}, {Parker}, {Reid}, {Seabroke}, {Siebert}, {Siviero}, {Steinmetz},
  {Watson}, {Williams}, \& {Zwitter}}]{Ruchti2011}
{Ruchti}, G.~R., {Fulbright}, J.~P., {Wyse}, R.~F.~G., {et~al.} 2011, \apj,
  743, 107

\bibitem[{{Siess} \& {Arnould}(2008)}]{Siess2008}
{Siess}, L. \& {Arnould}, M. 2008, \aap, 489, 395

\bibitem[{{Siess} {et~al.}(2000){Siess}, {Dufour}, \& {Forestini}}]{Siess2000}
{Siess}, L., {Dufour}, E., \& {Forestini}, M. 2000, \aap, 358, 593

\bibitem[{{Siess} \& {Livio}(1999)}]{Siess1999}
{Siess}, L. \& {Livio}, M. 1999, \mnras, 308, 1133

\bibitem[{{Singh} {et~al.}(2019){Singh}, {Reddy}, {Bharat Kumar}, \&
  {Antia}}]{Singh2019}
{Singh}, R., {Reddy}, B.~E., {Bharat Kumar}, Y., \& {Antia}, H.~M. 2019, \apjl,
  878, L21

\bibitem[{{Smiljanic} {et~al.}(2018){Smiljanic}, {Franciosini}, {Bragaglia},
  {Tautvai{\v{s}}ien{\.{e}}}, {Fu}, {Pancino}, {Adibekyan}, {Sousa}, {Randich},
  {Montalb{\'a}n}, {Pasquini}, {Magrini}, {Drazdauskas}, {Garc{\'\i}a},
  {Mathur}, {Mosser}, {R{\'e}gulo}, {de Assis Peralta}, {Hekker}, {Feuillet},
  {Valentini}, {Morel}, {Martell}, {Gilmore}, {Feltzing}, {Vallenari},
  {Bensby}, {Korn}, {Lanzafame}, {Recio-Blanco}, {Bayo}, {Carraro}, {Costado},
  {Frasca}, {Jofr{\'e}}, {Lardo}, {de Laverny}, {Lind}, {Masseron}, {Monaco},
  {Morbidelli}, {Prisinzano}, {Sbordone}, \& {Zaggia}}]{Smiljanic2018}
{Smiljanic}, R., {Franciosini}, E., {Bragaglia}, A., {et~al.} 2018, \aap, 617,
  A4

\bibitem[{{Stellingwerf}(1978)}]{Stellingwerf1978}
{Stellingwerf}, R.~F. 1978, \apj, 224, 953

\bibitem[{{Stock} {et~al.}(2018){Stock}, {Reffert}, \&
  {Quirrenbach}}]{Stock2018}
{Stock}, S., {Reffert}, S., \& {Quirrenbach}, A. 2018, \aap, 616, A33

\bibitem[{{Takeda} \& {Tajitsu}(2017)}]{Takeda2017}
{Takeda}, Y. \& {Tajitsu}, A. 2017, \pasj, 69, 74

\bibitem[{{Tautvai{\v{s}}ien{\.{e}}} {et~al.}(2013){Tautvai{\v{s}}ien{\.{e}}},
  {Barisevi{\v{c}}ius}, {Chorniy}, {Ilyin}, \& {Puzeras}}]{Tautvaisiene2013}
{Tautvai{\v{s}}ien{\.{e}}}, G., {Barisevi{\v{c}}ius}, G., {Chorniy}, Y.,
  {Ilyin}, I., \& {Puzeras}, E. 2013, \mnras, 430, 621

\bibitem[{{Tokovinin}(2014)}]{Tokovinin2014}
{Tokovinin}, A. 2014, \aj, 147, 87

\bibitem[{{Udry} {et~al.}(1999){Udry}, {Mayor}, \& {Queloz}}]{Udry1999}
{Udry}, S., {Mayor}, M., \& {Queloz}, D. 1999, in Astronomical Society of the
  Pacific Conference Series, Vol. 185, IAU Colloq. 170: Precise Stellar Radial
  Velocities, ed. J.~B. {Hearnshaw} \& C.~D. {Scarfe}, 367

\bibitem[{{Van Winckel}(2003)}]{VanWinckel2003}
{Van Winckel}, H. 2003, \araa, 41, 391

\bibitem[{{Vos} {et~al.}(2015){Vos}, {{\O}stensen}, {Marchant}, \& {Van
  Winckel}}]{Vos2015}
{Vos}, J., {{\O}stensen}, R.~H., {Marchant}, P., \& {Van Winckel}, H. 2015,
  \aap, 579, A49

\bibitem[{{Wallerstein} \& {Sneden}(1982)}]{Wallerstein1982}
{Wallerstein}, G. \& {Sneden}, C. 1982, \apj, 255, 577

\bibitem[{{Waters} {et~al.}(1999){Waters}, {Beintema}, {Cami}, {de Graauw},
  {Hony}, {de Jong}, {Justtanont}, {Kemper}, {de Koter}, {van Loon}, {Molster},
  {Tielens}, {Waelkens}, {Van Winckel}, \& {Yamamura}}]{Waters1999}
{Waters}, L.~B.~F.~M., {Beintema}, D.~A., {Cami}, J., {et~al.} 1999, in ESA
  Special Publication, Vol. 427, The Universe as Seen by ISO, ed. P.~{Cox} \&
  M.~{Kessler}, 219

\bibitem[{{Wilson} \& {Hilferty}(1931)}]{Wilson-hilferty1931}
{Wilson}, E.~B. \& {Hilferty}, M.~M. 1931, Proceedings of the National Academy
  of Science, 17, 684

\bibitem[{{Wright} {et~al.}(2010){Wright}, {Eisenhardt}, {Mainzer}, {Ressler},
  {Cutri}, {Jarrett}, {Kirkpatrick}, {Padgett}, {McMillan}, {Skrutskie},
  {Stanford}, {Cohen}, {Walker}, {Mather}, {Leisawitz}, {Gautier}, {McLean},
  {Benford}, {Lonsdale}, {Blain}, {Mendez}, {Irace}, {Duval}, {Liu}, {Royer},
  {Heinrichsen}, {Howard}, {Shannon}, {Kendall}, {Walsh}, {Larsen}, {Cardon},
  {Schick}, {Schwalm}, {Abid}, {Fabinsky}, {Naes}, \& {Tsai}}]{Wright2010}
{Wright}, E.~L., {Eisenhardt}, P.~R.~M., {Mainzer}, A.~K., {et~al.} 2010, \aj,
  140, 1868

\bibitem[{{Zuckerman} {et~al.}(1995){Zuckerman}, {Kim}, \&
  {Liu}}]{Zuckerman1995}
{Zuckerman}, B., {Kim}, S.~S., \& {Liu}, T. 1995, \apjl, 446, L79

\end{thebibliography}

\begin{appendix}

\section{Detailed properties of samples S and R}

Table~\ref{Tab:sample2}
lists all relevant properties for sample S2 of Li-rich K giants collected from \citet{BharatKumar2015} and \citet{Charbonnel2019}. In Table~\ref{Tab:binary_R}, all details are provided about sample R consisting in 190 Kepler and CoRoT K giants, serving as reference sample to which the binary frequency among Li-rich giants may be compared.

\begin{table*}
\caption[]{\label{Tab:sample2}
Sample S2 of Li-rich K giants collected from \citet{BharatKumar2015} and \citet{Charbonnel2019}. An asterisk in column `Name' denotes a star already present in sample S1. When $V-I$ is not available to convert Gaia $G$ magnitudes  into  
$G_{\rm RVS}$ magnitudes, a typical offset of 1 mag  was adopted. Values of the Gaia RV standard deviation $\sigma_{\rm RV}$ in bold face correspond to stars flagged as binaries. }
\begin{tabular}{lrrrrrrrrrrr}
\hline\\
Name            &$V-I$&$G$&$G_{\rm RVS}$&\multicolumn{2}{c}{$\epsilon_{\rm RV}$}&\multicolumn{1}{c}{$\sigma_{\rm RV}$} & $N_{\rm RV}$\\
\cline{5-6}
         & & & &$N_{\rm RV}=8$ & $N_{\rm RV}=40$\\
         & & & & (\kms) & (\kms) & (\kms) \\
\hline\\ 
HD 3750         &1.06&5.68&4.71&0.26&0.13& 0.12& 40 \\ 
HD 4042         &0.92&6.51&5.65&0.25&0.12& 0.13& 19 \\ 
HD 5395         &1.01&4.28&3.35&0.35&0.17& 0.42& 12 \\ 
HD 6665 *       &1.15&8.06&7.03&0.27&0.13& 0.19&  7 \\ 
HD 8676         &1.02&7.48&6.54&0.26&0.13& 0.13& 45 \\ 
HD 9746 *       &1.20&5.46&4.39&0.28&0.14& 0.14& 16 \\ 
HD 10437        &1.04&6.26&5.31&0.25&0.13& 0.12& 38  \\ 
HD 12203        &0.98&6.47&5.56&0.25&0.12& 0.13& 11  \\ 
HD 19745        &&8.78&7.78&0.29&0.15& 0.19& 10 \\ 
HD 30197        &1.10&5.62&4.63&0.27&0.13& {\bf 1.79}& 17\\  
HD 37719        &1.04&7.33&6.38&0.25&0.13& 0.16& 12 \\ 
HD 40168        &0.99&6.58&5.66&0.25&0.12& 0.20& 12 \\ 
HD 40827 *       &1.08&6.02&5.04&0.26&0.13& 0.18& 15  \\ 
HD 51367        &1.11&6.67&5.67&0.25&0.12& 0.15& 13  \\ 
HD 63798 *       &0.93&6.23&5.37&0.25&0.13& 0.15& 19 \\ 
HD 233517 *      &&9.27&8.27&0.31&0.16& {\bf 1.53}&  3 \\ 
HD 77361        &1.08&5.88&4.90&0.26&0.13& 0.12& 20 \\  
HD  83506       &1.00&4.84&3.92&0.31&0.15& 0.21& 16 \\ 
HD  85563       &1.12&5.25&4.24&0.29&0.14& {\bf 1.99}& 22 \\ 
HD  88476       &0.93&6.61&5.74&0.25&0.12& 0.15&  8  \\ 
HD  90507       &0.92&6.50&5.64&0.25&0.12& 0.14& 12  \\ 
HD  90633 *     &1.10&6.00&5.01&0.26&0.13& 0.13& 17 \\ 
HD  93859       &1.10&5.31&4.32&0.28&0.14& 0.14& 11 \\ 
HD 102845       &0.94&5.87&4.99&0.26&0.13& 0.14& 21 \\ 
HD 106574       &1.14&5.33&4.31&0.28&0.14& 0.15& 14 \\ 
HD 107484       &1.12&7.39&6.38&0.25&0.13& 0.15& 17 \\ 
HD 108471       &0.93&6.10&5.24&0.25&0.13& 0.16& 18  \\ 
HD 112127 *     &1.15&6.53&5.50&0.25&0.12& 0.14& 24  \\ 
HD 113049       &1.00&5.70&4.79&0.26&0.13& {\bf 0.49}& 22 \\ 
HD 115299       &1.08&7.21&6.23&0.25&0.13& 0.24& 11 \\ 
HD 116292 *     &0.97&5.05&4.15&0.29&0.15& 0.14& 11 \\ 
IRAS 13313-5838 & &12.34&11.34&1.49&0.75& 0.75& 21 \\ 
HD 118319       &1.00&6.18&5.26&0.25&0.13& 0.12&  4 \\ 
HD 120602       &0.90&5.77&4.93&0.26&0.13& 0.13&  8 \\ 
IRAS 13539-4153 & &12.16&11.16&1.31&0.66& 0.50&  7 \\ 
HD 133086       &0.97&6.56&5.66&0.25&0.12& 0.14& 13 \\ 
HD 145457       &1.01&6.28&5.35&0.25&0.13& 0.13& 24\\  
HD 148293       &1.08&4.92&3.94&0.30&0.15& 0.13& 10  \\ 
HD 150902       &1.04&7.63&6.68&0.26&0.13& 0.14& 14 \\ 
HD 160781       &1.23&5.54&4.45&0.28&0.14& 0.14& 12 \\ 
IRAS 17596-3952 & &11.80&10.80&1.01&0.51& {\bf 2.70}&  3 \\ 
HD 167304       &1.01&6.07&5.14&0.25&0.13& 0.14& 13 \\ 
HD 170527       &0.97&6.72&5.82&0.25&0.12& 0.32& 12 \\ 
TYC 3105-00152-1& &9.53&8.53&0.33&0.17& 0.23&  3 \\ 
HD 183492       &1.02&5.26&4.32&0.28&0.14& 0.17& 14 \\ 
V859 Aql        &&9.78&8.78&0.35&0.18& 0.19& 10 \\ 
HD 186815       &0.89&6.03&5.20&0.25&0.13& 0.15& 11\\  
HD 194937       &1.04&5.92&4.97&0.26&0.13& 0.15& 18 \\ 
HD 196857       &0.98&5.50&4.59&0.27&0.13& 0.15& 13 \\ 
HD 199437       &1.11&4.98&3.98&0.30&0.15& 0.15& 13 \\ 
HD 203136       &0.93&7.51&6.64&0.26&0.13& 0.16& 16 \\ 
HD 206078       &0.95&6.85&5.97&0.25&0.12& 0.13& 15 \\ 
HD 212430       &0.95&5.49&4.61&0.27&0.13& 0.15&  5 \\ 
HD 214995       &1.08&5.58&4.60&0.27&0.13& 0.23&  6 \\ 
HD 217352       &1.11&6.76&5.76&0.25&0.12& {\bf 0.78}&  7  \\ 
HD 219025       &1.17&7.33&6.29&0.25&0.13& 0.30& 16 \\ 
\hline\\
\end{tabular}
\end{table*}

\renewcommand{\tabcolsep}{3pt}
\begin{table*}
\caption[]{Sample R (Kepler and CoRoT K giants) and its binary properties (see Sect.~\ref{Sect:methodology}). 
The column labelled $\sigma_j(RV)$ lists the RV standard deviation for the corresponding star $j$. For systems with orbital solutions, the third column lists the standard deviation of the $O-C$ residuals. $N$obsH$_j$ is the number of HERMES RV observations for star $j$, and $\overline{RV}_H$ their average; $\Delta t$ is their time span. The next two columns list the reduced $\chi^2$, i.e. $\chi_j^2/\nu_j$ (with $\nu_j = N$obsH$_j - 1$), and the associated probability that the star has a variable RV. The column labelled `Type' mentions whether the star  is Li-rich or surrounded by dust. The column labelled `SB' marks the final decision about binarity, from HERMES  and Gaia DR2 \citep{Gaia2018} data (see text). In column labelled `SB', `outlier' means that one data point is responsible for the larger than expected $\sigma_j(RV)$ value. The Gaia DR2 RV and its associated standard deviation is listed in column $RV_G$. $\epsilon(RV_G)$ is the expected uncertainty on the Gaia DR2 RV (see Eq.~\ref{Eq:eps_RV}).  Values in bold face identify fulfilled binary criteria.
\label{Tab:binary_R}
}
{\fontsize{8}{10}\selectfont
\begin{tabular}{rrcrrrrlrrcccrrlll}
\hline\\
KIC/CoRoT &  \multicolumn{1}{c}{$\sigma_j(RV)$} & $\sigma(O-C)$ & \multicolumn{1}{c}{$\overline{RV}_H$} & $N$obsH & $\Delta t$  & $\chi_j^2/\nu_j$ & Prob. & $F2_j$ &  \multicolumn{2}{c}{$\overline{RV}_G\pm\sigma(RV_G)$} & \multicolumn{1}{c}{$\epsilon(RV_G)$} & $|\overline{RV}_G - \overline{RV}_H|$ &  \multicolumn{1}{c}{$G$} & $G_{\rm RVS}$ & SB \\ 
    & \multicolumn{1}{c}{(\kms)} & (\kms) & (\kms) & & (d) & &RV var & & \multicolumn{2}{c}{(\kms)} &\multicolumn{1}{c}{(\kms)} &  \multicolumn{1}{c}{(\kms)} \\
   \hline
   \medskip\\
1160789&0.05&-&-25.51&3&298&0.61&0.46&-0.12&&&&&8.86&\\
1161618&2.28&-&-51.05&6&1074&1060&{\bf 1.00}&43.84&$-47.14$&$0.18$&
0.37&{\bf 3.91}&10.17&8.96&Y\\
1433730&0.07&-&-2.21&5&753&1.13&0.66&0.41&$-1.82$&$0.35$&0.57&
0.39&10.87&9.92&\\
1572780&0.04&-&20.45&3&761&0.30&0.26&-0.67&$21.14$&$0.26$&
0.38&0.69&10.26&8.99&\\
1720425&0.02&-&-19.56&3&758&0.12&0.12&-1.18&       
&& &&9.69&\\
1724879&0.09&-&-16.99&6&1362&1.53&0.82&0.94&$-16.57$&$
0.36$&0.49&0.42&10.74&9.62&\\
1725190&0.06&-&-18.16&4&269&0.85&0.53&0.08&$-17.71$&$
0.31$&0.34&0.45&9.19&8.63&\\
1726211&0.15&-&-127.21&3&748&4.30&0.99&2.21&$-126.85$&$
0.36$&0.57&0.36&10.92&9.90&\\
1864959&0.04&-&-27.21&5&674&0.33&0.14&-1.08&$-27.37$&$0.20$&
0.28&0.16&8.75&7.61&\\
1865747&0.10&-&18.49&4&1071&1.89&0.87&1.14&$18.58$&$
0.75$&0.51&0.09&10.75&9.72&\\
1995859&0.08&-&17.21&5&679&1.42&0.77&0.76&$17.51$&$0.30$&
0.55&0.30&10.86&9.86&\\
2015820&0.09&-&-19.67&8&1084&1.70&0.90&1.26&$-19.18$&$0.25$&
0.34&0.49&9.51&8.59&\\
2016706&0.08&-&-7.06&3&760&1.20&0.70&0.52&$-6.72$&$0.33$&
0.48&0.34&10.59&9.59&\\
2140982&0.05&&-5.25&3&323&0.45&0.36&-0.37&-5.00&0.79&0.47&0.25&10.55&9.55& &\\
2141928&0.09&&-56.97&3&324&1.69&0.82&0.91&-56.49&0.34&0.51&0.48&10.70&9.70&&\\
2160572&0.09&&30.94&3&671&1.61&0.80&0.85&30.77&0.29&0.56&0.17&10.90&9.88&&\\
2164327&0.33&&-4.73&4&1070&21.89&{\bf 1.00}&6.88&-3.84&0.37&0.53&0.89&10.79&9.79&Y\\
2167774&5.66&&-25.76&7&1089&6533&{\bf 1.00}&92.14&-22.94&{\bf 4.34}&0.36&{\bf 2.82}&9.87&8.87&Y\\
2283075&8.67&0.107&15.47&18&1037&15331&{\bf 1.00}&208.65&17.88&{\bf 4.25}&0.56&{\bf 2.41}&10.88&9.88&Y ORB\\
2285898&0.09&&20.26&8&1101&1.63&0.88&1.17&20.96&0.37&0.55&0.70&10.82&9.83&\\
2303367&0.60&&-35.02&8&1118&73.01&{\bf 1.00}&18.02&       &&& &10.17&& outlier\\
2305930&0.21&&-119.66&10&1127&5.71&{\bf 1.00}&5.17&-119.78&0.51&0.39&0.12&10.78&9.08&outlier\\
2309550&0.03&&-10.11&3&289&0.16&0.15&-1.03&-10.13&0.21&0.28&0.02&9.19&7.33&&\\
2310129&0.02&&12.88&4&637&0.11&0.04&-1.66&13.43&0.57&0.57&0.55&10.90&9.91& &\\
2424934&0.09&&-13.37&7&758&1.57&0.85&1.03&-12.93&0.51&0.43&0.44&10.36&9.32&&\\
2447604&0.40&&20.27&4&1079&31.91&{\bf 1.00}&8.25&22.04&0.32&0.52&{\bf 1.77}&10.73&9.76&Y\\
2448225&0.08&&-21.09&4&1052&1.31&0.73&0.61&-20.67&0.14&0.52&0.42&10.78&9.76&\\
2583651&0.04&&-25.55&3&763&0.36&0.30&-0.53&-25.34&0.25&0.48&0.21&10.58&9.58&&\\
2583884&0.06&&-21.54&3&435&0.82&0.56&0.15&-21.12&0.22&0.45&0.42&10.44&9.44&&\\
2695267&0.16&&-5.22&11&1122&5.03&{\bf 1.00}&4.94&-4.42&{\bf 1.05}&0.33&{\bf 0.80}&9.48&8.48&Y\\
2714397&0.04&&-172.31&4&651&0.37&0.23&-0.76&-172.29&0.36&0.53&0.02&10.49&9.79&&\\
2831788&0.03&&-36.50&3&322&0.19&0.17&-0.94&-36.40&0.22&0.37&0.10&9.97&8.97&&\\
2845408&0.08&&-21.65&5&738&1.41&0.77&0.75&-21.16&0.37&0.51&0.49&10.64&9.71&\\
2858440&0.07&&13.93&3&310&0.91&0.60&0.24&13.91&0.61&0.50&0.02&10.70&9.69& &\\
2975717&0.11&&-17.96&5&1105&2.53&0.96&1.77&-17.78&0.26&0.46&0.18&10.48&9.47&\\
2987113&0.05&&-6.27&3&428&0.59&0.44&-0.15&-5.81&0.26&0.53&0.46&10.60&9.76&&\\
2988988&0.05&&-5.60&4&661&0.43&0.27&-0.62&-5.21&0.61&0.55&0.39&10.82&9.83& &\\
2992350&0.07&&-58.77&4&680&1.01&0.61&0.28&-58.08&0.30&0.34&{\bf 0.69}&9.47&8.60&Y?\\
3001851&0.11&&-78.84&6&1130&2.29&0.96&1.72&-78.69&0.71&0.40&0.15&10.14&9.14&\\
3101632&0.20&&8.64&8&1102&8.50&{\bf 1.00}&6.02&9.05&0.15&0.63&0.41&10.88&10.07&Y\\
3118806&0.07&&-12.82&3&1492&0.88&0.58&0.21&-12.61&0.19&0.59&0.21&10.96&9.96&&\\
3127825&0.06&&-25.20&3&760&0.70&0.50&-0.01&-25.04&0.38&0.54&0.16&10.80&9.80&&\\
3222670&0.11&&6.91&4&1105&2.44&0.94&1.54&7.65&0.35&0.41&0.74&10.22&9.25& \\
3231503&0.79&0.057&-23.93&8&1125&127.1&{\bf 1.00} &22.78&-24.82&0.39&0.50&0.89&10.65&9.66&Y ORB\\
3234396&0.03&&-19.35&3&761&0.13&0.12&-1.15&-19.23&0.38&0.53&0.12&10.78&9.78&&\\
3234597&0.09&&-12.06&6&958&1.62&0.85&1.04&-11.99&0.36&0.38&0.07&10.01&9.01&&\\
3234703&0.11&&-38.13&3&671&2.46&0.91&1.39&-37.72&0.20&0.46&0.41&10.48&9.48&&\\
3241374&0.09&&23.03&4&878&1.48&0.78&0.78&23.37&0.13&0.41&0.34&10.23&9.23&\\
3247016&0.07&&-12.33&5&774&0.87&0.52&0.04&-12.23&0.50&0.33&0.10&9.51&8.51&&\\
3324929&0.07&&6.94&5&674&1.07&0.63&0.34&6.76&0.63&0.43&0.18&10.25&9.32& &\\
3336731&0.13&&-37.18&6&1125&3.63&{\bf 1.00}&2.76&-36.12&0.27&0.43&{\bf 1.06}&10.29&9.36&Y\\
3337400&2.69&0.062&6.28&8&1132&1480&{\bf 1.00}&58.52&4.71&{\bf 1.34}&0.48&{\bf 1.57}&10.66&9.60&Y ORB\\
3341327&0.02&&-12.48&4&648&0.11&0.05&-1.62&-11.54&{\bf 1.29}&0.46&{\bf 0.94}&10.51&9.51&Y\\
3427850&0.07&&-35.75&3&311&0.87&0.58&0.19&-34.87&0.51&0.32&{\bf 0.88}&9.17&8.39&Y?\\
3429738&0.01&&-4.77&3&297&0.01&0.01&-2.13&-4.32&0.20&0.38&0.45&10.02&9.02&&\\
3455760&0.81&0.030&-47.02&10&1445&134.95&{\bf 1.00}&26.44&-45.74&0.38&0.57&{\bf 1.28}&10.91&9.91&Y ORB\\
3457190&0.09&&20.74&7&1131&1.74&0.89&1.25&21.29&0.22&0.39&0.55&10.08&9.08& &\\
3459109&0.05&&4.12&3&431&0.53&0.41&-0.23&4.43&0.72&0.59&0.31&10.98&9.98&&\\
\hline
\end{tabular}
}
\end{table*}

\addtocounter{table}{-1}
\begin{table*}
\caption[]{Continued.
}

{\fontsize{8}{10}\selectfont
\begin{tabular}{rrcrrrrlrrcccrrlll}
\hline\\
KIC/CoRoT &  \multicolumn{1}{c}{$\sigma_j(RV)$} & $\sigma(O-C)$ & \multicolumn{1}{c}{$\overline{RV}_H$} & $N$obsH & $\Delta t$  & $\chi_j^2/\nu_j$ & Prob. & $F2_j$ &  \multicolumn{2}{c}{$\overline{RV}_G\pm\sigma(RV_G)$} & \multicolumn{1}{c}{$\epsilon(RV_G)$} & $|\overline{RV}_G - \overline{RV}_H|$ &  \multicolumn{1}{c}{$G$} & $G_{\rm RVS}$ & SB \\ 
    & \multicolumn{1}{c}{(\kms)} & (\kms) & (\kms) & & (d) & &RV var & & \multicolumn{2}{c}{(\kms)} &\multicolumn{1}{c}{(\kms)} &  \multicolumn{1}{c}{(\kms)}\\
   \hline
   \medskip\\
3526061&0.20&&-27.51&12&2842&8.57&{\bf 1.00}&7.51&-27.33&0.19&0.39&0.18&9.99&9.06&Y\\
3529480&0.09&&-55.24&9&770&1.52&0.86&1.07&-54.92&0.33&0.35&0.32&9.71&8.75&&\\
3548732&0.02&&-14.74&3&773&0.12&0.11&-1.18&-14.21&0.37&0.37&0.53&9.90&8.90& &\\
3558705&8.53&&1.53&18&1050&14866&{\bf 1.00}&206.43&-4.27&{\bf 2.78}&0.27&{\bf 5.80}&8.25&7.25 & Y ORB\\
3558848&0.08&&-12.30&6&1040&1.18&0.69&0.48&-11.88&0.47&0.49&0.42&10.64&9.64&\\
3560093&1.88&&-20.70&4&1055&725&{\bf 1.00}&29.61&-15.93&0.68&0.59&{\bf 4.77}&10.96&9.96&Y\\
3631821&0.03&&10.02&3&297&0.18&0.17&-0.97&10.16&0.21&0.40&0.14&10.09&9.16&&\\
3634720&1.79&&-10.73&9&1107&657.0&{\bf 1.00}&46.33&-8.30&0.33&0.35&{\bf 2.43}&9.72&8.72&Y\\
3654420&0.11&&13.13&5&1132&2.64&0.97&1.86&13.43&0.16&0.40&0.30&10.19&9.19&\\
3730953&0.04&&11.57&6&711&0.34&0.11&-1.23&11.51&0.30&0.32&0.06&8.82&8.40&&\\
3736289&0.05&&-18.79&3&353&0.46&0.37&-0.36&-18.25&0.58&0.55&0.54&10.84&9.85& &\\
3742673&0.05&&-20.62&3&382&0.50&0.39&-0.29&-20.71&0.33&0.63&0.09&10.96&10.09&&\\
3744043&0.02&&-36.82&5&1571&0.12&0.02&-1.93&-36.20&0.22&0.35&0.62&9.65&8.74& &\\
3744681&0.08&&-22.61&5&1087&1.16&0.67&0.45&-21.78&0.28&0.46&0.83&10.41&9.48& &\\
3749487&0.09&&-30.67&3&678&1.57&0.79&0.82&-29.78&0.29&0.53&0.89&10.77&9.77& &\\
3759654&0.05&&-21.23&3&614&0.54&0.42&-0.22&-20.79&0.25&0.41&0.44&10.23&9.23&&\\
3763790&0.04&&-0.43&8&396&0.40&0.10&-1.30&-0.10&0.51&0.38&0.33&10.00&9.00&&\\
3833399&0.09&&-62.14&12&1420&1.53&0.89&1.22&-61.73&0.15&0.27&0.41&9.22&7.27&&\\
3934458&0.06&&-56.87&3&337&0.66&0.48&-0.06&-56.62&0.40&0.45&0.25&10.42&9.44&&\\
3935726&0.06&&-26.95&3&352&0.86&0.58&0.19&-26.70&0.21&0.55&0.25&10.86&9.85&&\\
3936921&0.17&&-35.90&10&1129&5.71&{\bf 1.00}&5.17&-35.16&0.40&0.49&0.74&10.70&9.63&Y \\
3938291&0.08&&9.23&4&1122&1.39&0.76&0.70&9.46&0.24&0.48&0.23&10.51&9.58&&\\
3955590&0.09&&-52.89&4&600&1.62&0.82&0.91&-52.56&0.27&0.43&0.33&10.36&9.36&&\\
3958400&0.06&&-22.26&4&797&0.74&0.47&-0.08&-21.88&0.39&0.45&0.38&10.46&9.46&&\\
4072740&0.06&&-17.13&6&1026&0.76&0.42&-0.21&-17.28&0.43&0.34&0.15&9.67&8.67&&\\
4139805&0.06&&-21.96&5&687&0.75&0.44&-0.15&-21.59&0.27&0.50&0.37&10.56&9.65&&\\
4149966&0.14&&-12.24&7&1077&4.11&{\bf 1.00}&3.32&-12.27&0.55&0.38&0.03&10.04&9.00&Y\\
4180705&0.09&&-22.95&3&617&1.53&0.78&0.79&-22.32&0.32&0.45&0.63&10.42&9.42&\\
4180903&0.04&&-9.76&4&601&0.34&0.20&-0.85&-9.53&0.18&0.32&0.23&9.38&8.38&&\\
4241369&0.05&&-34.51&3&338&0.57&0.43&-0.18&-34.00&0.14&0.46&0.51&10.43&9.49& &\\
4243796&0.15&&5.53&9&1122&4.45&{\bf 1.00}&4.04&5.59&0.65&0.52&0.06&10.77&9.75&Y\\
4262505&0.20&&-10.10&6&1118&8.15&{\bf 1.00}&5.01&-9.44&0.39&0.43&0.66&10.34&9.34&Y? (outlier?)\\
4351319&0.07&&-18.74&5&1009&1.12&0.65&0.40&-18.39&0.14&0.39&0.35&9.95&9.06&&\\
4378473&0.20&&7.51&4&1072&8.01&{\bf 1.00}&3.95&8.81&0.46&0.58&{\bf 1.30}&10.94&9.94&Y\\
4445711&0.07&&3.36&3&346&0.92&0.60&0.25&3.42&0.23&0.51&0.06&10.72&9.72&&\\
4476422&0.16&&0.99&5&1337&5.38&{\bf 1.00}&3.43&0.29&0.39&0.59&0.70&10.97&9.97&Y\\
4557817&0.04&&13.49&6&820&0.32&0.10&-1.28&14.46&0.64&0.56&0.97&10.89&9.89&\\
4770846&0.06&&-38.87&10&2577&0.67&0.26&-0.65&-38.72&0.38&0.33&0.15&9.56&8.56&&\\
4902641&0.07&&13.24&3&337&0.97&0.62&0.30&13.50&0.66&0.53&0.26&10.70&9.76& &\\
5128171&0.04&&-8.99&4&861&0.38&0.23&-0.75&-8.78&0.46&0.39&0.21&10.08&9.08&&\\
5175152&0.09&&2.40&7&1107&1.52&0.83&0.97&3.34&0.22&0.45&{\bf 0.94}&10.42&9.42&Y?\\
5184199&0.10&&-26.09&5&1055&2.23&0.94&1.53&-26.05&0.25&0.36&0.04&9.87&8.87&&\\
5266416&0.06&&-37.63&3&311&0.84&0.57&0.16&-37.21&0.35&0.48&0.42&10.60&9.60&&\\
5307747&0.07&&6.63&9&2918&0.89&0.48&-0.06&6.69&0.31&0.28&0.06&8.40&7.40&&\\
5353108&0.03&&-50.96&4&264&0.14&0.06&-1.49&-50.51&0.20&0.31&0.45&9.24&8.24&&\\
5457811&0.09&&-55.56&3&619&1.69&0.81&0.90&-55.74&0.30&0.50&0.18&10.68&9.68&\\
5515314&0.09&&-51.96&4&442&1.56&0.80&0.86&-51.98&0.35&0.55&0.02&10.83&9.86&&\\
5530598&0.24&&-16.21&14&2644&12.00&{\bf 1.00}&9.99&-16.36&0.15&0.29&0.15&8.63&7.63&Y? (jitter?)\\
5546141&0.07&&-17.61&4&595&0.99&0.60&0.26&-17.31&0.18&0.44&0.30&10.39&9.39&&\\
5612549&0.03&&-2.89&3&322&0.23&0.21&-0.82&-2.30&0.34&0.42&0.59&10.30&9.30& &\\
5648894&0.06&&-17.21&5&2203&0.76&0.45&-0.14&-17.07&0.18&0.28&0.14&8.60&7.60&&\\
5700368&0.05&&-32.18&20&2880&0.61&0.10&-1.31&-32.20&0.14&0.27&0.02&8.16&7.16&&\\
5706341&0.04&&-53.01&2&329&0.33&0.43&-0.19&-52.97&0.20&0.44&0.04&10.42&9.42&&\\
5723165&0.04&&-55.03&3&825&0.30&0.26&-0.66&-54.90&0.40&0.48&0.13&10.59&9.59&&\\
5770923&0.08&&5.93&3&337&1.25&0.71&0.56&6.16&0.51&0.52&0.23&10.68&9.74&&\\
5782127&0.05&&-34.69&7&662&0.58&0.25&-0.68&-34.18&0.21&0.55&0.51&10.84&9.84& &\\
5795626&2.24&0.057&-84.14&14&2301&1024&{\bf 1.00}&69.58&-83.10&{\bf 1.20}&0.31&{\bf 1.04}&9.16&8.16&Y ORB\\
5854239&2.42&0.050&7.49&11&1156&1197.77&{\bf 1.00}&64.68&-0.49&{\bf 0.76}&0.40&{\bf 7.98}&10.23&9.18&Y ORB\\
5866737&0.08&&13.39&7&1107&1.47&0.82&0.90&14.13&0.22&0.52&0.74&10.75&9.75& &\\
5872509&0.06&&-11.05&3&710&0.66&0.48&-0.06&-10.34&0.26&0.57&0.71&10.91&9.91& &\\
5981666&4.13&0.122&-21.80&7&1126&3486.&{\bf 1.00}&73.79&-9.72&{\bf 1.02}&0.34&{\bf 12.08}&9.61&8.61&Y ORB\\
5990753&0.12&&-27.23&4&1112&2.99&0.97&1.89&-26.84&0.26&0.56&0.39&10.90&9.90&&\\
6037858&0.06&&-19.84&5&707&0.85&0.50&0.01&-19.71&0.39&0.36&0.13&9.81&8.81&\\
6045299&0.11&&-48.24&5&1068&2.52&0.96&1.77&-47.86&0.19&0.49&0.38&10.63&9.63&\\
6117517&0.05&&1.18&3&371&0.60&0.45&-0.14&1.29&0.37&0.48&0.11&10.57&9.57&&\\
\hline
\end{tabular}
}
\end{table*}

\addtocounter{table}{-1}
\begin{table*}
\caption[]{Continued.
}

{\fontsize{8}{10}\selectfont
\begin{tabular}{rrcrrrrlrrcccrrlll}
\hline\\
KIC/CoRoT &  \multicolumn{1}{c}{$\sigma_j(RV)$} & $\sigma(O-C)$ & \multicolumn{1}{c}{$\overline{RV}_H$} & $N$obsH & $\Delta t$  & $\chi_j^2/\nu_j$ & Prob. & $F2_j$ &  \multicolumn{2}{c}{$\overline{RV}_G\pm\sigma(RV_G)$} & \multicolumn{1}{c}{$\epsilon(RV_G)$} & $|\overline{RV}_G - \overline{RV}_H|$ &  \multicolumn{1}{c}{$G$} & $G_{\rm RVS}$ & SB \\ 
    & \multicolumn{1}{c}{(\kms)} & (\kms) & (\kms) & & (d) & &RV var & & \multicolumn{2}{c}{(\kms)} &\multicolumn{1}{c}{(\kms)} &  \multicolumn{1}{c}{(\kms)}\\
   \hline
   \medskip\\
6118479&0.02&&-6.50&3&376&0.12&0.12&-1.18&-6.19&0.25&0.55&0.31&10.86&9.86&&\\
6144777&0.07&&-30.83&4&651&0.98&0.60&0.25&-30.79&0.19&0.50&0.04&10.65&9.65&&\\
6276948&0.10&&-22.50&6&749&2.04&0.93&1.49&-22.27&0.28&0.54&0.23&10.80&9.80&\\
6290627&0.37&&-24.91&6&1121&27.21&{\bf 1.00}&9.73&-24.54&0.94&0.59&0.37&10.97&9.97&Y\\
6356581&0.02&&-2.15&3&315&0.05&0.05&-1.58&-2.16&0.31&0.54&0.01&10.80&9.80&&\\
6631489&0.51&0.023&-12.90&9&1057&52.88&{\bf 1.00}&16.69&-13.27&0.14&0.32&0.37&9.40&8.40&Y ORB\\
6677653&0.10&&-59.26&6&1105&2.24&0.95&1.67&-58.74&0.28&0.54&0.52&10.83&9.83& \\
6700113&0.06&&-12.35&3&784&0.66&0.48&-0.05&-11.94&0.41&0.46&0.41&10.48&9.48&&\\
6779699&0.08&&-4.49&3&763&1.21&0.70&0.53&-4.72&0.50&0.54&0.23&10.82&9.82&&\\
6849167&0.14&&-9.39&4&726&3.90&0.99&2.38&-8.83&0.33&0.58&0.56&10.93&9.93& & \\
7205067&0.12&&-12.53&4&1077&2.74&0.96&1.74&-11.79&0.21&0.51&0.74&10.69&9.69& &\\
7467630&4.79&&15.04&6&1074&4686.&{\bf 1.00}&74.85&17.71&{\bf 1.68}&0.39&{\bf 2.67}&10.08&9.08&Y\\
7628676&0.02&&-25.24&3&422&0.10&0.09&-1.28&-25.16&0.62&0.53&0.08&10.79&9.79& &\\
7936407&0.01&&-17.00&3&264&0.01&0.01&-2.00&-16.77&0.24&0.40&0.23&10.13&9.13&&\\
7985438&0.09&&-18.88&3&284&1.48&0.77&0.75&-18.73&0.30&0.45&0.15&10.44&9.44&&\\
8007217&3.31&&-5.27&12&1148&2236&{\bf 1.00}&85.11&-5.84&0.51&0.34&0.57&9.69&8.69&Y\\
8106525&0.32&&3.88&6&1100&20.94&{\bf 1.00}&8.54&4.27&0.14&0.32&0.39&9.34&8.34&Y\\
8110811&1.12&&-22.83&4&1066&256.32&{\bf 1.00}&19.94&-25.00&0.54&0.49&{\bf 2.17}&10.63&9.63&Y\\
8181509&1.78&&-32.90&5&1073&649.26&{\bf 1.00}&32.73&-31.46&0.39&0.43&{\bf 1.44}&10.37&9.37&Y\\
8188772&0.04&&-33.49&3&641&0.27&0.23&-0.74&-33.82&0.40&0.57&0.33&10.90&9.90&&\\
8299922&0.08&&-23.48&8&1121&1.25&0.73&0.61&-22.93&0.19&0.39&0.55&10.12&9.12& &\\
8506295&0.13&&-8.86&5&1053&3.31&0.99&2.32&-8.74&0.25&0.42&0.12&10.29&9.29&\\
8631401&0.02&&0.85&3&352&0.12&0.11&-1.18&0.97&0.14&0.28&0.12&8.42&7.42&&\\
8708536&0.05&&-18.63&3&309&0.50&0.40&-0.28&-18.23&0.22&0.46&0.40&10.50&9.50&&\\
8718745&0.07&&-20.17&3&423&1.08&0.66&0.41&-20.32&0.18&0.51&0.15&10.72&9.72&&\\
8872979&0.08&&-7.21&5&1122&1.22&0.70&0.52&-5.96&0.92&0.52&{\bf 1.25}&10.73&9.73&Y?\\
9145955&0.05&&16.74&3&342&0.49&0.39&-0.30&17.95&0.44&0.35&{\bf 1.21}&9.77&8.77&Y?\\
9151007&0.06&&-11.59&3&372&0.63&0.47&-0.10&-11.42&0.45&0.48&0.17&10.58&9.58&&\\
9173371&0.02&&-0.40&7&1858&0.13&0.01&-2.40&-0.37&0.16&0.31&0.03&9.27&8.27&&\\
9244428&0.04&&-42.48&3&253&0.40&0.33&-0.46&-42.21&0.20&0.31&0.27&9.23&8.23&&\\
9327993&0.06&&-54.36&4&337&0.73&0.46&-0.10&-54.22&0.31&0.35&0.14&9.77&8.77&&\\
9346602&0.05&&-11.84&5&1470&0.44&0.22&-0.78&-11.40&0.26&0.46&0.44&10.51&9.51&&\\
9409513&0.04&&8.79&3&764&0.40&0.33&-0.47&9.04&0.26&0.41&0.25&10.24&9.24&&\\
9705687&0.08&&8.61&5&786&1.17&0.68&0.46&9.16&0.44&0.33&0.55&9.56&8.56& &\\
9716522&0.06&&7.14&6&761&0.78&0.44&-0.16&8.04&0.31&0.44&{\bf 0.90}&10.41&9.41&Y?\\
9812421&0.07&&1.08&3&337&1.05&0.65&0.38&0.93&0.32&0.40&0.15&10.18&9.18&&\\
9836930&0.05&&-5.08&5&832&0.52&0.28&-0.60&-5.78&{\bf 0.66}&0.36&{\bf 0.70}&9.82&8.82&Y\\
9838510&0.09&&-38.40&4&1130&1.72&0.84&1.00&-37.75&0.33&0.42&0.65&10.29&9.29& \\
100901998&0.03&&70.68&3&746&0.13&0.12&-1.15&71.31&0.34&1.02&0.63&11.81&10.81& &\\
100958710&0.05&&-21.91&3&430&0.51&0.40&-0.27&       &&& &12.00&11.00&&\\
101037205&0.04&&-4.93&4&785&0.41&0.26&-0.67&-4.90&0.61&1.22&0.03&12.06&11.06& &\\
101043587&0.03&&27.77&2&375&0.17&0.32&-0.48&29.29&1.11&0.95&1.52&11.71&10.71& &\\
101092813&0.07&&-68.60&3&753&0.91&0.60&0.24&-68.28&0.26&0.85&0.32&11.54&10.54&&\\
101101456&0.17&&-38.05&5&787&6.09&{\bf 1.00}&3.74&-38.37&0.97&1.10&0.32&11.92&10.92&Y?\\
101232297&0.06&&-17.22&4&387&0.73&0.46&-0.10&-16.67&0.42&1.22&0.55&12.06&11.06& &\\
101238328&0.09&&9.11&5&1082&1.50&0.80&0.85&10.00&0.49&1.23&0.89&12.08&11.08& &\\
101411168&0.01&&-45.83&3&758&0.04&0.04&-1.68&-45.39&0.60&1.12&0.44&11.95&10.95&&\\
101442365&0.12&&-17.23&5&1071&2.95&0.98&2.08&-16.30&0.52&1.16&0.93&12.00&11.00&\\
101561050&0.05&&-15.20&3&743&0.54&0.42&-0.22&-14.86&0.45&1.10&0.34&11.91&10.91&&\\
102632259&0.06&&9.60&9&1168&0.66&0.27&-0.61&9.36&0.89&0.99&0.24&11.77&10.77&&\\
102639822&7.48&&13.57&14&1235&11424&{\bf 1.00}&164.75&11.75&{\bf 5.77}&1.10&1.82&12.05&10.92&Y \\
102706346&0.05&&51.39&6&1108&0.48&0.21&-0.81&       &&& &11.74&&&\\
102726093&0.26&&29.68&11&1259&14.17&{\bf 1.00}&9.67&30.36&0.92&1.12&0.68&11.95&10.95&Y? (outlier)\\
102748522&0.33&&41.42&11&1261&22.30&{\bf 1.00}&12.32&41.27&0.15&0.76&0.15&11.38&10.38&Y? (jitter)\\
102791704&0.12&&53.47&5&740&2.77&0.97&1.95&54.25&0.56&1.15&0.78&11.99&10.99& &\\
110663970&0.06&&65.96&6&1170&0.73&0.40&-0.27&65.75&0.52&0.94&0.21&11.69&10.69&&\\
110667090&0.08&&55.28&6&1171&1.24&0.71&0.56&55.86&1.35&1.05&0.58&11.85&10.85& &\\
110672452&0.05&&25.38&7&677&0.60&0.27&-0.62&25.38&0.61&1.19&0.00&12.03&11.03& &\\
110679004&0.07&&61.02&8&1115&1.14&0.66&0.42&61.87&0.73&0.77&0.85&11.39&10.39& &\\
110679775&0.08&&1.76&7&1164&1.41&0.79&0.83&2.41&0.92&1.17&0.65&12.01&11.01& \\
110741575&0.06&&89.70&9&1170&0.80&0.40&-0.26&90.38&0.25&0.87&0.68&11.57&10.57& &\\
110770047&0.15&&32.17&7&742&4.42&{\bf 1.00}&3.52&       &&& &11.91&&Y?\\
110774446&0.08&&15.92&6&676&1.19&0.69&0.49&16.64&0.36&0.99&0.72&11.77&10.77& &\\
110827535&0.05&&24.30&8&1192&0.47&0.14&-1.07&24.13&0.74&0.92&0.17&11.67&10.67&&\\
110828605&0.07&&41.01&5&673&0.95&0.57&0.17&41.20&0.21&1.14&0.19&11.97&10.97&&\\
110847547&0.07&&52.97&7&1115&0.97&0.55&0.13&53.97&0.40&0.73&1.00&11.32&10.32& &\\
110858607&0.05&&8.19&5&1176&0.42&0.21&-0.83&9.01&0.88&1.17&0.82&12.01&11.01& &\\
\hline
\end{tabular}
}
\end{table*}

\section{SED of dusty stars}

\begin{figure}
    \includegraphics[width=9.5cm]{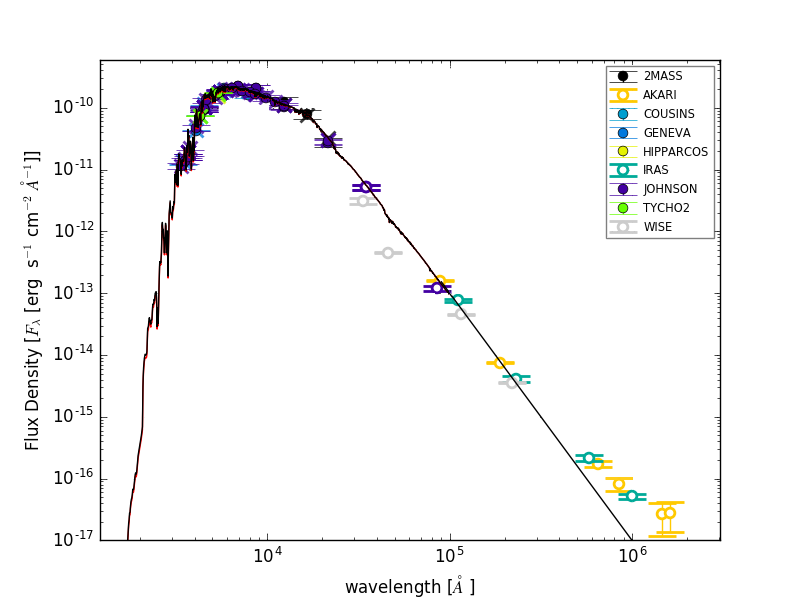}
    \caption{Spectral energy distribution for HD 3627 with photometry extracted from the SIMBAD data base, with the best-fitting MARCS model superimposed (red curve, barely visible). The black curve corresponds to the dereddened SED. Far-IR excess has been detected by IRAS and AKARI.}
    \label{Fig:SED3627}
\end{figure}

\begin{figure}
    \includegraphics[width=9.5cm]{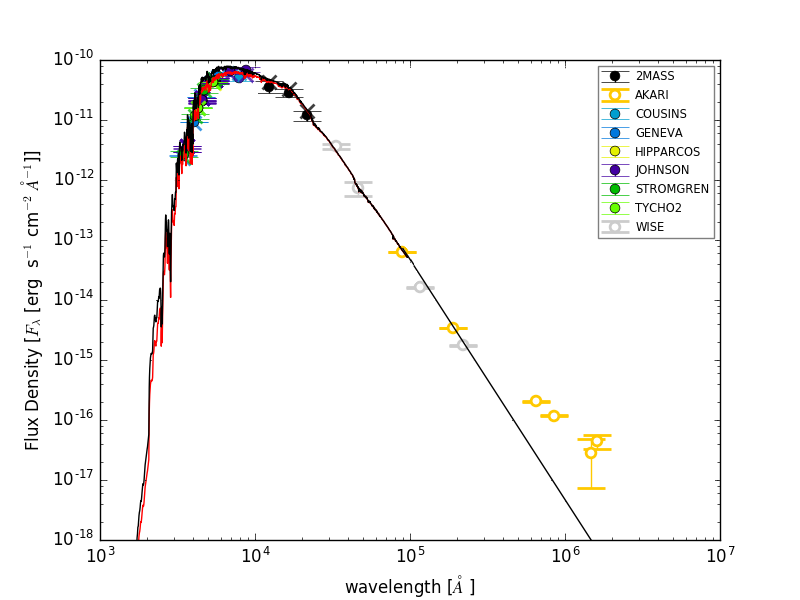}
    \caption{Same as Fig.~\ref{Fig:SED3627} for HD 30834.  }
    \label{Fig:SED30834}
\end{figure}

\begin{figure}
    \includegraphics[width=9.5cm]{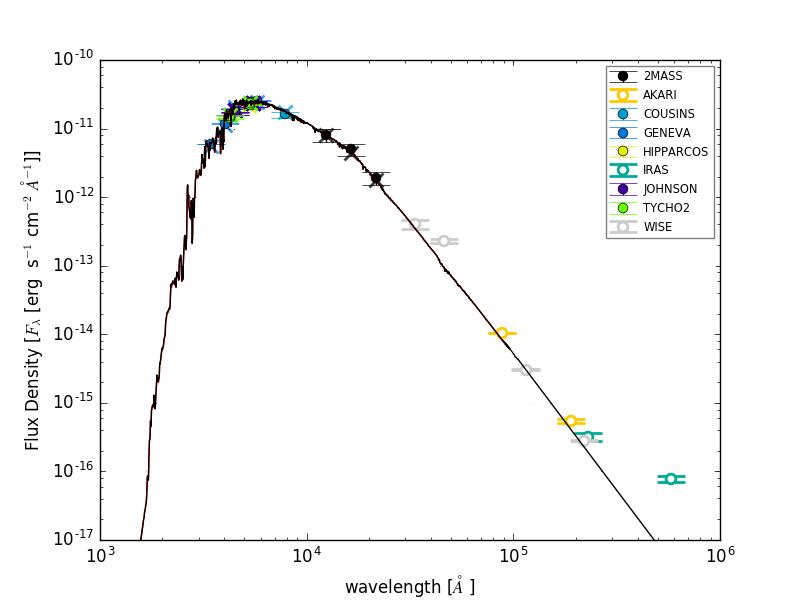}
    \caption{Same as Fig.~\ref{Fig:SED3627} for HD 119853. }
    \label{Fig:SED119853}
\end{figure}

\begin{figure}
    \includegraphics[width=9.5cm]{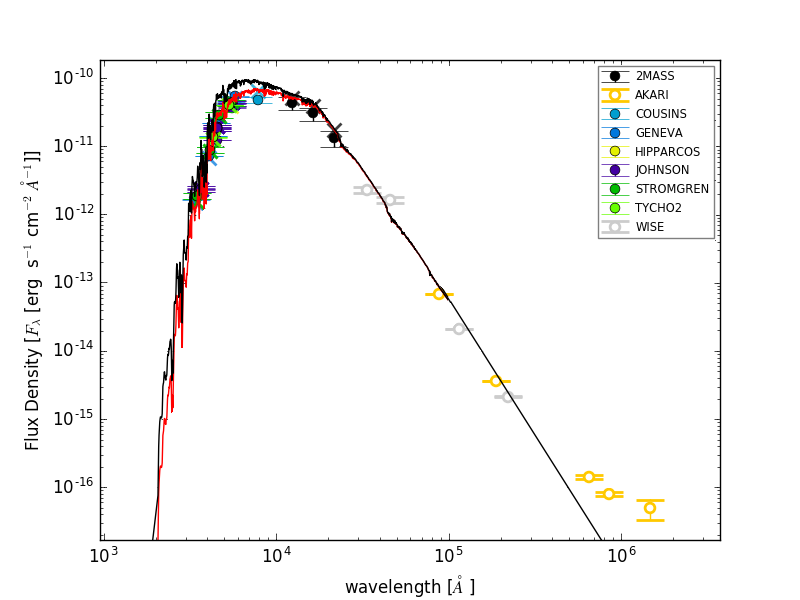}
    \caption{Same as Fig.~\ref{Fig:SED3627} for HD 153687.}
    \label{Fig:SED153687}
\end{figure}

\begin{figure}
    \includegraphics[width=9.5cm]{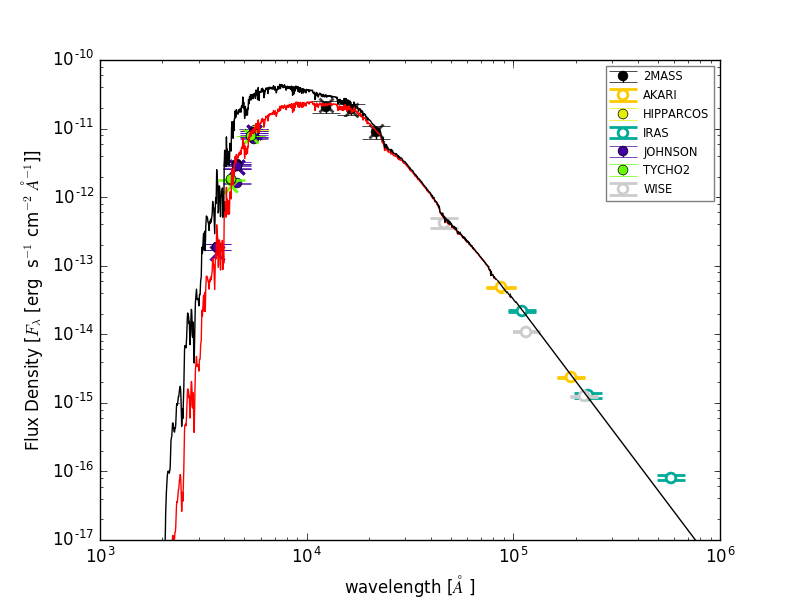}
    \caption{Same as Fig.~\ref{Fig:SED3627} for HD 156115. The SED temperature is 3900~K  and the IR excess is doubtful.}
    \label{Fig:SED156115}
\end{figure}

\begin{figure}
    \includegraphics[width=9.5cm]{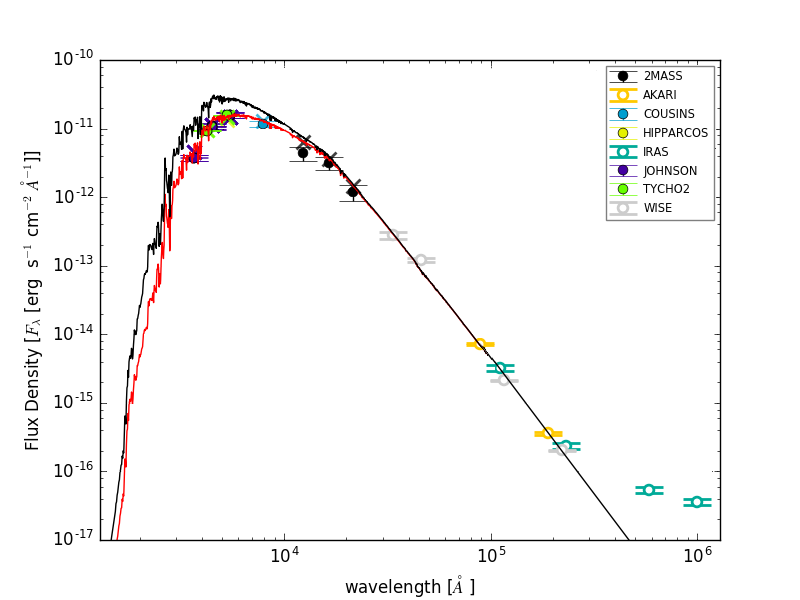}
    \caption{Same as Fig.~\ref{Fig:SED3627} for HD 212320. }
    \label{Fig:SED212320}
\end{figure}

\begin{figure}
    \includegraphics[width=9.5cm]{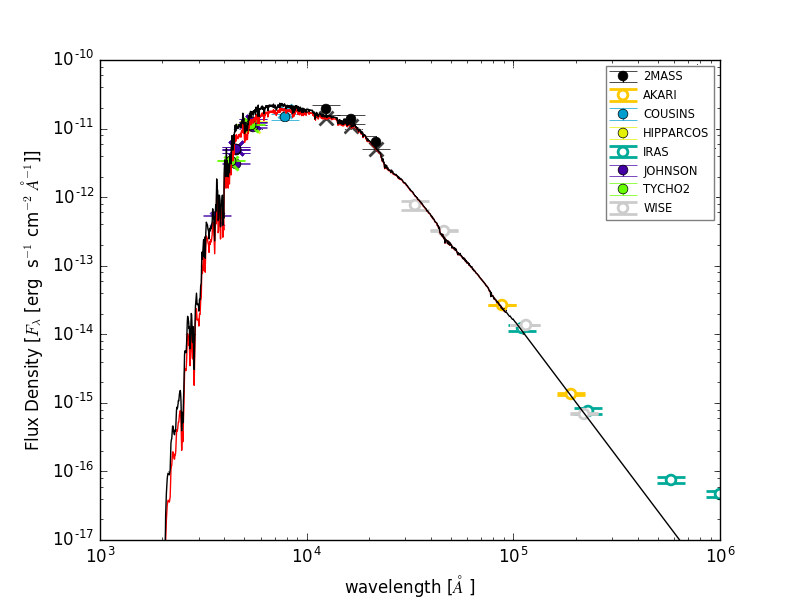}
    \caption{Same as Fig.~\ref{Fig:SED3627} for HD 221776. }
    \label{Fig:SED221776}
\end{figure}

\begin{figure}
    \includegraphics[width=9.5cm]{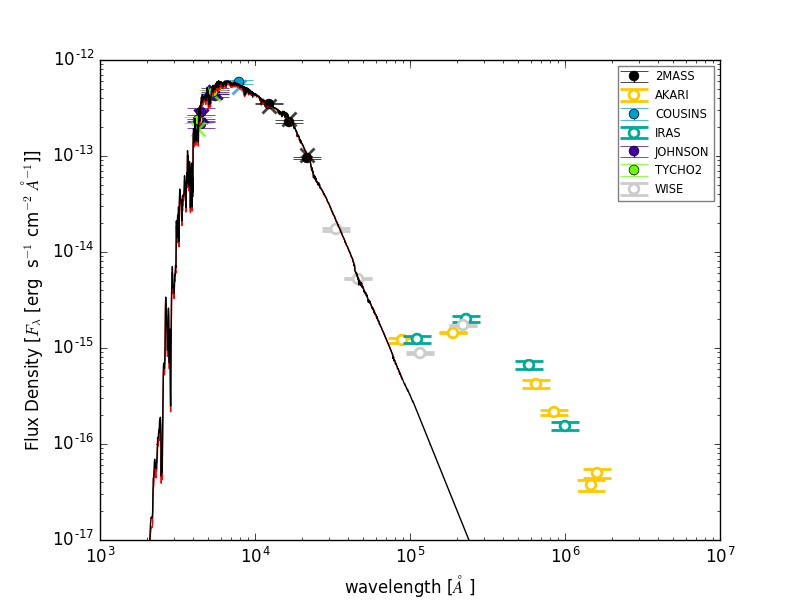}
    \caption{Same as Fig.~\ref{Fig:SED3627} for HDE 233517. The SED temperature is 4211~K.}
    \label{Fig:SED233517}
\end{figure}

This section presents the SED of dusty stars (Figs.~\ref{Fig:SED3627} -- \ref{Fig:SED233517}).

\begin{figure}
    \includegraphics[width=9.5cm]{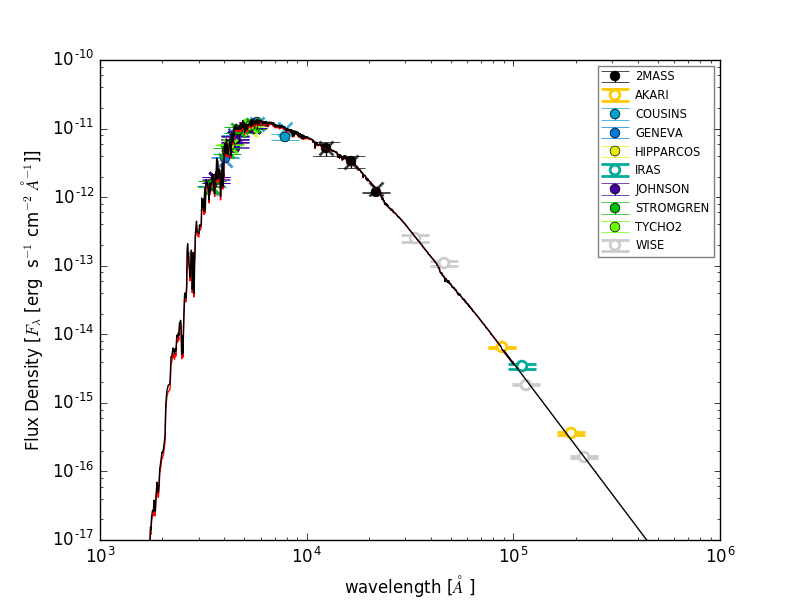}
    \caption{Same as Fig.~\ref{Fig:SED3627} for HD~6, which does not exhibit any IR excess. }
    \label{Fig:SED6}
\end{figure}
\section{RV curves}
\begin{figure}
    \includegraphics[width=9.5cm]{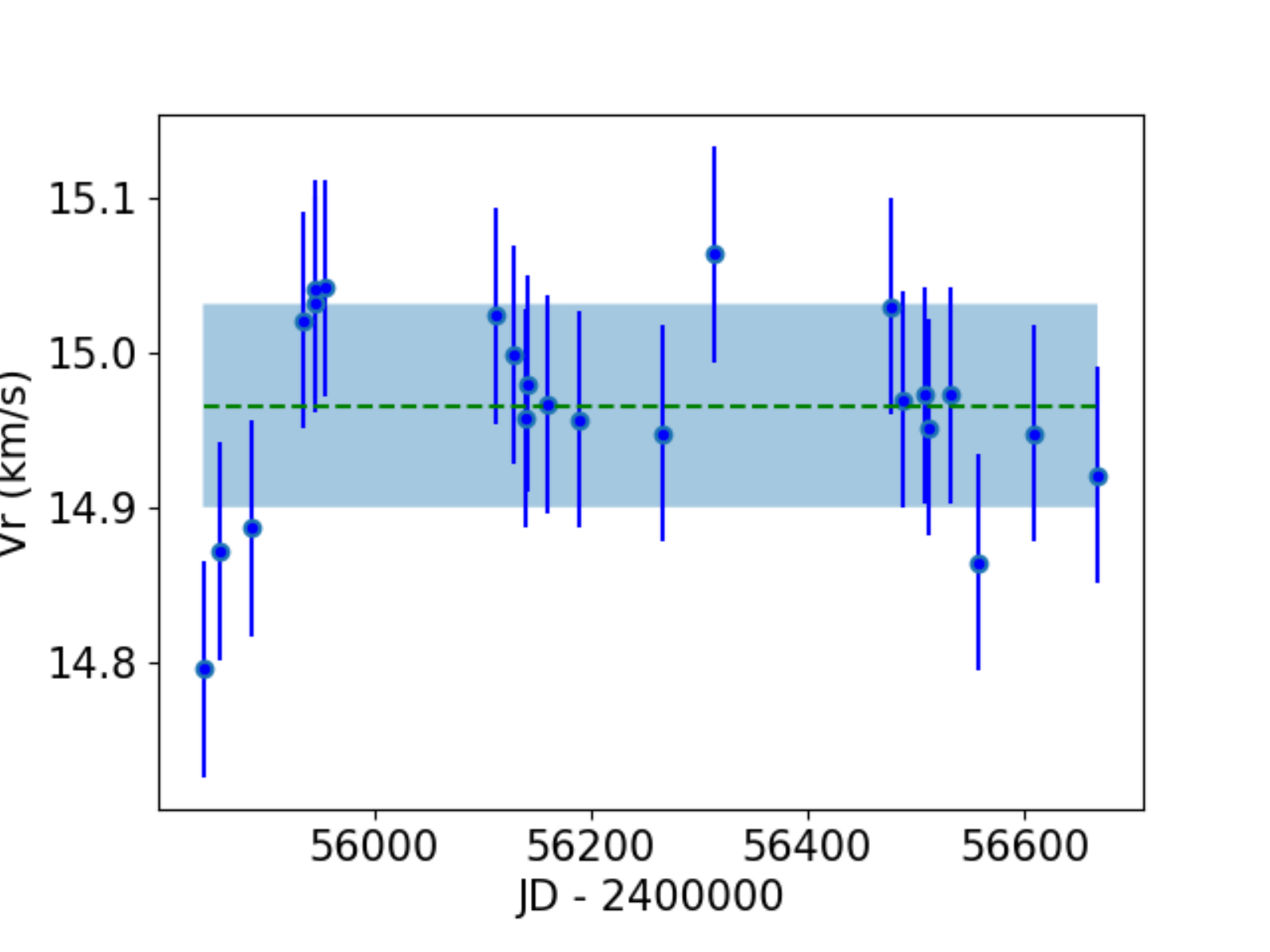}
    \caption{Radial-velocity data for HD~6. The dashed green line marks the average velocity, and the shaded blue zone corresponds to $\pm 1 \sigma$.}
    \label{Fig:6}
\end{figure}

\begin{figure}
       \includegraphics[width=9.5cm]{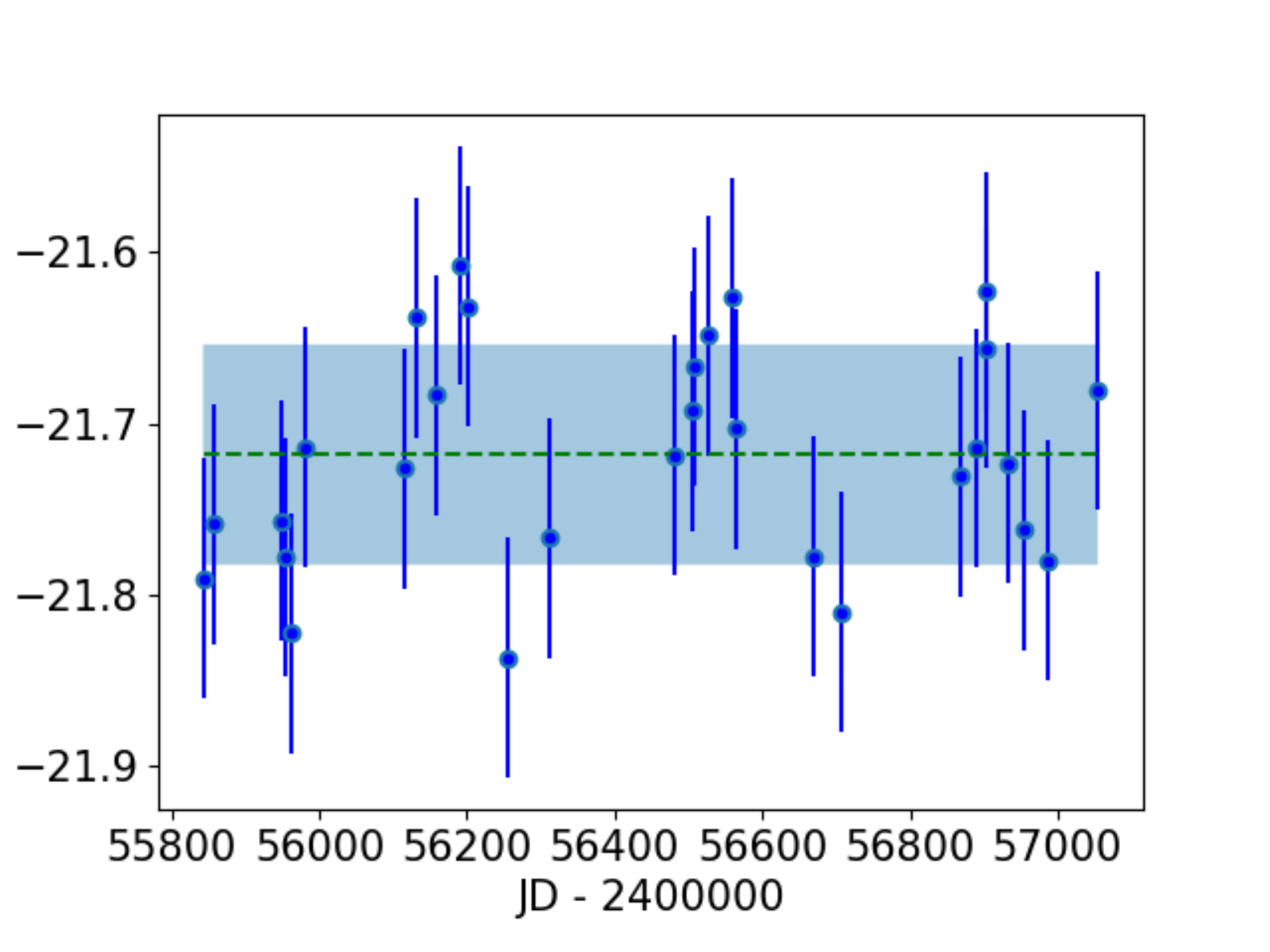}
    \caption{Same as Fig.~\ref{Fig:6} for HD~6665. 
    }
    \label{Fig:6665}
\end{figure}

\begin{figure}
    \includegraphics[width=9.5cm]{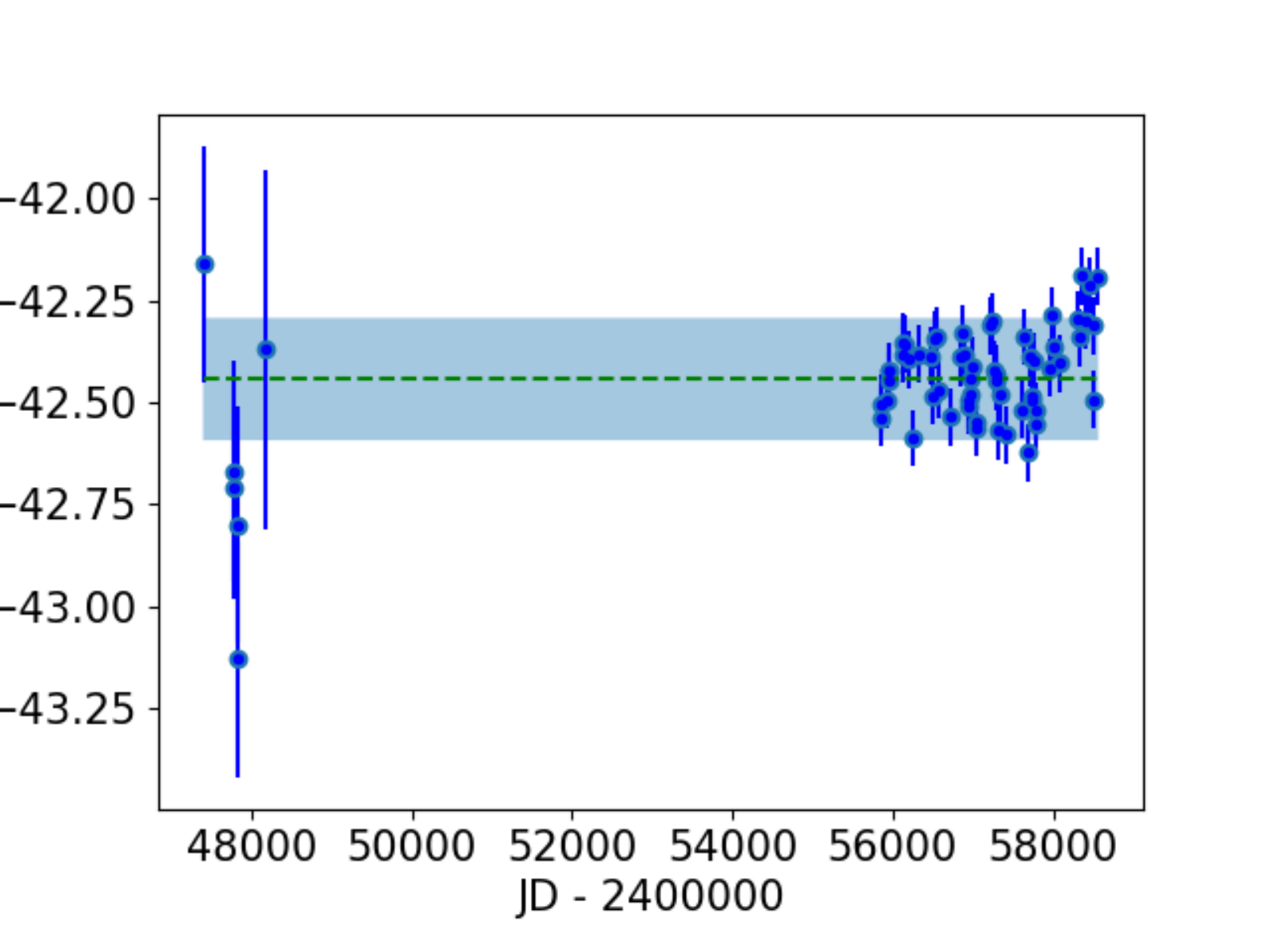}
       \includegraphics[width=9.5cm]{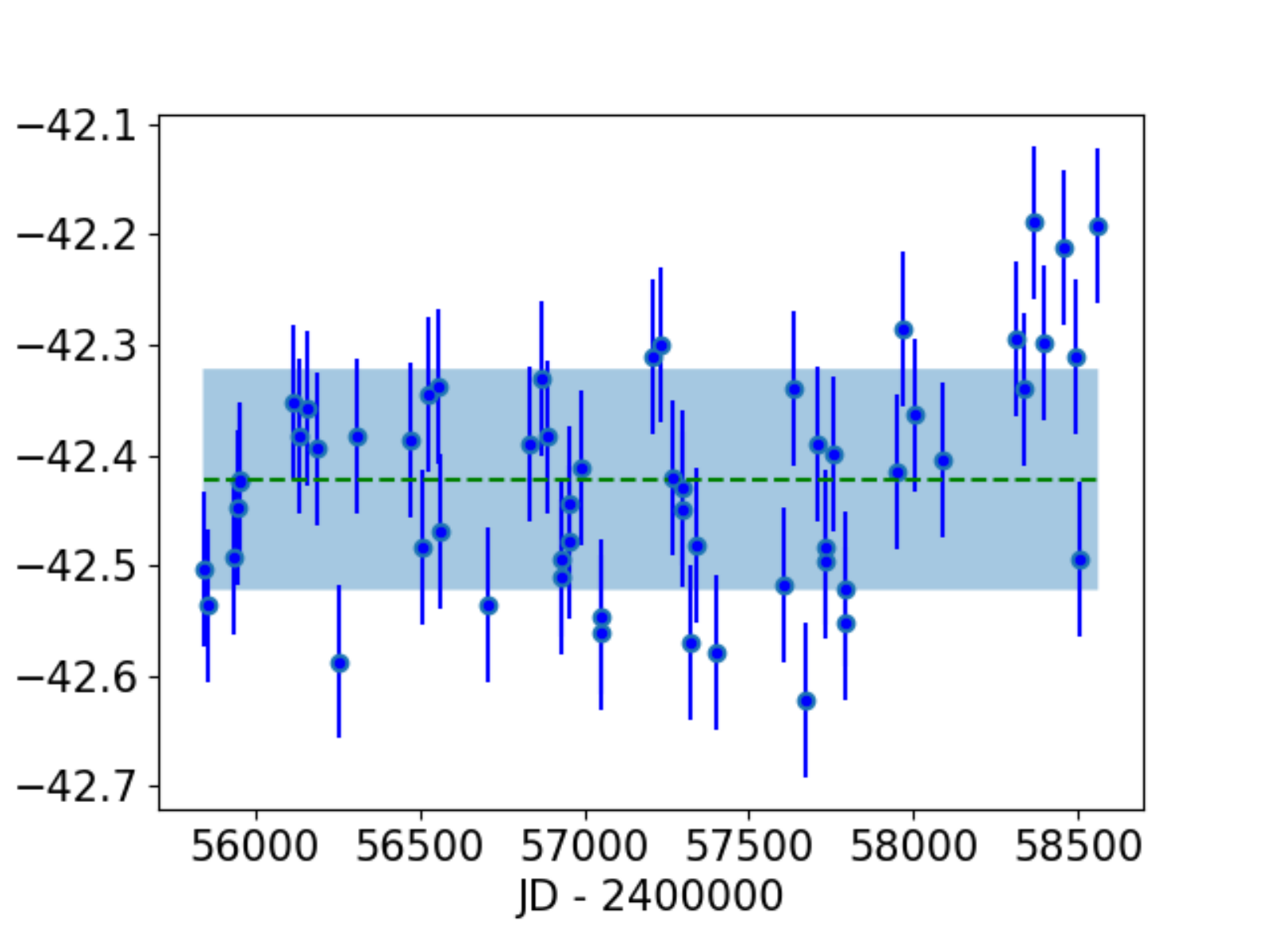}
    \caption{Top panel: Same as Fig.~\ref{Fig:6} for HD~9746. Bottom panel: Zoom on the HERMES data.}
    \label{Fig:9746}
\end{figure}

\begin{figure}
    \includegraphics[width=9.5cm]{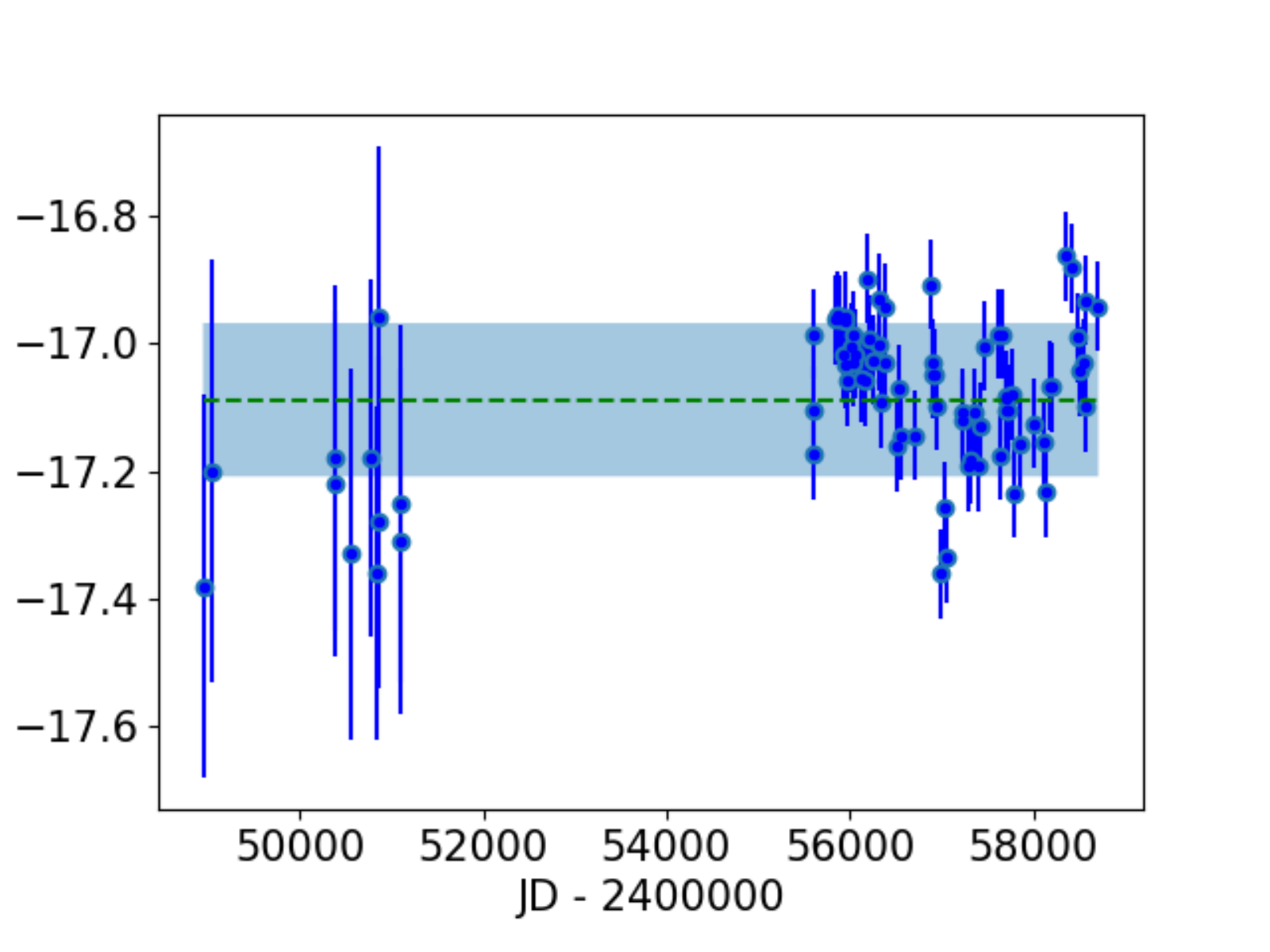}
    \includegraphics[width=9.5cm]{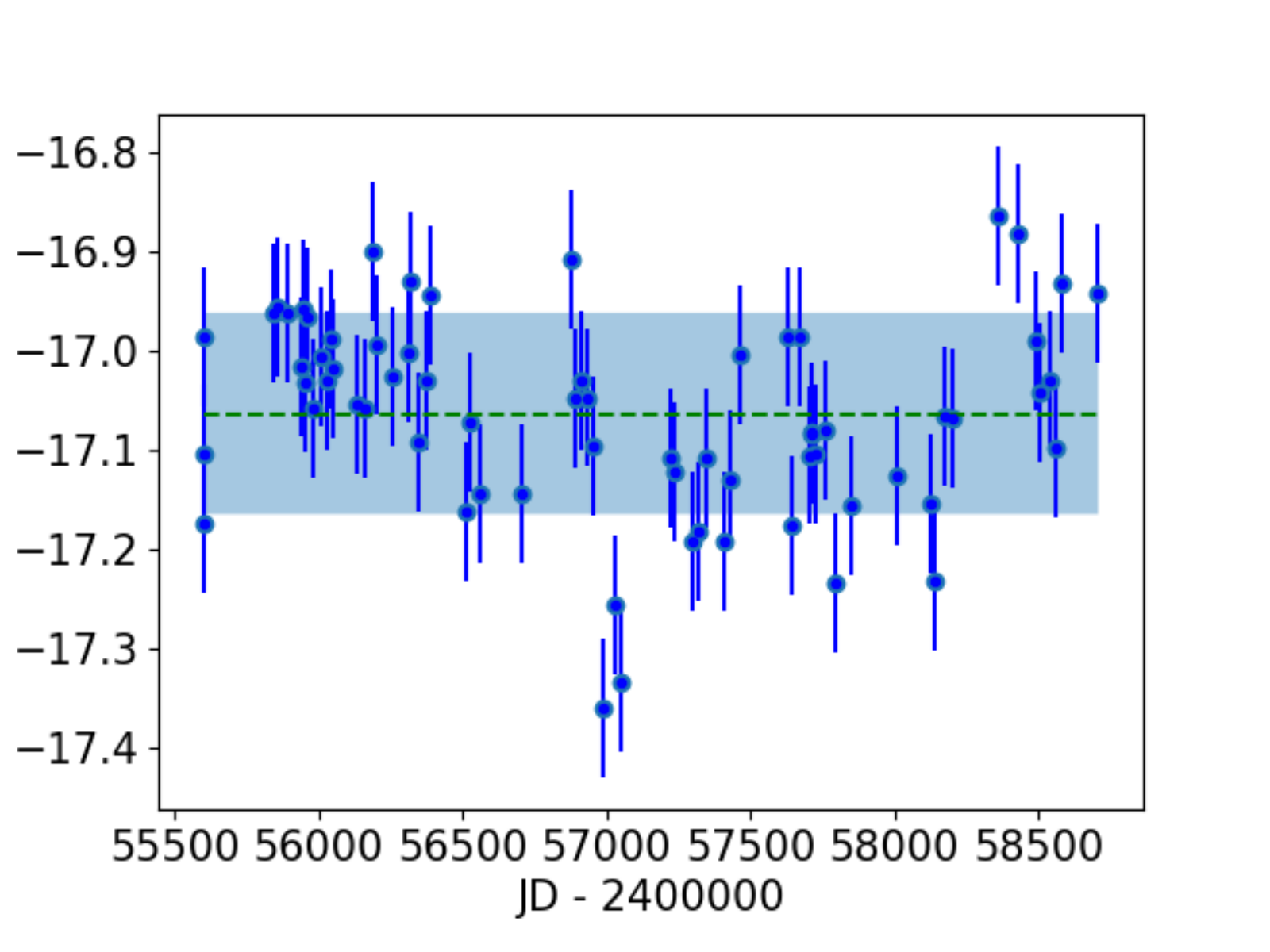}
    \caption{Top panel: Radial-velocity data for HD~30834. 
Early measurements with large error bars are from \citet{Famaey2005}. Bottom panel: Zoom on the HERMES velocities.  }
    \label{Fig:30834}
\end{figure}

\begin{figure}
    \includegraphics[width=9.5cm]{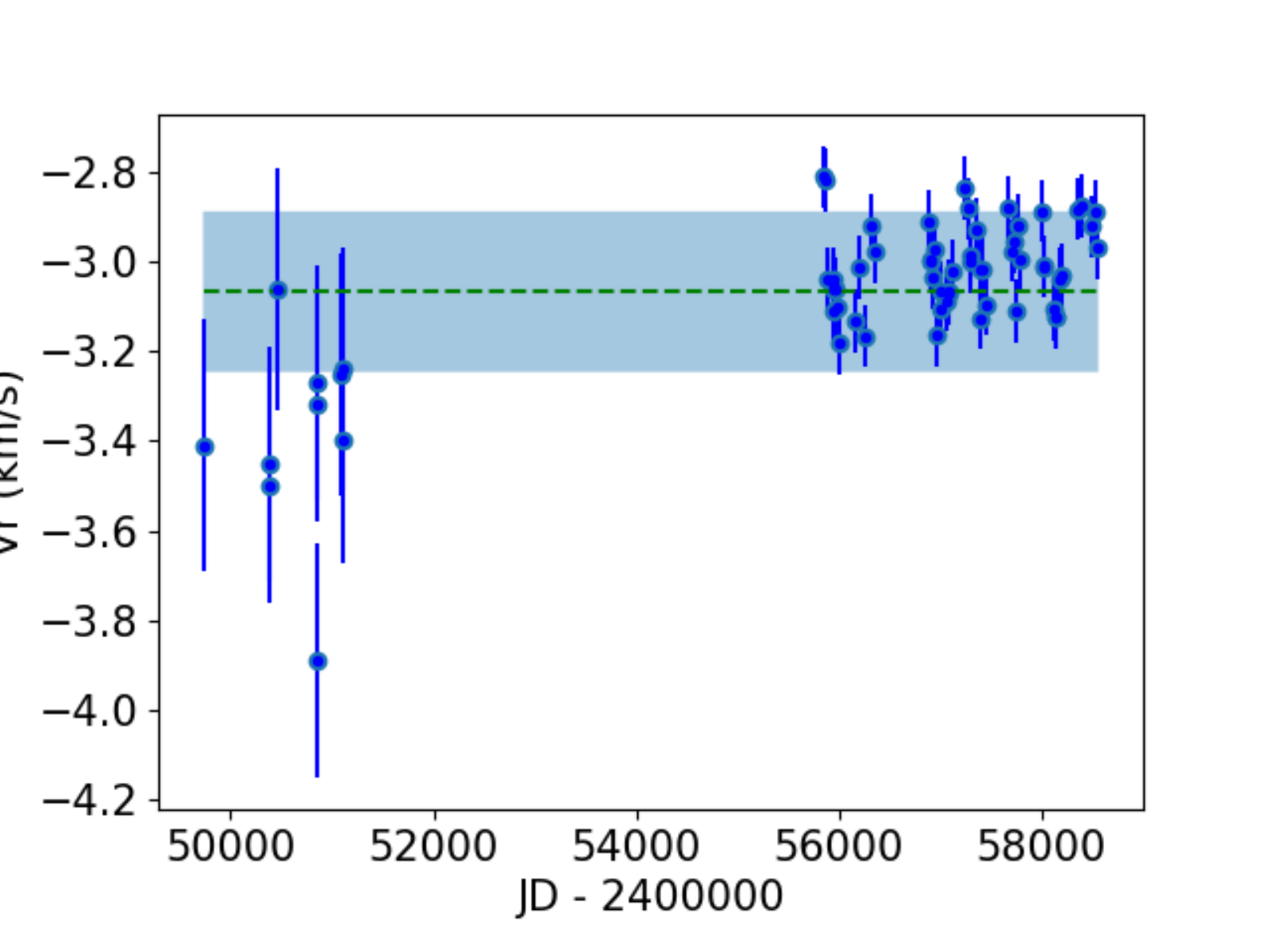}
    \includegraphics[width=9.5cm]{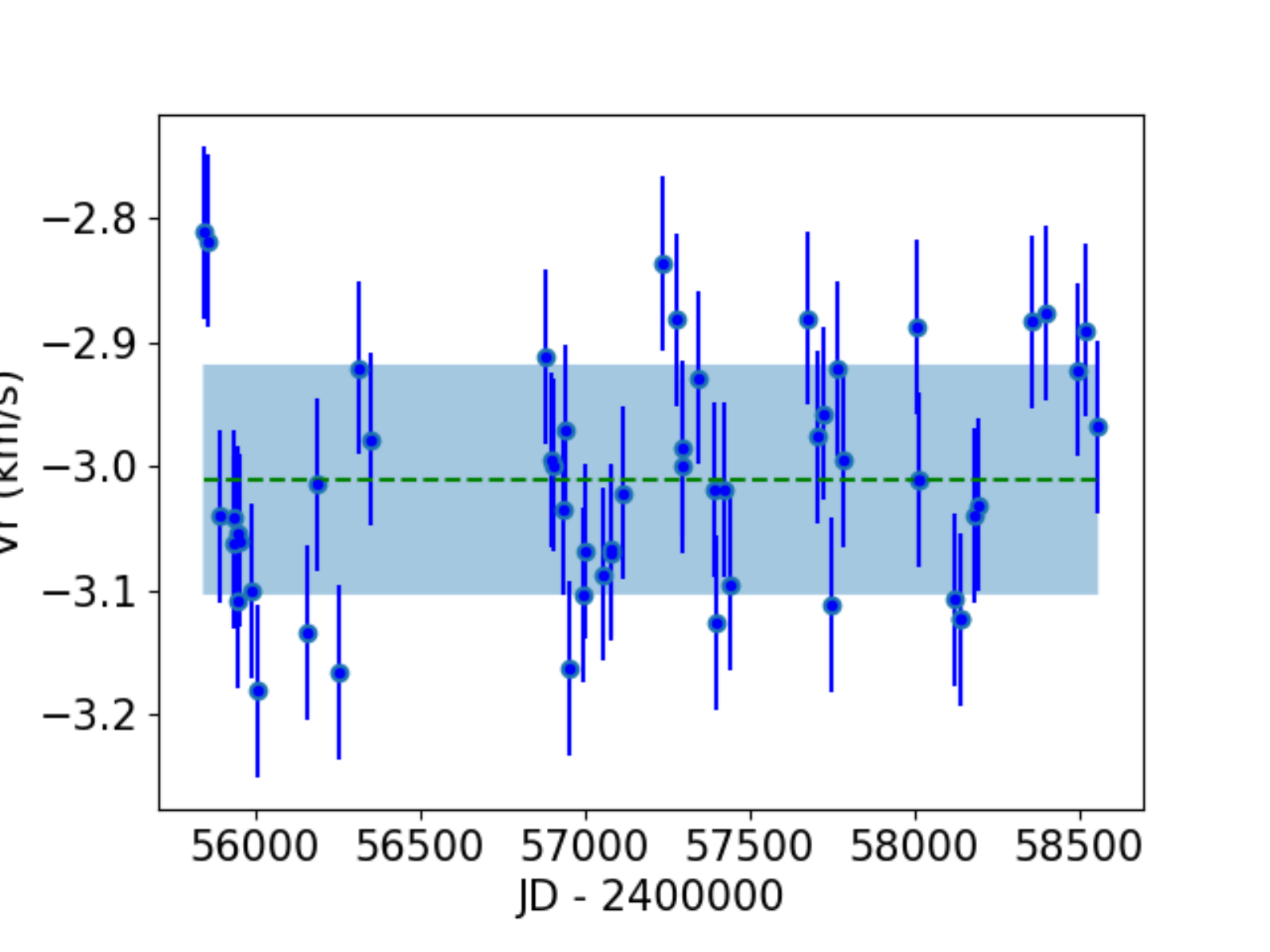}
    \caption{Same as Fig.~\ref{Fig:30834} for HD~34043.}
    \label{Fig:34043}
\end{figure}

\begin{figure}
    \includegraphics[width=9.5cm]{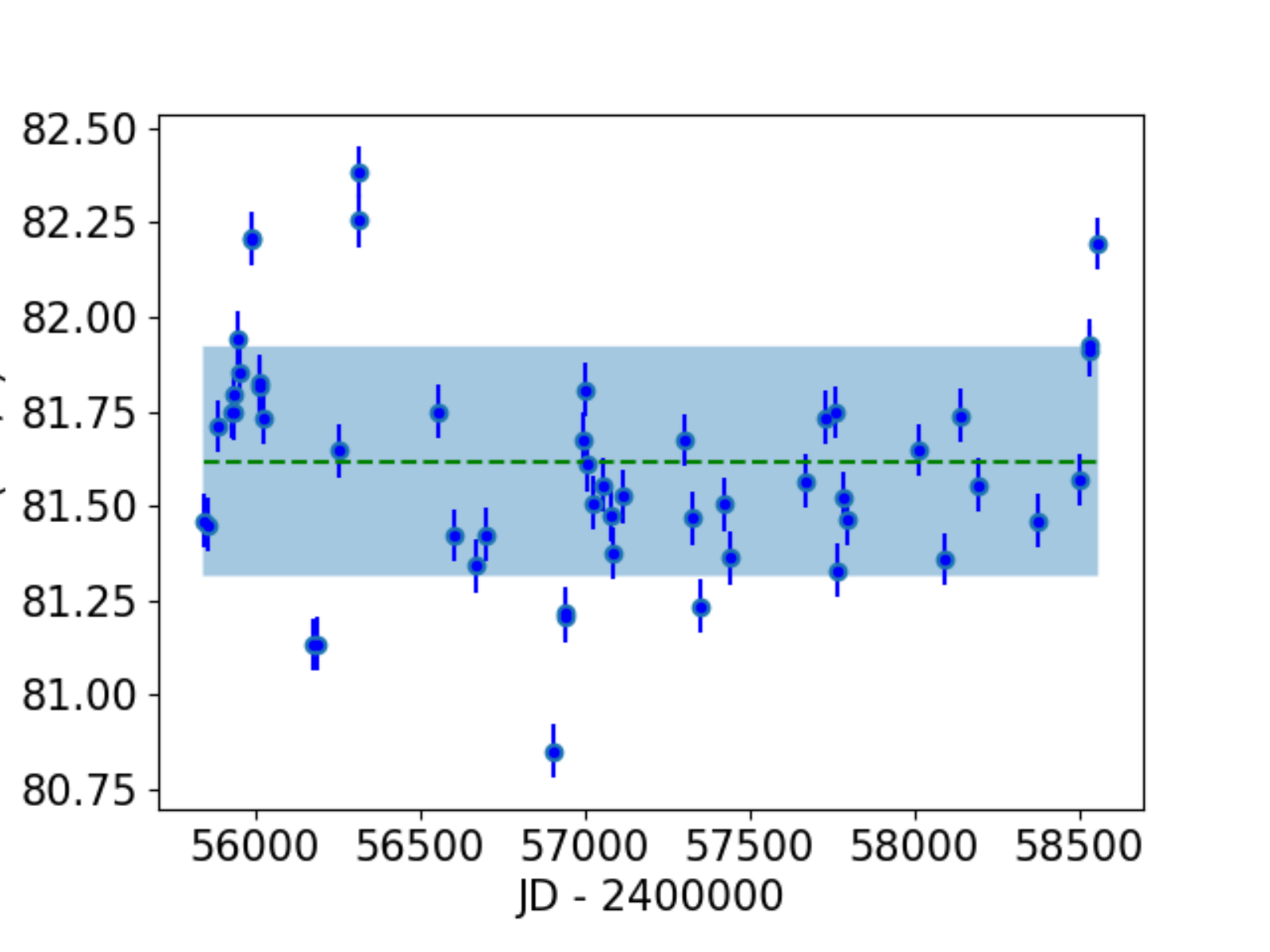}
    \caption{Radial-velocity data for HD~39853, flagged as a binary by the $\chi^2$ test. }
    \label{Fig:39853a}
\end{figure}

\begin{figure}
        \includegraphics[width=9.5cm]{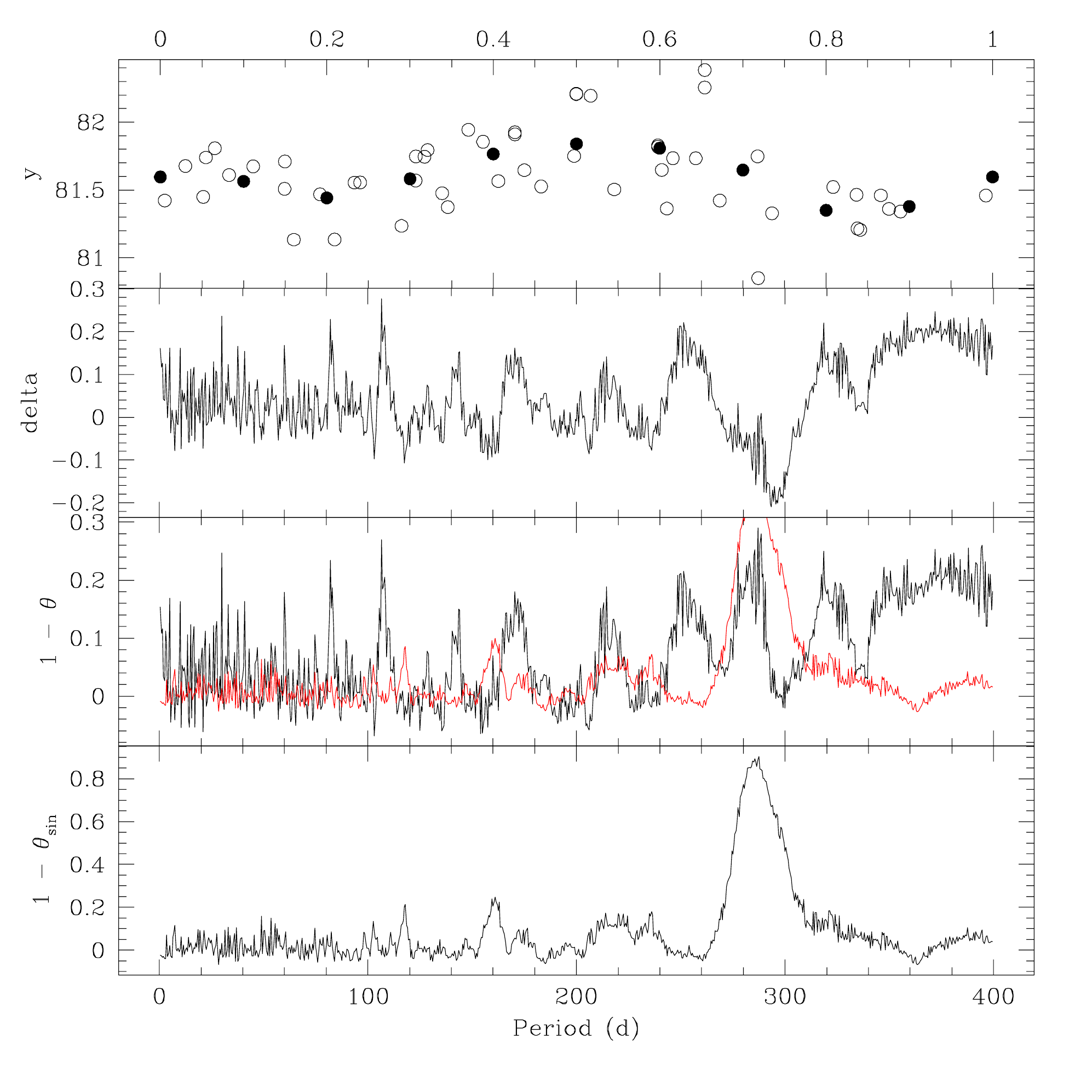}
    \caption{Period analysis of the RVs using the Stellingwerf statistics. Top panel: RVs plotted in phase with a period of 282~d. Bottom panel: The Stellingwerf $1 - \theta$ statistical indicator for a sine signal of period 282~d sampled the same way as HD 39853. Middle bottom panel: Same as bottom for the 282~d sine signal (red curve) and for the HD 39853 data (black curve). Middle top panel: The difference between the red and black curve of the previous panel, showing possible residual signals (here one around 110~d).}
    \label{Fig:39853b}
\end{figure}

\begin{figure}
    \includegraphics[width=9.5cm]{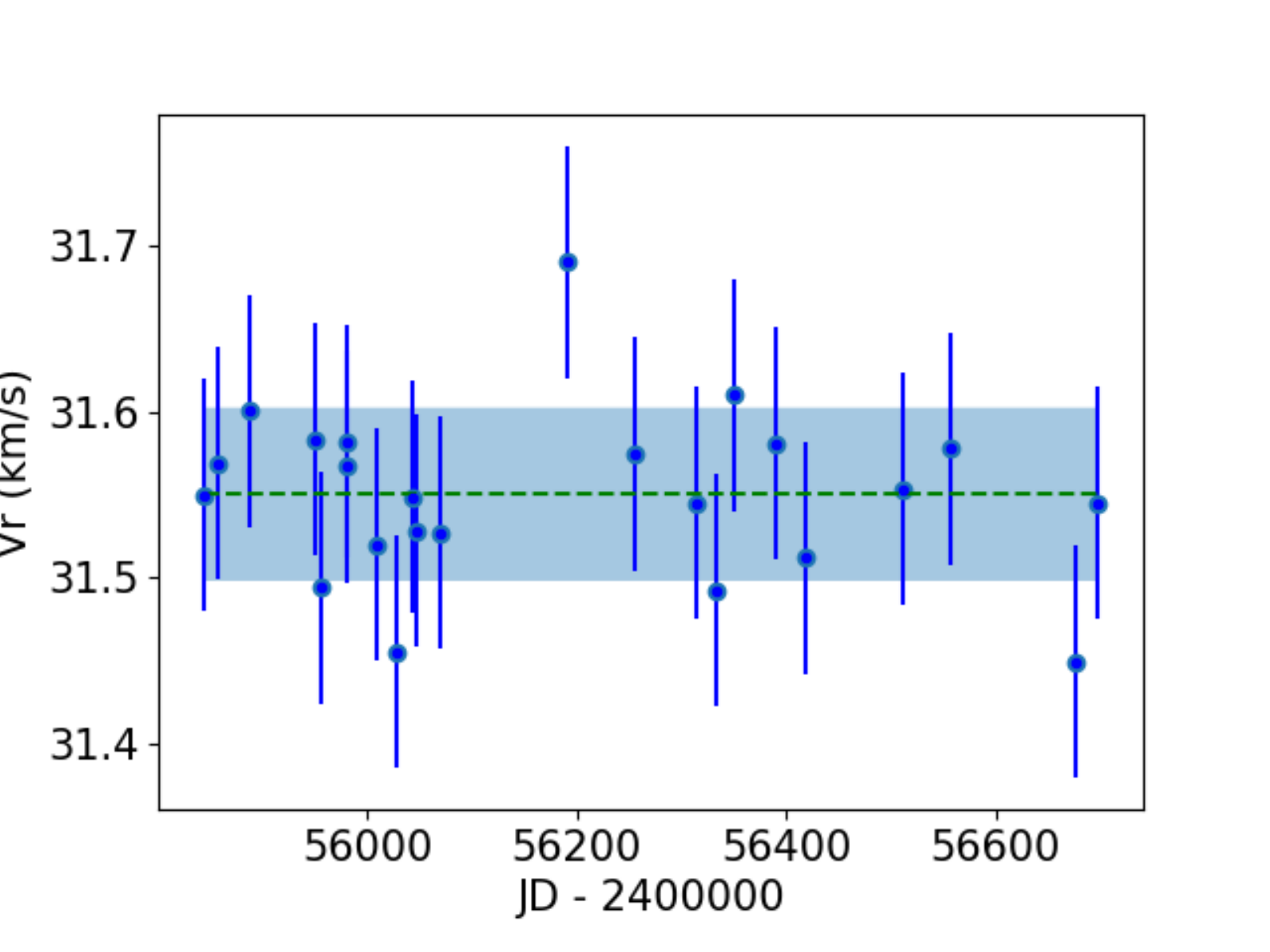}
    \caption{Same as Fig.~\ref{Fig:6} for HD~40827.}
    \label{Fig:40827}
\end{figure}

\begin{figure}
    \includegraphics[width=9.5cm]{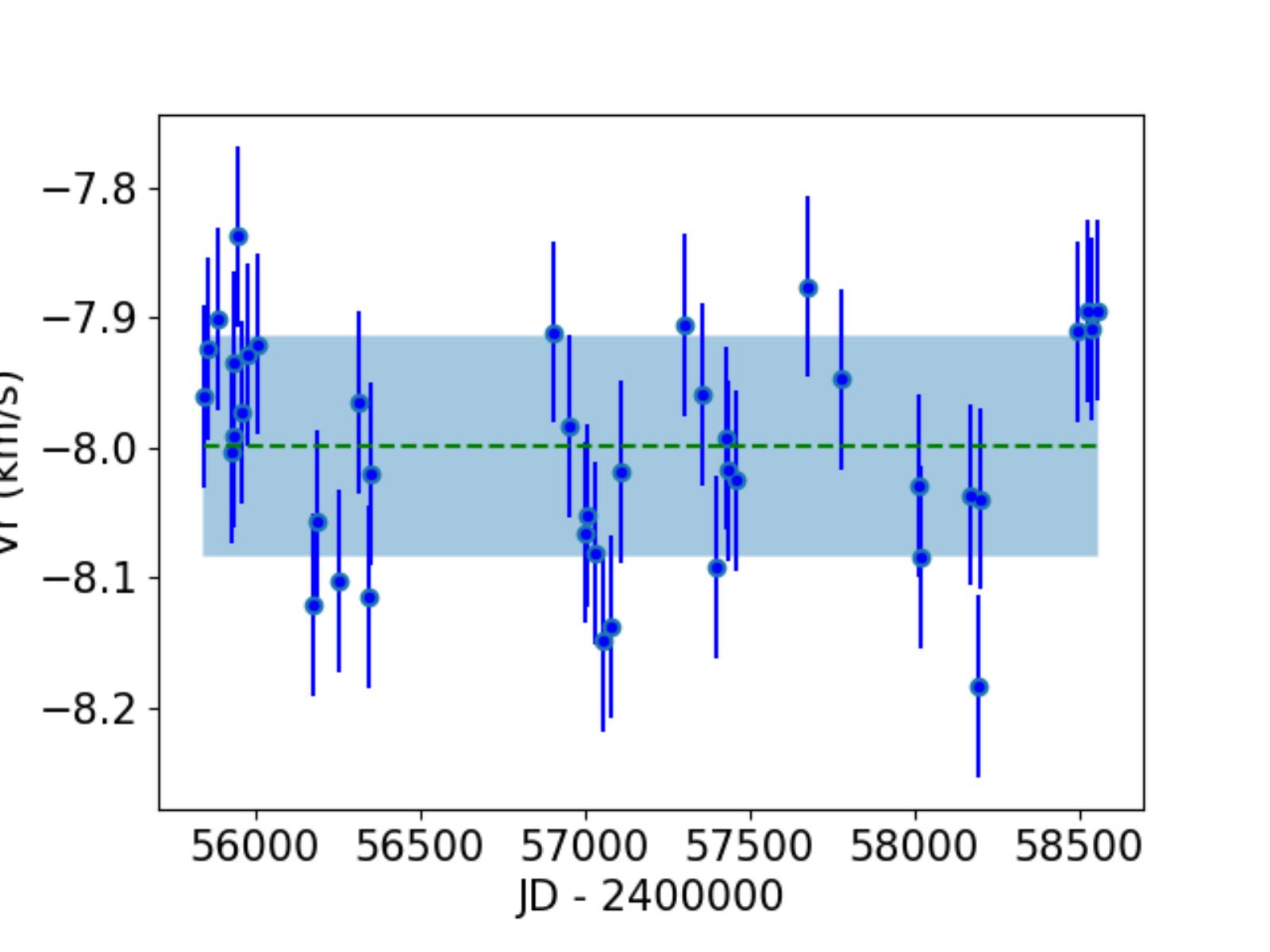}
    \caption{Same as Fig.~\ref{Fig:6}  for HD~43827.}
    \label{Fig:43827}
\end{figure}

\begin{figure}
    \includegraphics[width=9.5cm]{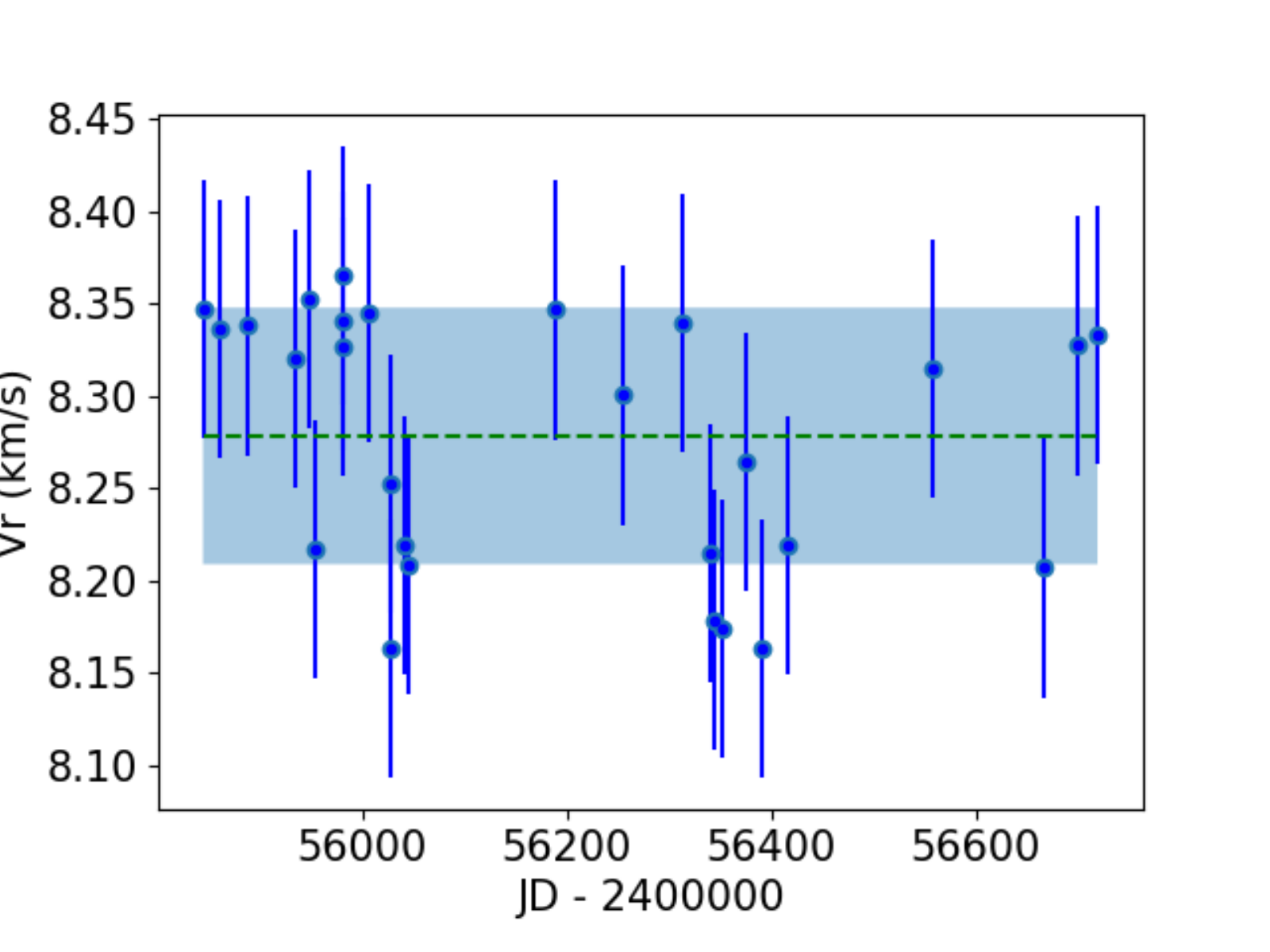}
    \caption{Same as Fig.~\ref{Fig:6}  for HD~63798. 
    }
    \label{Fig:63798}
\end{figure}

\begin{figure}
    \includegraphics[width=9.5cm]{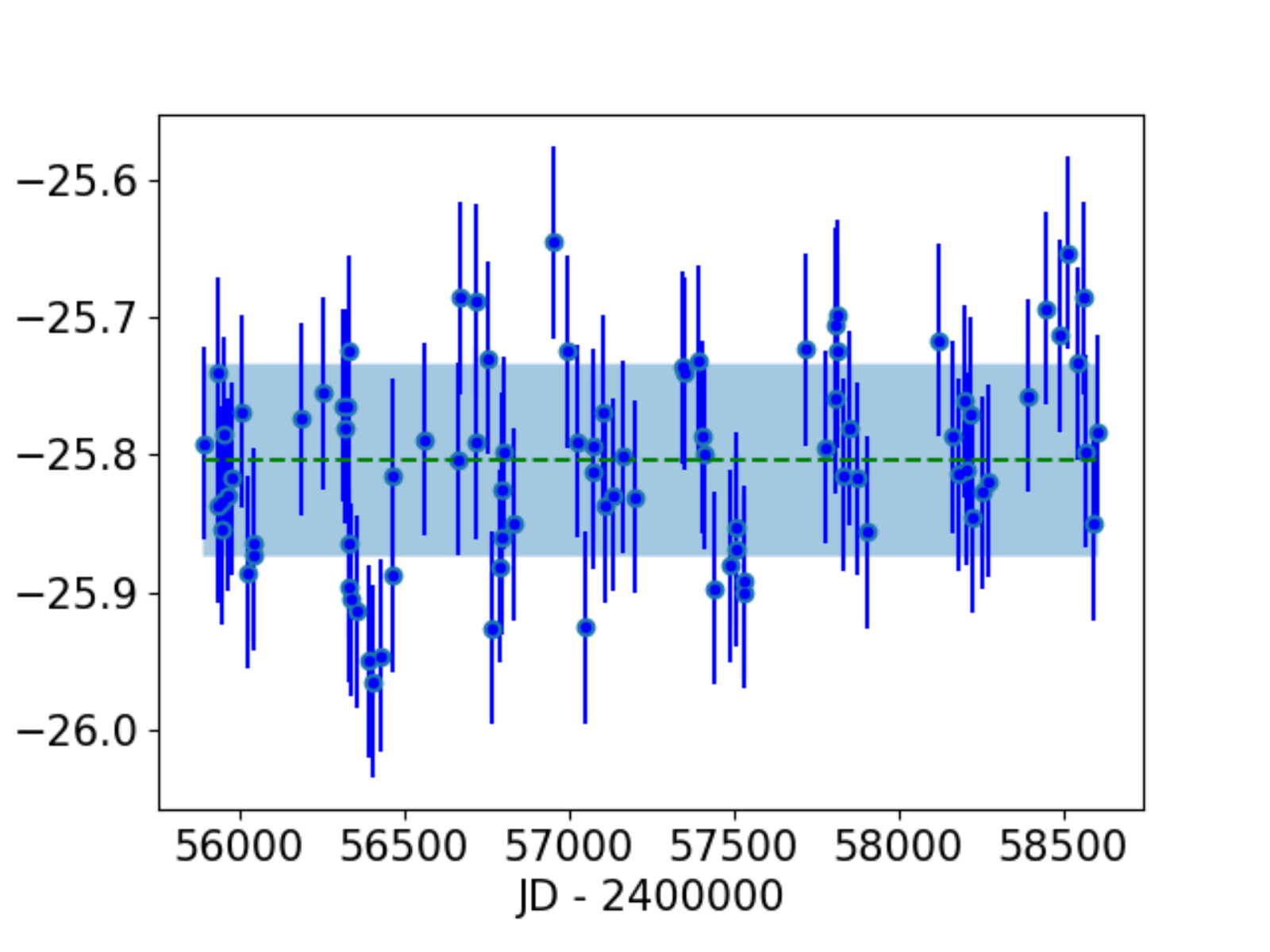}
    \caption{Same as Fig.~\ref{Fig:6}  for HD~90633. }
    \label{Fig:90633}
\end{figure}

\begin{figure}
    \includegraphics[width=9.5cm]{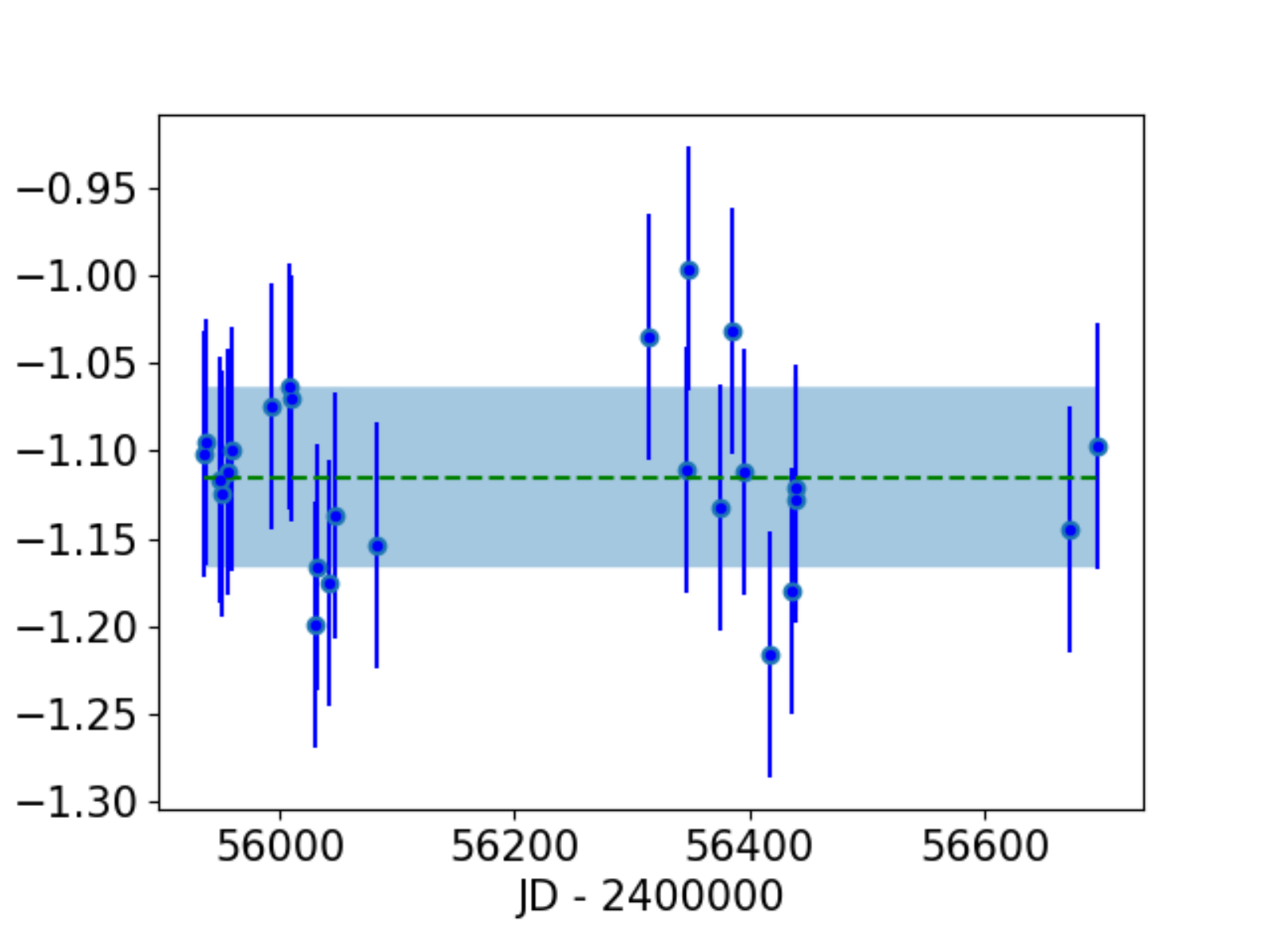}
    \caption{Same as Fig.~\ref{Fig:6}  for HD~108741. 
    }
    \label{Fig:108741}
\end{figure}

\begin{figure}
    \includegraphics[width=9.5cm]{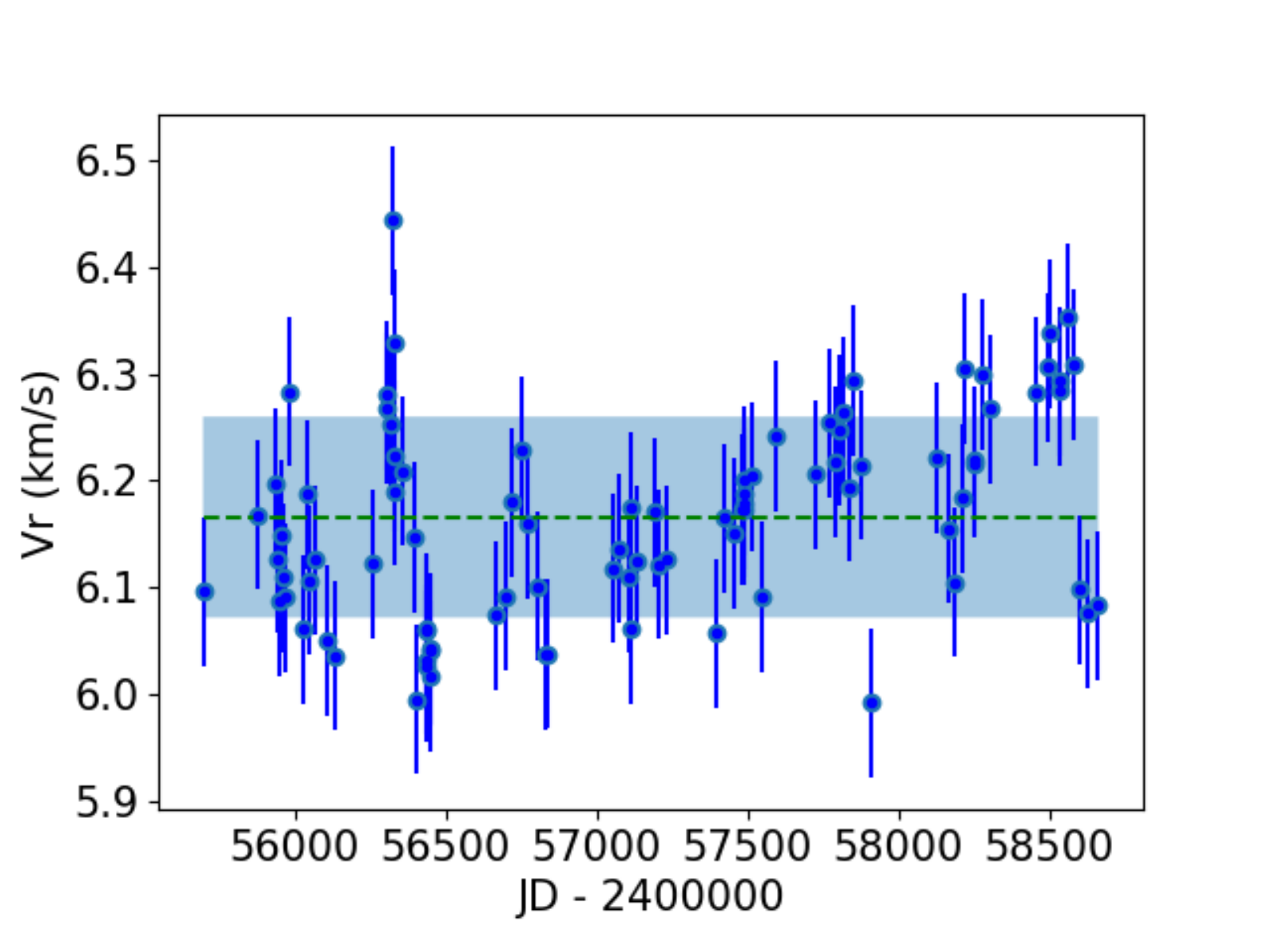}
    \caption{Same as Fig.~\ref{Fig:6}  for HD~112127. 
    This star is sometimes classified as a carbon star of type R \protect\citep{Barnbaum1996}.}
    \label{Fig:112127}
\end{figure}

\begin{figure}
    \includegraphics[width=9.5cm]{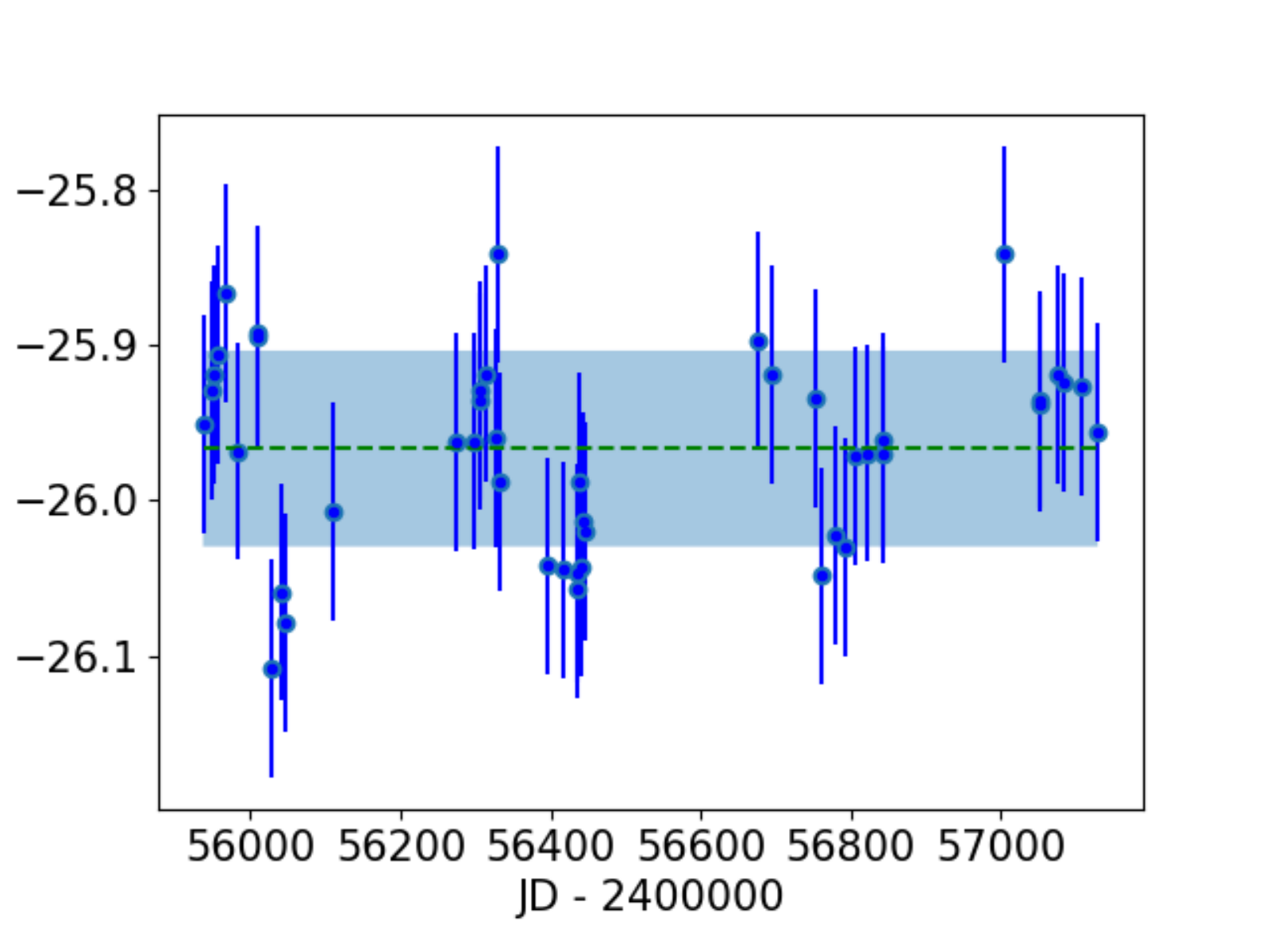}
    \caption{Same as Fig.~\ref{Fig:6}  for HD~116292.}
    \label{Fig:116292}
\end{figure}

\begin{figure}
    \includegraphics[width=9.5cm]{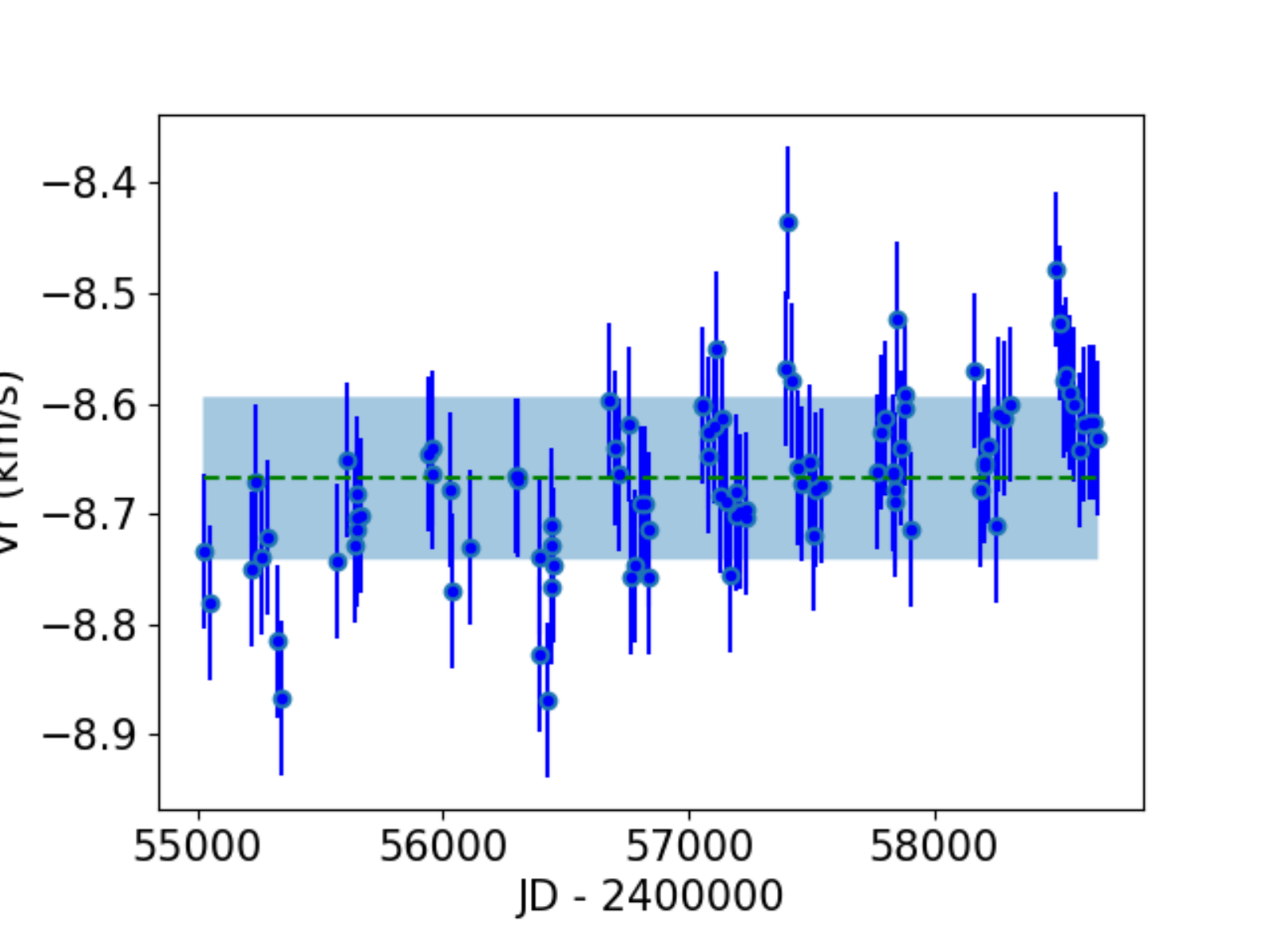}
    \caption{Same as Fig.~\ref{Fig:6}  for HD~119853, showing a long-term trend.}
    \label{Fig:119853}
\end{figure}

\begin{figure}
    \includegraphics[width=9.5cm]{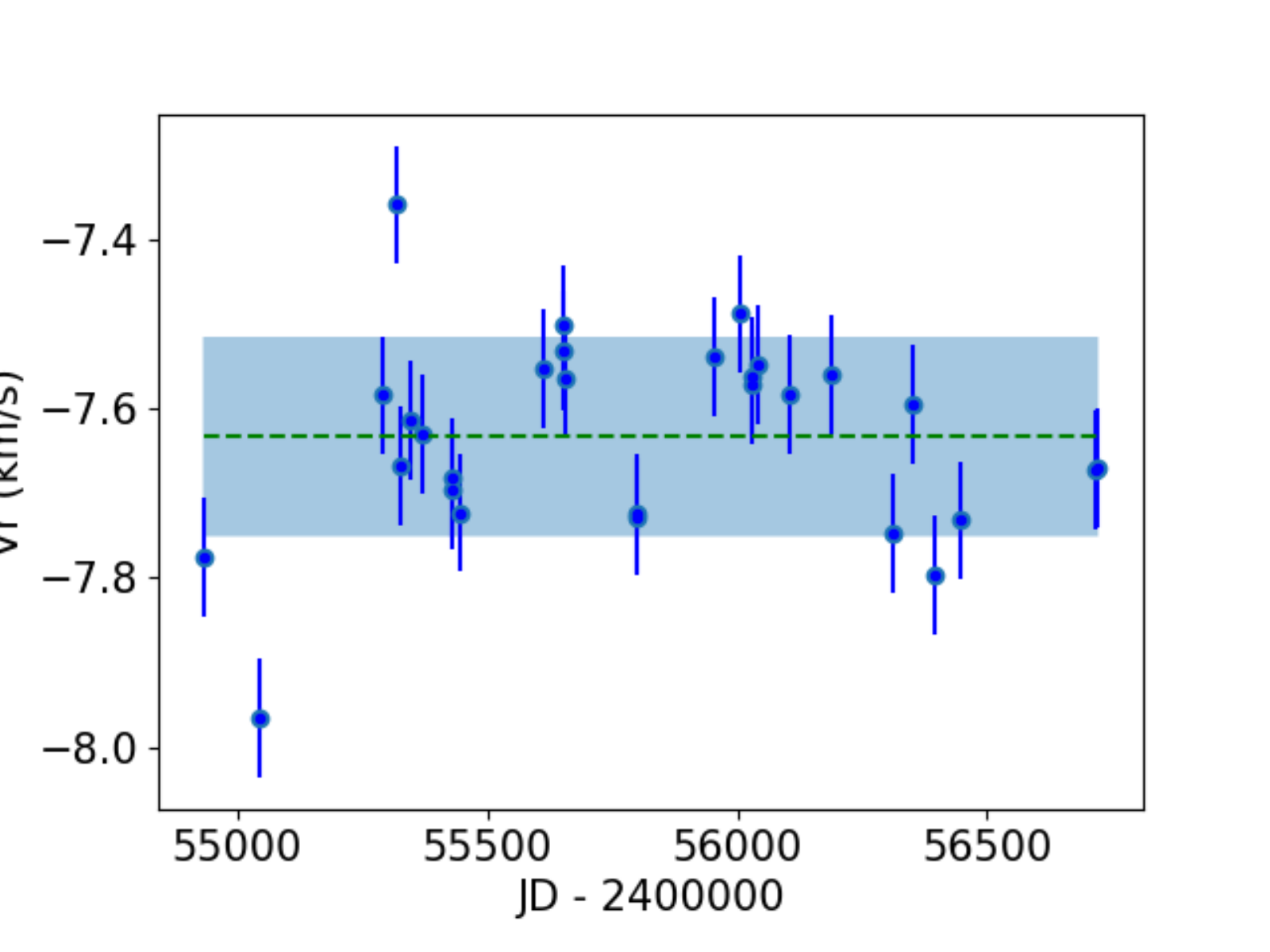}
    \caption{Same as Fig.~\ref{Fig:6}  for HD~153687.}
    \label{Fig:153687}
\end{figure}

\begin{figure}
    \includegraphics[width=9.5cm]{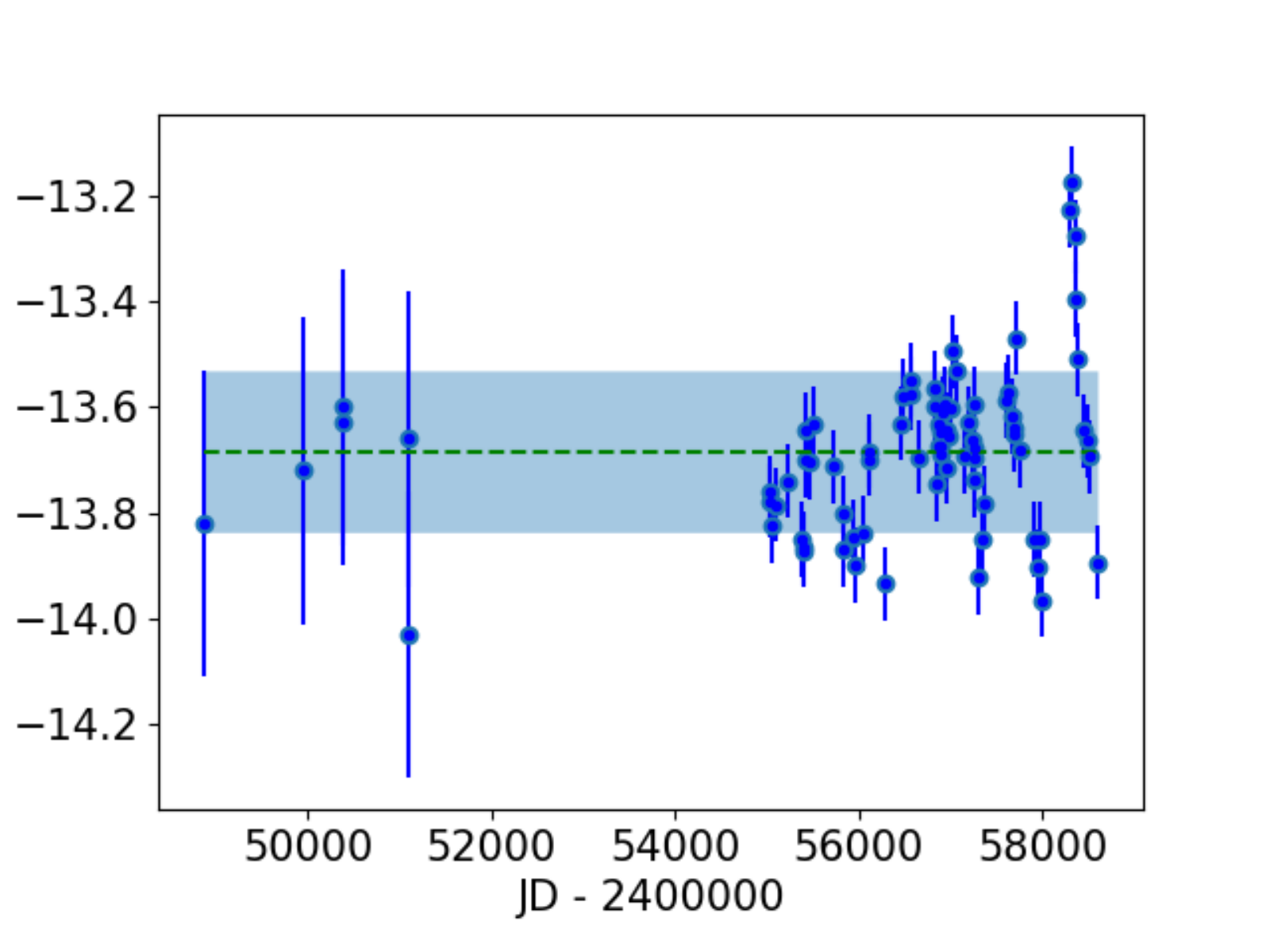}
    \includegraphics[width=9.5cm]{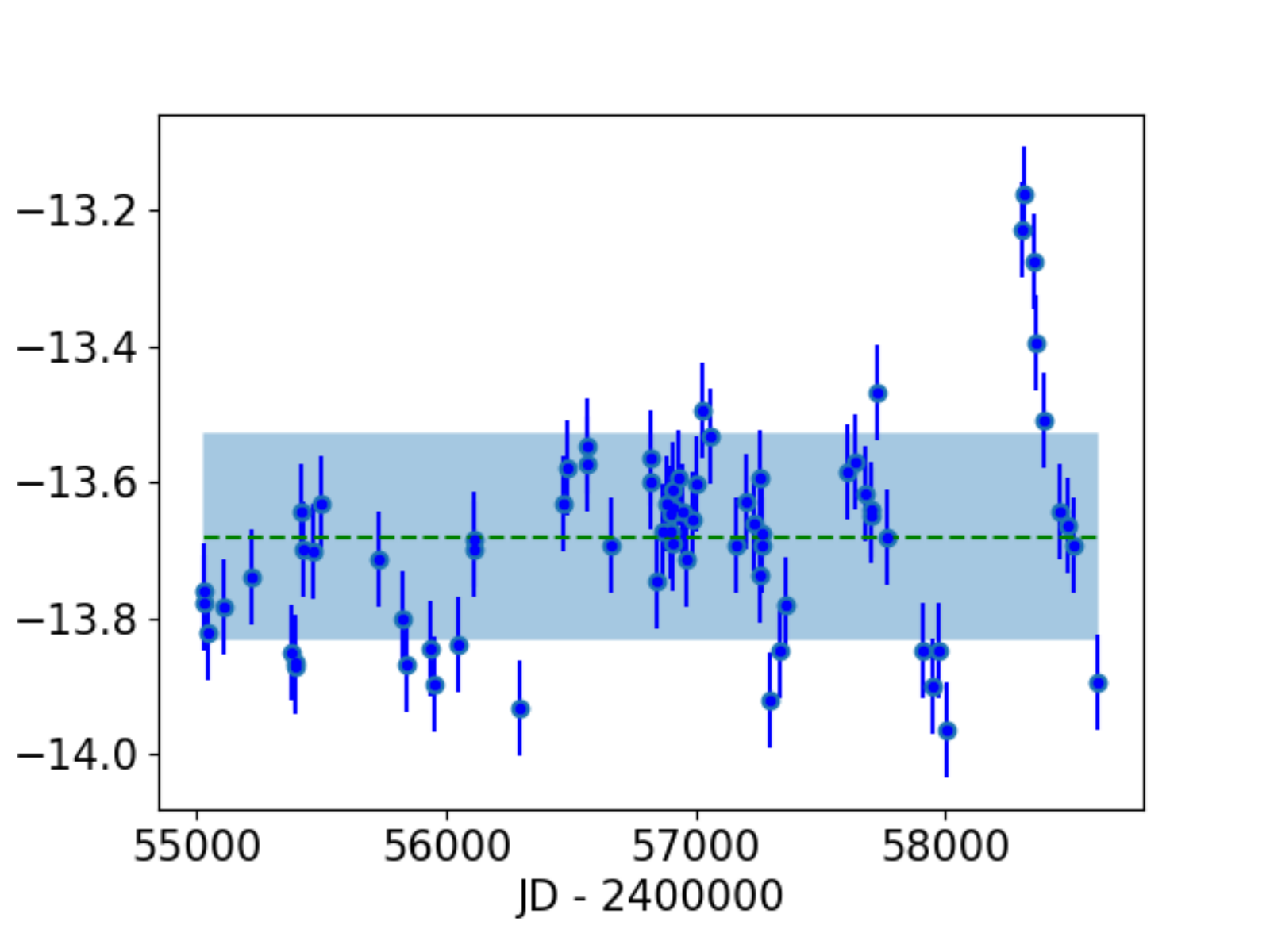}
    \caption{Same as Fig.~\ref{Fig:6}  for HD~221776, showing no sign for binarity despite a probability unity of having a variable RV (Table~\protect\ref{Tab:binary}). The top panel also displays  six RV measurements
     from \citet{Famaey2005}. The bottom panel shows only the HERMES RV.}
    \label{Fig:221776}
\end{figure}

\begin{figure}
    \includegraphics[width=9.5cm]{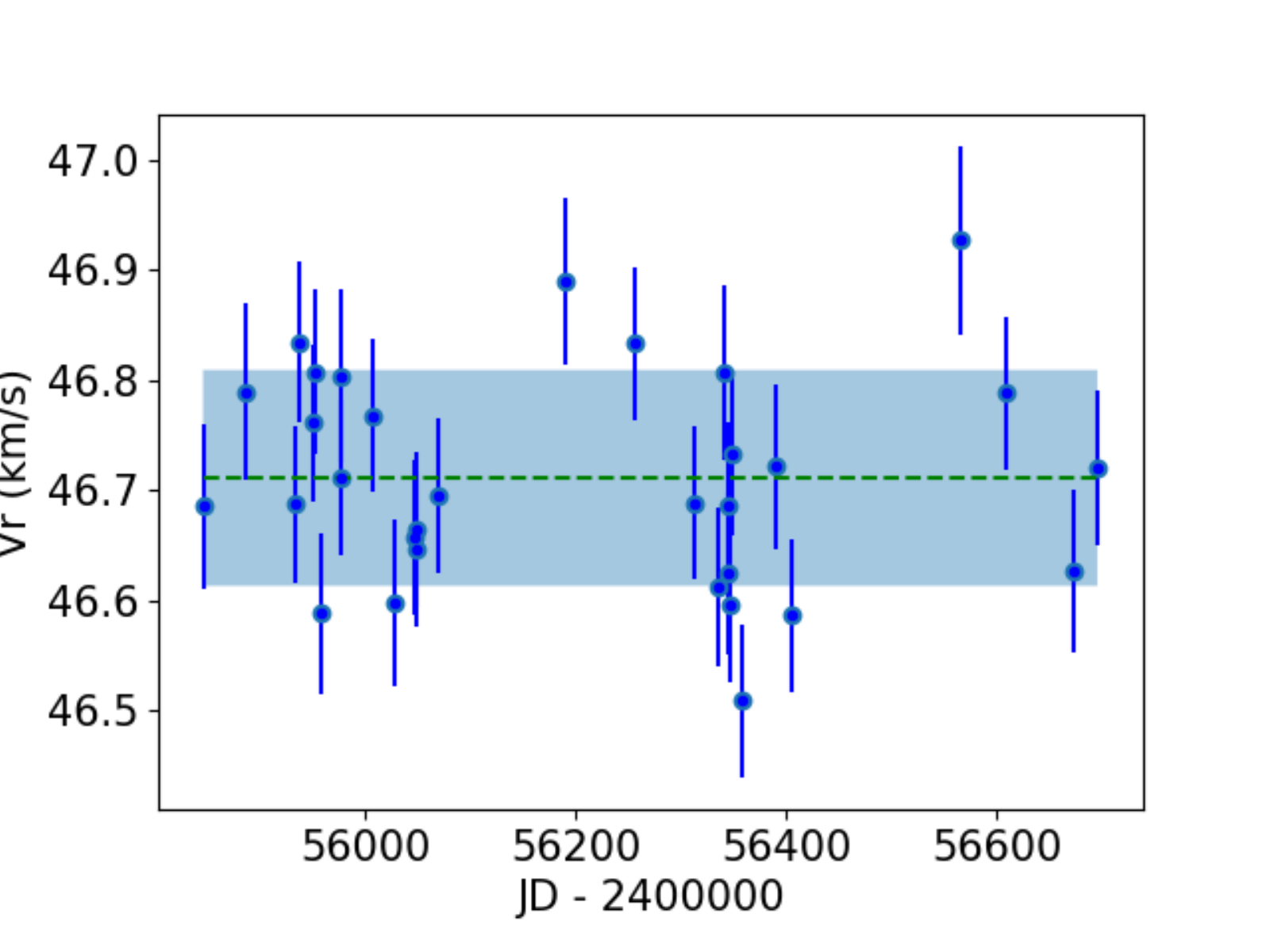}
    \caption{Same as Fig.~\ref{Fig:6}  for HD~233517.}
    \label{Fig:233517}
\end{figure}

This section presents the RV curves of all target  stars (Figs.~\ref{Fig:6} -- \ref{Fig:233517}).

\section{Orbital solutions}
\label{Appendix:orbits}

This section presents the orbital solutions found so far, displayed in Figs.~\ref{Fig:Orb_787} -- \ref{Fig:Orb_212320}, as well as notes on individual stars.

\citet{Famaey2005} data have been used to improve the orbit of the very long-period binary HD~3627. \citet{Bakos1976} proposed a preliminary orbit based on observations spanning 75 years, but only provides a preliminary period of 15\ts000~ d, whereas the Ninth Catalogue of Orbits of Spectroscopic Binaries \citep{Pourbaix2004} rederived an orbital period of 20158~d from the observations collected by \citet{Bakos1976}.

\citet{Griffin2013} provides for HD 27497 an orbit very similar to that derived from our HERMES data, as listed in Table~\ref{Tab:orbits}.
For HD~31553, measurements from \citet{Fekel1998} have been added to the HERMES data set in order to better constrain the orbital period, but no zero-point offset has been considered.

HD~21078 (= HIP~15769) is a high proper-motion star (LTT~1604) flagged as an acceleration solution (DMSA/G) in the Double and Multiple Star Annex of the Hipparcos catalogue  \citep{ESA1997}.
It is also a so-called $\Delta\mu$ binary (long-term Tycho-2 and short-term Hipparcos proper motions are discrepant) flagged by \citet{Makarov2005} and \citet{Frankowski2007}.
We computed a combined HERMES/Hipparcos spectroscopic/astrometric solution  following the method outlined by \citet{Jancart2005}. All significance tests described in the latter paper are satisfied, indicating the good quality of the combined orbit whose elements are listed in Table~\ref{Tab:orbits}c. We also note the good agreement between the dynamical parallax of  $13.3\pm1.1$~mas (Table~\ref{Tab:orbits}c) and the Gaia DR2 parallax of  $12.5\pm0.1$~mas (Table~\ref{Tab:parallax}).
The knowledge of the orbital inclination allows us to derive pairs of possible ($M_1, M_2$) values (listed in Table~\ref{Tab:orbits}c) from the spectroscopic mass function. 
The location of HD~21078 in the HR diagram (Sect.~\ref{Sect:HRD}) points at a mass $\sim 1.2$~\Msun\ for component A, and therefore the companion must have a mass of 0.71~\Msun. Such a mass is compatible with the companion being either a main-sequence star or a white dwarf \citep[the latter possibility is less likely since the K giant should then appear as a barium star, which it is not; see][]{Merle2016}. The low-mass of the companion implies that it contributes negligible light to the system, and therefore the photocentre location must be identical 
with the giant location. Hence the semi-major axis of the giant around the centre of mass of the system ($a_1$) must be identical with the semi-major axis of the photocentric orbit ($a_0$), as confirmed from Tables~\ref{Tab:orbits}ac.

\begin{figure}
\includegraphics[width=9cm]{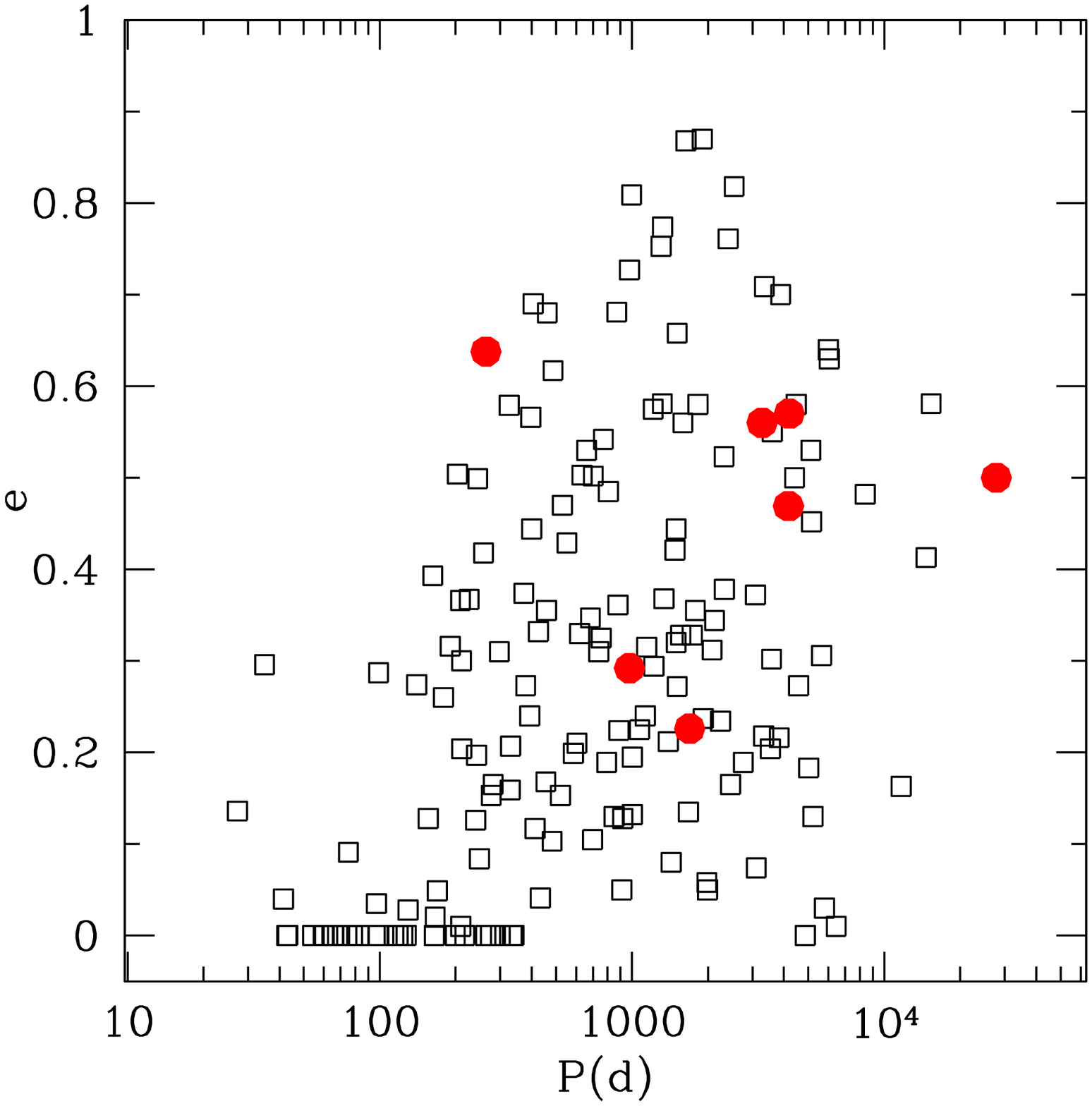}
\caption{\label{Fig:eP}
Eccentricity--period diagram for the new orbits derived in this paper (red filled circles), as compared to the sample of K giants in open clusters (open squares) from \citet{Mermilliod2007}.
}
\end{figure}

Figure~\ref{Fig:eP} presents the eccentricity--period diagram for the new orbits derived in this paper (red filled circles), and is compared to the sample of K giants in open clusters from \citet{Mermilliod2007}.  All binaries from sample S1 fall within the locus occupied by the binary K giants in open clusters (even though HD~21078 lies along its boundary; this is compatible with the fact that HD~21078 is a subgiant rather than a giant -- as revealed by Fig.~\ref{Fig:HRD} -- and therefore the tidal orbital circularisation has not yet operated). Therefore, there is nothing peculiar to report about the orbital dynamics of our sample stars.

\begin{figure}
    \includegraphics[width=9cm]{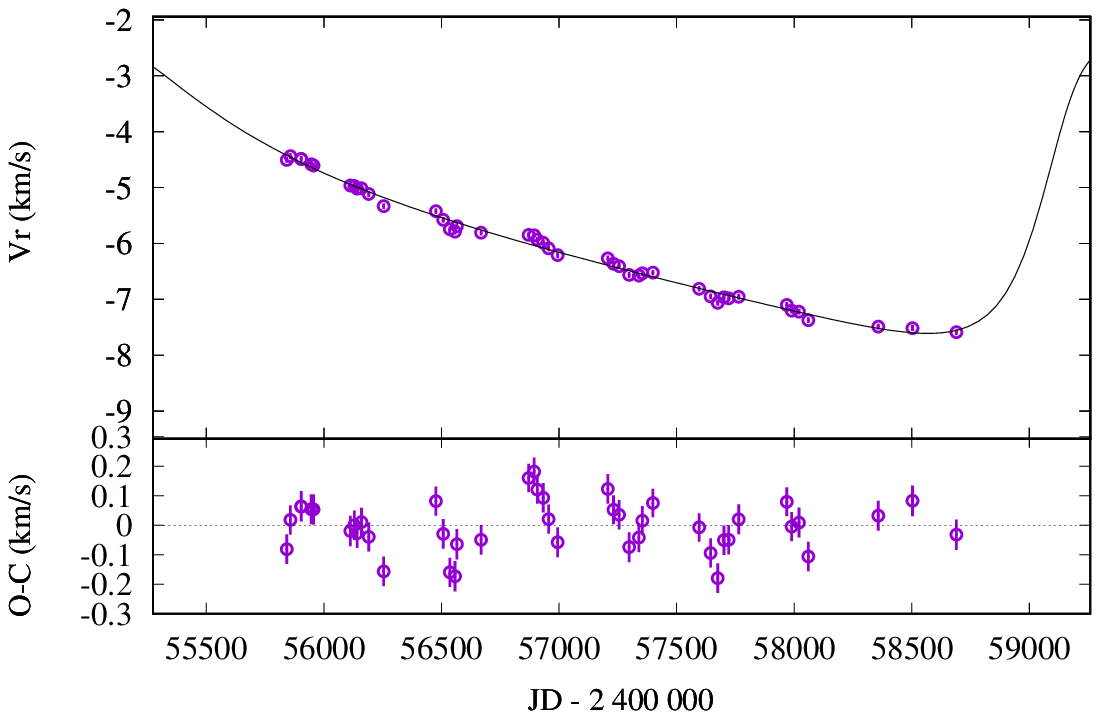}
        \includegraphics[width=9cm]{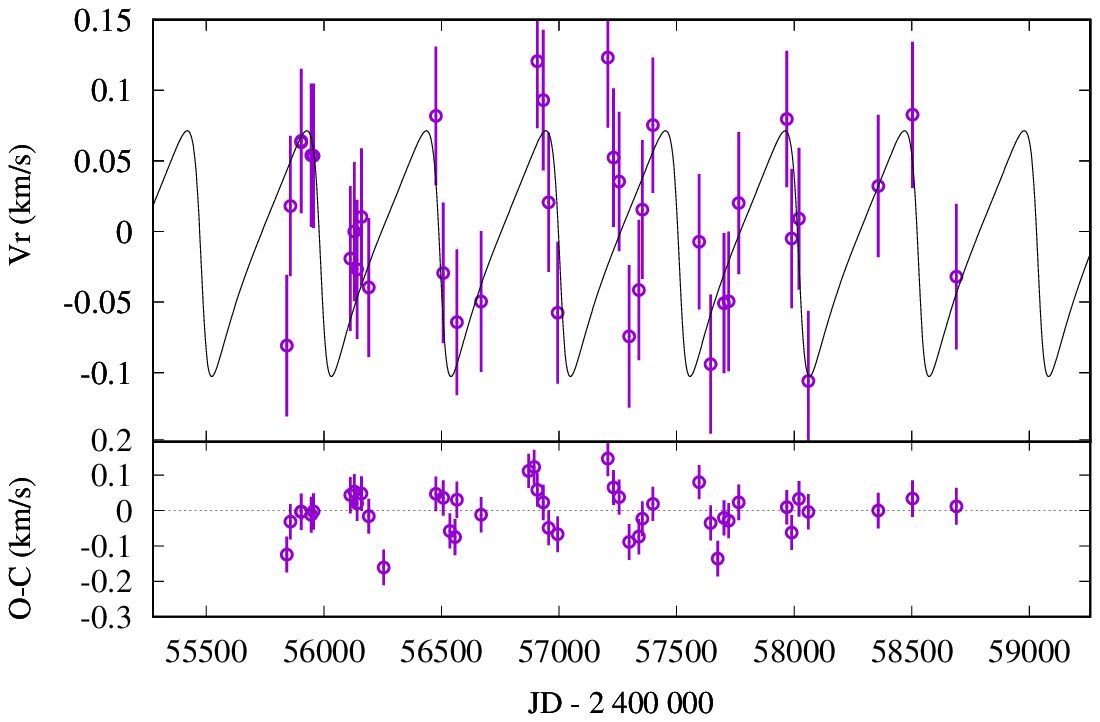}
    \caption{Top panel: Preliminary orbital solution for HD~787 (with the shortest possible orbital period of 4200~d). Bottom panel: Orbital solution fitted to the residuals of the top panel.}
    \label{Fig:Orb_787}
\end{figure}

\begin{figure}
    \includegraphics[width=9cm]{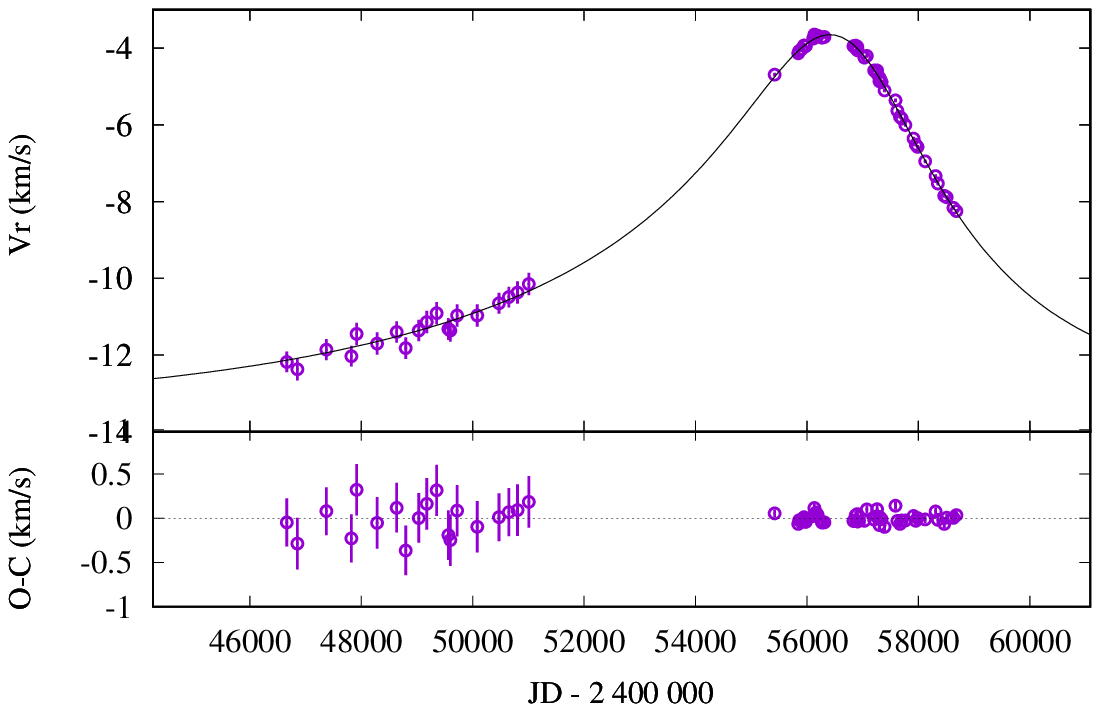}
    \caption{Same as Fig.~\ref{Fig:Orb_787} for HD~3627, with a tentative period of the order of 76~yr. The older measurements are from \citet{Famaey2005}.}
    \label{Fig:Orb_3627}
\end{figure}

\begin{figure}
    \includegraphics[width=9cm]{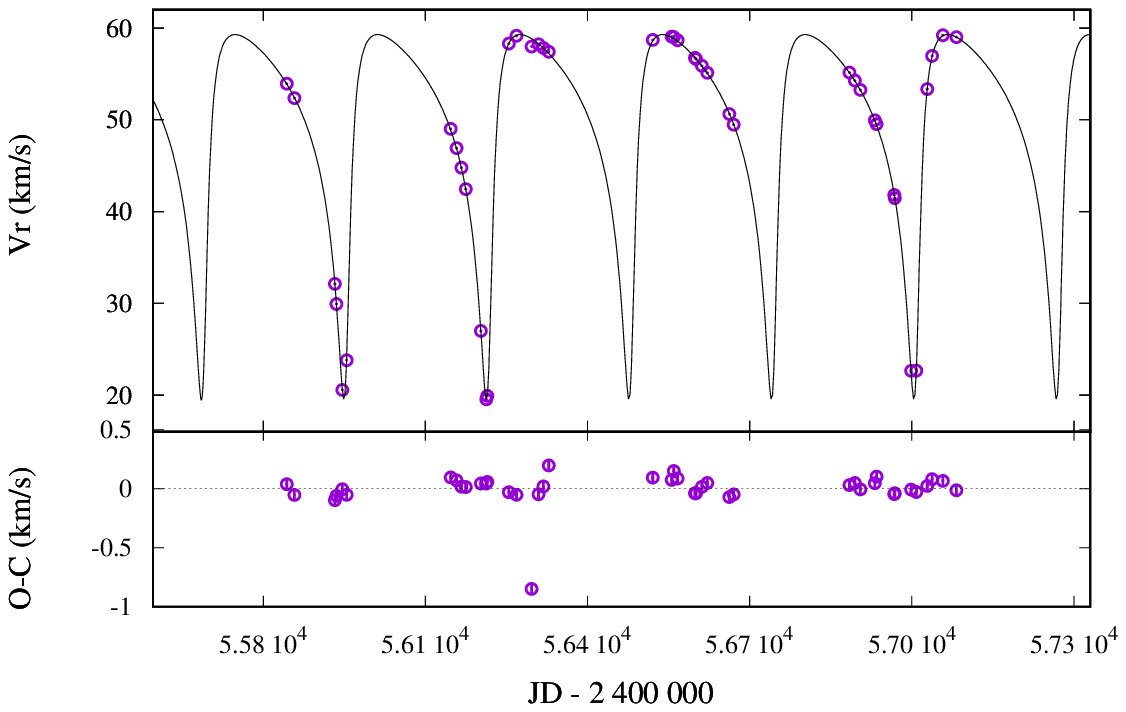}
    \caption{Same as Fig.~\ref{Fig:Orb_787} for HD~21078. }
    \label{Fig:Orb_21078}
\end{figure}

\begin{figure}
    \includegraphics[width=9cm]{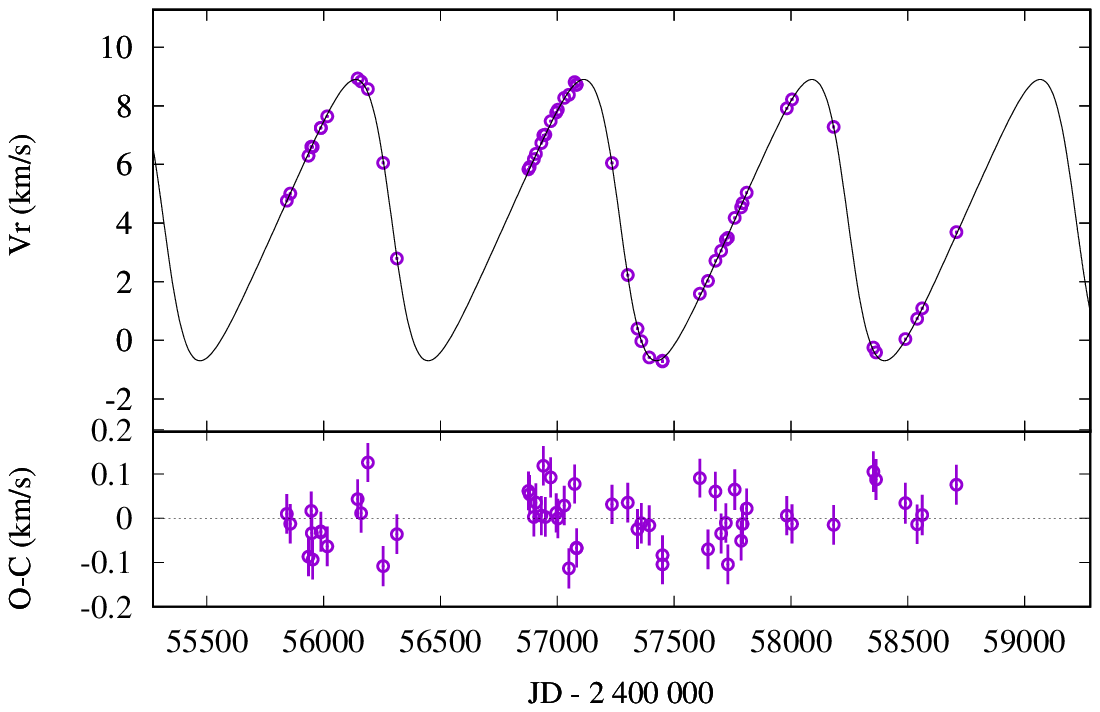}
    \caption{Same as Fig.~\ref{Fig:Orb_787} for HD~27497. }
    \label{Fig:Orb_27497}
\end{figure}

\begin{figure}
    \includegraphics[width=9cm]{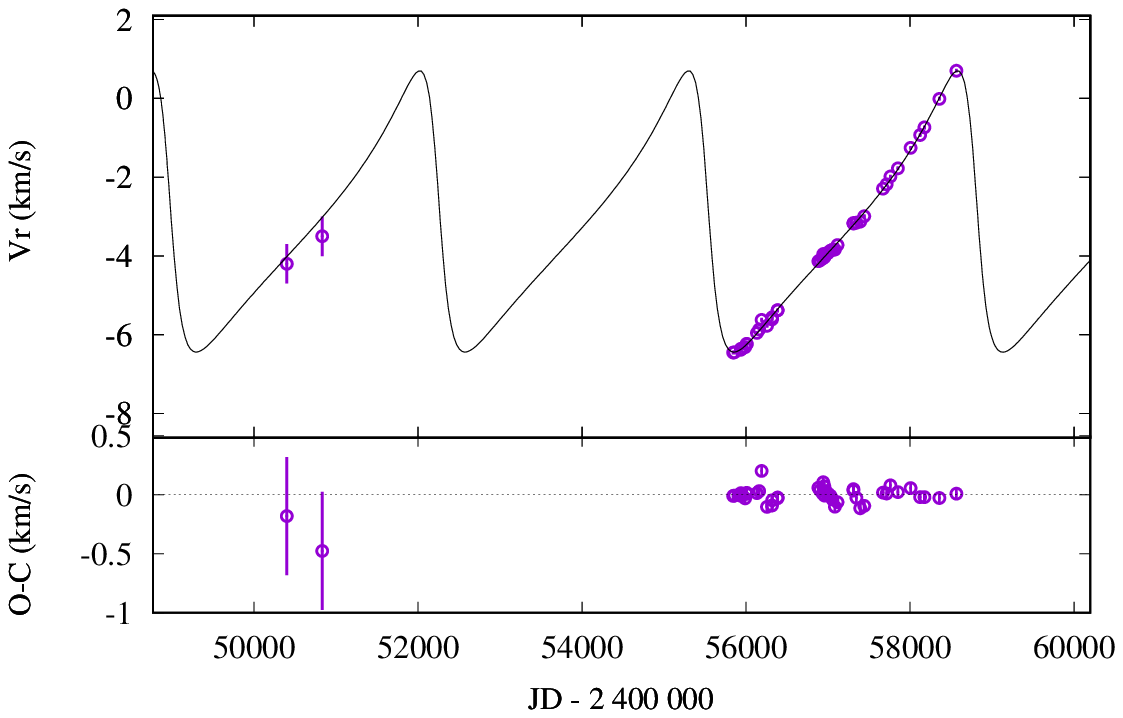}
    \caption{Same as Fig.~\ref{Fig:Orb_787} for HD~31553, including \protect\citet{Fekel1998} measurements. }
    \label{Fig:Orb_31553}
\end{figure}

\begin{figure}
    \includegraphics[width=9cm]{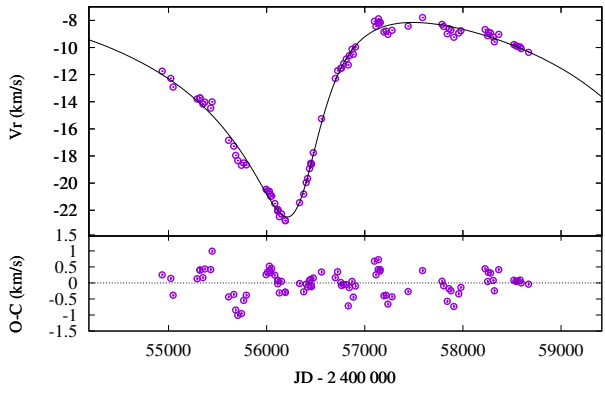}
    \includegraphics[width=9cm]{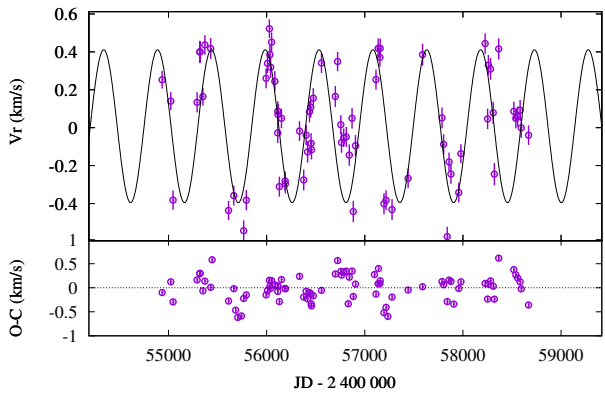}
    \caption{Top panel: Same as Fig.~\ref{Fig:Orb_787} for HD~156115. Bottom panel: Orbital solution fitted to the residuals of the top panel.}
    \label{Fig:Orb_156115}
\end{figure}

\begin{figure}
    \includegraphics[width=9cm]{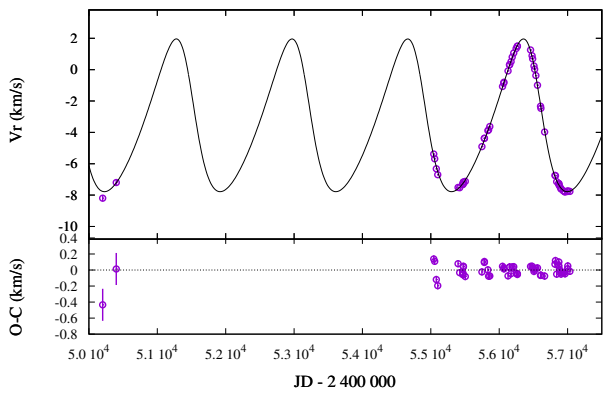}
    \caption{Same as Fig.~\ref{Fig:Orb_787} for HD~212320. }
    \label{Fig:Orb_212320}
\end{figure}

\end{appendix}

\end{document}